\documentclass[aps,prx,twocolumn,nofootinbib,longbibliography]{revtex4-1}
\usepackage{times}
\usepackage{graphicx}
\usepackage{tabularx}
\usepackage{amsmath}
\usepackage{amstext}
\usepackage{amssymb}
\usepackage{xfrac}
\usepackage[colorlinks,citecolor=blue]{hyperref}
\usepackage{graphicx}
\usepackage{amsmath}
\usepackage{amstext}
\usepackage{amssymb}
\usepackage{amsfonts}
\usepackage{longtable,booktabs}
\usepackage{hyperref}
\usepackage{url}
\usepackage{subfigure}
\usepackage{dsfont}
\usepackage{booktabs}
\usepackage{amsbsy}
\usepackage{dcolumn}
\usepackage{amsthm}
\usepackage{bm}
\usepackage{esint}
\usepackage{multirow}
\usepackage{hyperref}
\usepackage{cleveref}
\usepackage{mathrsfs}
\usepackage{amsfonts}
\usepackage{amsbsy}
\usepackage{dcolumn}
\usepackage{bm}
\usepackage{multirow}
\usepackage{color}
\usepackage{extarrows}
\usepackage{datetime}
\usepackage{comment}
\usepackage[super]{nth}
\usepackage{tikz-cd}

\hypersetup{
    colorlinks=magenta,
    linkcolor=blue,
    filecolor=magenta,
        urlcolor=magenta,
}

\newcommand{\comments}[1]{}

\def\Z{\mathbb{Z}}

\begin{document}
\title{ Compatible  braidings    with Hopf Links, multiloop, and Borromean rings in (3+1)-dimensional spacetime}
\author{Zhi-Feng Zhang}
\author{Peng Ye}
\email{yepeng5@mail.sysu.edu.cn}
\affiliation{School of Physics and State Key Laboratory of Optoelectronic Materials and
Technologies, \\ Sun Yat-sen University, Guangzhou, 510275, China}
\date{\today}
\begin{abstract}
 Braiding phases among topological excitations are key data for physically characterizing  topological orders. In this paper, we provide a field-theoretical approach towards a complete list of mutually compatible braiding phases of  topological orders  in (3+1)D spacetime. More concretely, considering a discrete gauge group   as input data, topological excitations in this paper are bosonic \textit{particles} carrying gauge charges  and \textit{loops} carrying gauge fluxes. Among these excitations, there are three classes of   \textit{root braiding processes}: particle-loop braidings  (i.e., the familiar Aharonov-Bohm phase of winding  an electric charge around a thin magnetic solenoid),   multi-loop braidings [\href{https://dx.doi.org/10.1103/PhysRevLett.113.080403}{Phys. Rev. Lett. 113, 080403 (2014)}], and particle-loop-loop braidings [i.e., Borromean Rings braiding in \href{https://doi.org/10.1103/PhysRevLett.121.061601}{Phys. Rev. Lett. 121, 061601 (2018)}]. A naive way to exhaust all topological orders is to arbitrarily combine these root braiding processes.  Surprisingly, we find  that there exist illegitimate combinations in which certain  braiding phases cannot coexist, i.e., are mutually incompatible. Thus, the resulting topological orders are illegitimate and must be excluded. It is not obvious to identify these illegitimate combinations. But with the help of the powerful (3+1)D topological quantum field theories (TQFTs),  we  find that   illegitimate combinations    violate gauge invariance.  In this way, we are able to obtain  all     sets   of mutually compatible braiding phases and  all legitimate topological orders.  To illustrate, we work out all details when gauge groups are $\mathbb{Z}_{N_1},\mathbb{Z}_{N_1}\times\mathbb{Z}_{N_2},\mathbb{Z}_{N_1}\times\mathbb{Z}_{N_2}\times\mathbb{Z}_{N_3}$, and $\mathbb{Z}_{N_1}\times\mathbb{Z}_{N_2}\times\mathbb{Z}_{N_3}\times\mathbb{Z}_{N_4}$.  Finally, we concisely discuss compatible braidings and TQFTs in (4+1)D spacetime.
\end{abstract}

\maketitle
\tableofcontents

\section{Introduction}\label{sec_introduction}

The order parameter, which is designed for characterizing orders, is one of fundamental concepts of many-body  physics. Symmetry-breaking orders are characterized by local order parameters---local functions of spacetime. However, topological orders (e.g., fractional quantum Hall states) \cite{Wen1995,wen2004quantum,string1.5,Modular_Extension_LT_17} in gapped systems  are characterized by intrinsically nonlocal order parameters, such as adiabatic quantum phases accumulated by braiding topological excitations (e.g., anyons)\cite{2DbTO_wen_15}.   In topological orders, topological excitations are usually geometrically compact manifold-like after taking continuum limit, such as point-like particle excitations, string-like loop excitations, etc.\footnote{For non-manifold-like excitations, Ref.~\cite{LiYe2020PRB} provides some examples in a class of exotic stabilizer codes that support spatially extended excitations with restricted mobility and deformability.}
Braiding phases of topological excitations are   proportional to   integer-valued invariants of knots or links formed by world-lines of particles and world-sheets of loops, thereby  being quantized and   robust against local perturbations. In addition to topological orders,    braiding phases have also been applied to characterization of symmetry protected topological (SPT) phases \cite{levin_gu_12,chenguliuwenprb,chenguliuwenscience,PhysRevB.85.075125,1DSPT} despite that SPT bulk excitations are topologically trivial. The core reason is that SPTs can be properly dualized to specific topological orders \cite{levin_gu_12}.

 One  quantitatively efficient and powerful approach to braiding phases is topological quantum field theory (TQFT) \cite{sarma_08_TQC}. For example, braiding data of two-dimensional topological orders are encoded in $(2+1)$D Chern-Simons theory \cite{wen2004quantum,wen90a,2DbTO_wen_15}. In this paper, we focus on  deconfined phases of $(3+1)$D discrete   gauge theories\footnote{In order to reconcile different conventions in condensed matter physics and high energy physics, we take the following convention: gauge theories/field theories are always associated with spacetime dimensions but quantum states, ``Hamiltonian-type'' lattice model, or topological phases of matter (topological order, SPTs, etc.) are associated with spatial dimensions only.}   \cite{fradkin2013field,PhysRevLett.62.1221,hansson2004superconductors,delcamp2018gauge,delcamp20192}   with Abelian gauge group $G=\prod^n_{i=1}\Z_{N_i}$ where $n$ denotes the total number of cyclic subgroups. Such phases of matter are usually called ``$\prod^n_i\Z_{N_i}$ topological order''. For example, the ground state of the three-dimensional toric code model \cite{string2,Kitaev2006}  admits $\Z_2$ topological order. By using  group representation and conjugacy class \cite{preskill1999lecture}, we may label topological excitations via gauge charges and gauge fluxes. More specifically,  there are totally $\prod_{i=1}^nN_i$ distinct bosonic particles carrying gauge charges and $\prod_{i=1}^n N_i$ distinct loops carrying gauge fluxes.
  Without loss of generality, it is enough to consider   braiding phases among $n$ distinct  elementary particles (denoted as $e_1,e_2,\cdots,e_i,\cdots,e_n$) carrying the unit gauge charge of a specific gauge subgroup and $n$ distinct  elementary loops (denoted as $m_1,m_2,\cdots,m_i,\cdots,m_n$) carrying the unit gauge flux  of a specific gauge subgroup. When more than one particle  (loop) are simultaneously involved in the same braiding process, superscripts  will be added properly to $e_i$ ($m_i$).

Among these elementary excitations, there exist three classes of { braiding processes} that have been studied before:   particle-loop braiding \cite{hansson2004superconductors,abeffect,PRESKILL199050,PhysRevLett.62.1071,PhysRevLett.62.1221,ALFORD1992251}, multi-loop braiding \cite{wang_levin1,PhysRevLett.114.031601,2016arXiv161209298P,YeGu2015,ye16_set,string4,2016arXiv161008645Y,2018arXiv180101638N,PhysRevB.99.235137,jian_qi_14,string5,PhysRevX.6.021015,string6,YeGu2015,corbodism3,ye16_set,2016arXiv161008645Y,string4,PhysRevLett.114.031601,3loop_ryu,string10,2016arXiv161209298P,Tiwari:2016aa,peng2020gauge,PhysRevB.99.205120}, and particle-loop-loop braiding (i.e., Borromean Rings braiding) \cite{ye17b}. For the purpose of this paper, we regard these braidings as \textit{root braiding processes}. Within each class of root braiding processes, depending on gauge group assignment, there are still many different braiding phases among which compatibility is crucial.  To proceed further, let us briefly introduce the three classes of root braidings.

  \textit{In the first class}, within each gauge subgroup, e.g., $\Z_{N_i}$, there is a well-defined particle-loop braiding phase $\mathsf{\Theta}^{\text{H}}_{i}=\frac{2\pi}{N_i}\text{ mod }2\pi$ when the Hopf linking invariant is one\footnote{The superscript $H$ in $\Theta^\text{H}_i$ stands for Hopf. As an angle, the $2\pi$ period is important but obvious, so we will not write it explicitly hereafter. To characterize topological orders, it is sufficient to consider braiding processes in which the linking number is unit.}.  Here, the Hopf link is formed by  an elementary particle's trajectory $\gamma_{e_i}$ and an elementary loop $m_i$.  This braiding phase always exists since it physically encodes  the cyclic group structure of $\Z_{N_i}$. For the whole gauge group, the root braiding phases of the first class form a set $\{\mathsf{\Theta}_i^{\text{H}}\}_G$ with $i=1,\cdots, n$. A subscript $G$ is added for specifying the gauge group $G$. Apparently, all braiding phases in the set belong to distinct gauge subgroups,  thereby being mutually compatible and linearly independent. In the language of TQFT, one may compute braiding phases from gauge-invariant  correlation functions of Wilson operators of the topological $BF$ theories with action
$S_{\rm BF}=\int\sum_i\frac{N_i}{2\pi} B^i  \wedge dA^i\!$ (abbreviated as $BdA$) \cite{Witten1989,horowitz89,hansson2004superconductors,Baez2011,bti6}. Here,  the $1$-form $A^i$ and $2$-form $B^i$ are compact $\mathbb{U}(1)$ gauge fields describing the loop current ($\frac{1}{2\pi}*dA^i$) and particle current ($\frac{1}{2\pi}*dB^i$) degrees of freedom, respectively.   As a natural higher-dimensional generalization of the Chern-Simons theory,  the $BF$ theory has been broadly applied to condensed matter systems, such as superconductors \cite{hansson2004superconductors}, bosonic and fractional topological  insulators and more general $3$D SPTs \cite{bti2,bti1,Ye:2017aa,ye16a,bti6}.

     \textit{In the second class}, i.e., multi-loop (three or four) braidings, all objects involved in the braidings are loops.  More specifically, a three-loop braiding \cite{wang_levin1} consists of three elementary loops, and lead to a set of braiding phases $\{\mathsf{\Theta}^{\text{H}}_{i}\,;\mathsf{\Theta}^{\text{3L}}_{j,k|l}\}_G$, where $G=\prod_{i=1}^n\Z_{N_i}$ with $n\geq 2$. As mentioned above, when $G$ is given, $\mathsf{\Theta}^{\text{H}}_{i}$ always exists. Here  $j,k,l$ indicate that  three elementary loops (denoted by $m^1_j,m^2_k,m^b_l$) respectively carry the elementary gauge flux of $\Z_{N_j},\Z_{N_k},\Z_{N_l}$ gauge subgroups. Geometrically, the loop $m^b_l$ right after the symbol ``$|$'', which carries elementary gauge flux of $\Z_{N_l}$ gauge subgroup, is called ``base loop''\cite{wang_levin1}. The latter   is simultaneously \textit{hopfly} linked to the other two loops, i.e.,  $m^1_j,m^2_k$. Under this geometric setting, the three-loop braiding can be regarded as an anyonic braiding process on the Seifert surface bounded by the base loop.  Braiding phases in the set $\{\mathsf{\Theta}^{\text{H}}_{i}\,;\mathsf{\Theta}^{\text{3L}}_{j,k|l}\}_G$ satisfy  a series of remarkably elegant  constraints such that mutually compatible  braiding datasets can be unambiguously  determined. The result    was obtained in Ref.~\cite{wang_levin1} by means of  general properties of discrete gauge group and adiabaticity of braiding processes. The same result can also be obtained from TQFTs with topological terms of $BdA+AAdA$ form  \cite{PhysRevLett.114.031601,2016arXiv161209298P,YeGu2015,ye16_set,string4,2016arXiv161008645Y,PhysRevB.99.235137,2018arXiv180101638N,PhysRevB.99.205120}.
   Moreover, in the second class, if we consider four loops from four distinct gauge subgroups,  the four-loop braiding phases form a set $\{\mathsf{\Theta}^{\text{H}}_{i}\,;\mathsf{\Theta}^{\text{4L}}_{j,k,l,m}\}_G$, where $G=\prod_{i=1}^n\Z_{N_i}$ with $n\geq 4$, and the four loops carry  elementary gauge fluxes of four different gauge subgroups $\Z_{N_j},\Z_{N_k},\Z_{N_l},\Z_{N_m}$ respectively.   The four-loop braiding is associated with the quadruple linking number of surfaces, thereby being quantized \cite{2016arXiv161209298P}.  The corresponding TQFTs can be symbolically expressed as $BdA+AAAA$. \cite{PhysRevLett.114.031601,2016arXiv161209298P,YeGu2015,ye16_set,string4,2016arXiv161008645Y,PhysRevB.99.235137,2018arXiv180101638N,PhysRevB.99.205120}.

     \textit{In the third class}, i.e., the particle-loop-loop braiding or Borromean Rings (BR) braiding \cite{ye17b}, an elementary particle carrying unit gauge charge of $\Z_{N_k}$ gauge subgroup moves around two loops  (denoted by $m^1_i,m^2_j$)  that respectively carry  unit gauge fluxes of $\Z_{N_i}$ and $\Z_{N_j}$, such that the particle's trajectory $\gamma_{e_k}$ and the two loops together form a Borromean Rings link, or general Brunnian link. The corresponding braiding phase is denoted as $\mathsf{\Theta}^{\text{BR}}_{i,j | k}$, which is proportional to the Milnor's triple linking number $\bar{\mu}$ \cite{ye17b,milnor1954link,mellor2003geometric}. Likewise, one may define a set of braiding phases: $\{ \mathsf{\Theta}^{\text{H}}_{i}\,; \mathsf{\Theta}^{\text{BR}}_{j ,k| l} \}_G$, where $G=\prod_{i=1}^n\Z_{N_i}$ with $n\geq 3$. The corresponding TQFTs can be symbolically expressed as $BdA+AAB$ \cite{ye17b}.

In this paper, we put all root braiding processes,  which are denoted by $\{\mathsf{\Theta}_i^{\text{H}}\,;\mathsf{\Theta}^{\text{3L}}_{j,k|l}\,;\mathsf{\Theta}^{\text{4L}}_{m,n,o,p}\,;\mathsf{\Theta}^{\text{BR}}_{q,r | s} \}_{G}$,  in arbitrary combinations and try to exhaust all topological orders. We find that, not all possible gauge group assignments (i.e., the subscripts $i,j,k,l,\cdots$) are realizable. Not all braiding phases in a given set are linearly independent. As we reviewed above, within each class, compatible braiding phases have been studied via various approaches.
 By ``compatible'', we mean that these braiding processes can be supported in the \emph{same} system. In other words, the compatible braiding phases together as a set of braidings characterize a legitimate topological order. If there are two mutually incompatible braiding processes in the set, then both braiding processes \textit{must} always lead to two trivial   braiding phases, i.e., $0\text{ mod }2\pi$ regardless the values of linking numbers of the braidings. As we will show, gauge invariance is broken if any one of two braidings has a nontrivial braiding phase.    Therefore,  in order to exhaust all legitimate topological orders, it is sufficient to find all  sets of braidings formed by mutually compatible braiding processes.
For this purpose, in this paper, through TQFT approach, we compute all   braiding processes in a unified framework, and figure out all cases of incompatibility that are tightly related to gauge noninvariance. Especially, we focus on cases of incompatibility that occur when two or three distinct classes of root braiding processes have nontrivial braiding phases. Compatible braiding phases for different gauge groups are summarized  in Table~\ref{table_zn1zn2} ($G\!=\!\Z_{N}$ and $(G\!=\!\Z_{N_1}\!\!\times \!\Z_{N_2}$), Table~\ref{table_zn1zn2zn3} ($G\!=\!\Z_{N_1}\!\!\times \!\Z_{N_2}\!\!\times \!\Z_{N_3}$), and Table~\ref{table_zn1zn2zn3zn4} ($G\!=\! \Z_{N_1}\!\!\times \! \Z_{N_2}\!\!\times\!  \Z_{N_3}\!\!\times\!  \Z_{N_4}$). More general cases with more than four $\Z_N$ gauge subgroups can be straightforwardly analyzed  by applying the results in these four tables, as shown in Sec.~\ref{sec_general_gauge_groups}. In addition to braiding phases, in these tables, we also provide the corresponding TQFTs and   definitions of gauge transformations therein.  All other sets of braiding phases are incompatible and not realizable. Some typical examples of incompatibility will be analyzed in details in this paper.

The remainder of this paper is structured as follows. In Sec.~\ref{section_root}, we concretely analyze root braiding processes one by one, in order to lay the foundation for the forthcoming discussions on compatibility. In Sec.~\ref{section_compatible}, by combining all root braiding processes together, we study the corresponding TQFTs and extract all sets of compatible braiding processes. Then, in Sec.~\ref{section_incompatible}, some sets of incompatible braiding processes are illustrated in some concrete examples. In Sec.~\ref{section_4d}, along the same line,   we concisely discuss compatible braidings and TQFTs in $(4+1)$-dimensional topological orders. Conclusions are made in Sec.~\ref{section_conclusion}. Several technical details are collected in Appendices.

\section{Review on root braiding processes and gauge transformations \label{section_root}}

In this section, we review TQFTs of root braiding processes.
 We emphasize the correspondence between root braiding processes, topological terms and braiding phases, which is illustrated as the following triangle:
\[
\begin{tikzcd} [row sep=5.5em,column sep=-1em]
& \textbf{Braiding Process} \arrow[Leftrightarrow]{dr} \\
\textbf{Topological Term} \arrow[Leftrightarrow]{ur} && \textbf{Braiding Phase} \arrow[Leftrightarrow]{ll}
\end{tikzcd}
\]
 A braiding process can be identified from a topological term or a braiding phase, and vice versa. In this manner, we can study the braiding processes within the framework of TQFT. More concretely, in order to extract braiding phases from a TQFT, one can either add gauge-invariant source terms \cite{ye17b,ye16_set} or study algebra of Wilson operators \cite{3loop_ryu,PhysRevB.99.235137}, such that the braiding phases of a given braiding process  are connected to a linking / knot invariant formed by the spacetime trajectories of particles and loops.

\subsection{Microscopic origins of discrete gauge groups, topological excitations, and TQFT actions}\label{section_microscopic_derivation}

As our basic goal is to utilize TQFTs and braiding processes to  characterize and classify topological orders of underlying quantum many-body systems, it is very important to identify microscopic origins of input data of TQFTs and braiding processes.

  Hamiltonian realization, especially exactly solvable model proposal, is always the most powerful stimulus of the progress of topological orders. For example, toric code model proposed by Kitaev \cite{Kitaev2006} elegantly unveils all key properties of nonchiral Abelian topological orders in $(2+1)$D. String-net models constructed by Levin and Wen \cite{string_net_levin_wen_2005} are applied to exhaust non-Abelian topological orders in $(2+1)$D. In $(3+1)$D, topological orders that are within Dijkgraaf-Witten cohomology classification \cite{dwitten} have Hamiltonian realization in Ref.~\cite{string6}, which uses $4$-cocycles to cover all particle-loop and multi-loop braidings. However, exactly solvable models for Borromean Rings braiding \cite{ye17b} are not known, which is an interesting future direction.  Ref \citep{Murdy2016} and \citep{Murdy2019} present strategies for constructing higher-dimensional Abelian and non-Abelian
topological phases via coupling quantum wires, which may shed light on constructing exactly solvable lattice model for BR braiding.

While   exactly solvable models have Hamiltonian form, the Hamiltonians often look very intricate and unrealistic (four-spin, six-spin interactions, and more). Furthermore, it is unclear for us to rigorously connect the Hamiltonians to TQFTs. The latter have been proved to be a very powerful machine to study   topological orders since the discovery of the fractional quantum Hall effect. But, to the best of our knowledge, it is highly impossible to perform a standard perturbation theory to renormalize the intricate interacting electron system of the $\nu=1/3$ Laughlin state to a beautiful Chern-Simons gauge theory $\int\frac{3}{4\pi}AdA$ where $A$ is an emergent gauge field. Although there are many effective ways, e.g., parton construction and hydrodynamical approach, to handle strongly-correlated physics, it is still kind of mysterious to derive emergent dynamical gauge fields from the very beginning.

In order to identify microscopic origins, below we will provide an effective way of thinking, following the spirit of the previous works \cite{YeGu2015,PhysRevB.93.115136,bti2,Ye:2017aa,hansson2004superconductors,PhysRevB.99.235137}. From this effective derivation, we can find how discrete gauge groups, topological terms, quantized Wilson integral arise from a quantum many-body system. We can also find particle excitations and loop excitations are gauge charges and gauge fluxes of gauge groups. Below, we take $\int BdA+AAB$ as an example by means of exotic boson condensate and vortexline condensate. The detailed derivation is given
in Appendix \ref{appendix_microscopic_derivation}. \textit{Here we just briefly sketch
the key idea}.

We start from a multi-layer condensate in $3$D space in
which one layer is in vortexline condensation phase (from disordering a 3D superfluid) while the others are
in charge condensation phases (``bosonic superconductors''): The former can be regarded as the Higgs phase of two-form gauge fields while the latter can be regarded as the usual Higgs phase of 1-form gauge fields.
\begin{equation}
\begin{alignedat}{1}\mathcal{L}= & \frac{\rho_{3}}{2}\left(\partial_{[\mu}\Theta_{\nu]}-N_{3}B_{\mu\nu}^{3}\right)^{2} +\frac{\rho_{1}}{2}\left(\partial_{\mu}\theta^{1}-N_{1}A_{\mu}^{1}\right)^{2}\\
 & +\frac{\rho_{2}}{2}\left(\partial_{\mu}\theta^{2}-N_{2}A_{\mu}^{2}\right)^{2}\\
 & +i\Lambda\varepsilon^{\mu\nu\lambda\rho}\left(\partial_{[\mu}\Theta_{\nu]}-N_{3}B_{\mu\nu}^{3}\right)\left(\partial_{\lambda}\theta^{1}-N_{1}A_{\lambda}^{1}\right)\\
 &\times\left(\partial_{\rho}\theta^{2}-N_{2}A_{\rho}^{2}\right),
\end{alignedat}
\label{eq_Lagrangian_SF}
\end{equation}
where $\partial_{[\mu}\Theta_{\nu]}=\partial_{\mu}\Theta_{\nu}-\partial_{\nu}\Theta_{\mu}$. The vector-like phase angle $\Theta_\mu$ describes the phase field of vortexline condensation \cite{bti2}, while $\theta^1$ and $\theta^2$ are the usual phase angles of boson condensation. The coefficients $\rho_1,\rho_2,\rho_3$ represent phase rigidity of condensates. The last term of the above Lagrangian couple three condensates together in a gauge-invariant fashion.
By introducing Hubbard-Stratonovich fields $\Sigma_{\mu\nu}^{3}$,
$j^{1}$, $j^{2}$ and Lagrange multiplier fields $\xi^{I}$ and
$\eta^{I}$, we obtain:
\begin{widetext}
\begin{equation}
\begin{alignedat}{1}\mathcal{L}= & \frac{1}{2\rho_{1}}\left(j^{1}\right)^{2}-i\theta^{1}\partial_{\lambda}j_{\lambda}^{1}-iN_{1}A_{\lambda}^{1}j_{\lambda}^{1}+\frac{1}{2\rho_{2}}\left(j^{2}\right)^{2}-i\theta^{2}\partial_{\rho}j_{\rho}^{2}-iN_{2}A_{\rho}^{2}j_{\rho}^{2}+\frac{1}{8\rho_{3}}\left(\Sigma_{\mu\nu}^{3}\right)^{2}-i\Theta_{\mu}\partial_{\nu}\Sigma_{\mu\nu}^{3}-i\frac{1}{2}N_{3}B_{\mu\nu}^{3}\Sigma_{\mu\nu}^{3}\\
 & +i\Lambda\varepsilon^{\mu\nu\lambda\rho}\left[2\Theta_{\mu}\partial_{\nu}\left(N_{1}N_{2}A_{\lambda}^{1}A_{\rho}^{2}\right)+\theta^{1}\partial_{\lambda}\left(N_{2}A_{\rho}^{2}N_{3}B_{\mu\nu}^{3}\right)+\theta^{2}\partial_{\rho}\left(N_{3}B_{\mu\nu}^{3}N_{1}A_{\lambda}^{1}\right)\right]\\
 & +i\eta_{\lambda}^{1}\left[\xi_{\lambda}^{1}-\Lambda\varepsilon^{\mu\nu\lambda\rho}\cdot\frac{1}{2}\partial_{\rho}\theta^{2}N_{3}B_{\mu\nu}^{3}\right]+i\theta^{1}\partial_{\lambda}\xi_{\lambda}^{1}+i\eta_{\rho}^{2}\left[\xi_{\rho}^{2}-\Lambda\varepsilon^{\mu\nu\lambda\rho}\cdot\frac{1}{2}\partial_{\lambda}\theta^{1}N_{3}B_{\mu\nu}^{3}\right]+i\theta^{2}\partial_{\rho}\xi_{\rho}^{2}\\
 & +i\eta_{\mu\nu}^{3}\left[\xi_{\mu\nu}^{3}-\Lambda\varepsilon^{\mu\nu\lambda\rho}\left(2\partial_{\lambda}\theta^{1}N_{1}A_{\rho}^{2}+2\partial_{\rho}\theta^{2}N_{1}A_{\lambda}^{1}\right)\right]+i\Theta_{\mu}\partial_{\nu}\xi_{\mu\nu}^{3}-iN_{1}N_{2}N_{3}\Lambda\varepsilon^{\mu\nu\lambda\rho}A_{\lambda}^{1}A_{\rho}^{2}B_{\mu\nu}^{3}+\text{boundary terms}.
\end{alignedat}
\label{eq_Lagrangian_dualed}
\end{equation}
Integrating out $\Theta_{\mu}$, $\theta^{1}$ and $\theta^{2}$
yields constraints in the path-integral measure. These constraints
can be solved by introducing $1$-form gauge field $A^{3}$, $2$-form
gauge fields $B^{1}$ and $B^{2}$ respectively:
\begin{equation}
\begin{alignedat}{1}\Sigma_{\mu\nu}^{3}= & \frac{1}{2\pi}\varepsilon^{\mu\nu\lambda\rho}\partial_{\lambda}A_{\rho}^{3}+\xi_{\mu\nu}^{3}-\Lambda\varepsilon^{\mu\nu\lambda\rho}\cdot2N_{1}N_{2}A_{\lambda}^{1}A_{\rho}^{2},\\
j_{\lambda}^{1}= & \frac{1}{4\pi}\varepsilon^{\lambda\rho\mu\nu}\partial_{\rho}B_{\mu\nu}^{1}+\xi_{\lambda}^{1}-\Lambda\varepsilon^{\mu\nu\lambda\rho}\left(N_{2}A_{\rho}^{2}N_{3}B_{\mu\nu}^{3}-\frac{1}{2}\eta_{\rho}^{2}N_{3}B_{\mu\nu}^{3}-2\eta_{\mu\nu}^{3}N_{2}A_{\rho}^{2}\right),\\
j_{\rho}^{2}= & \frac{1}{4\pi}\varepsilon^{\rho\lambda\mu\nu}\partial_{\lambda}B_{\mu\nu}^{2}+\xi_{\rho}^{2}-\Lambda\varepsilon^{\mu\nu\lambda\rho}\left(N_{3}B_{\mu\nu}^{3}N_{1}A_{\lambda}^{1}-\frac{1}{2}\eta_{\lambda}^{1}N_{3}B_{\mu\nu}^{3}-2\eta_{\mu\nu}^{3}N_{1}A_{\lambda}^{1}\right).
\end{alignedat}
\end{equation}
\end{widetext}
The physical meaning of $\Sigma_{\mu\nu}^{3}$ is the
current of loop while those of $j_{\lambda}^{1}$ and $j_{\rho}^{2}$
are the currents of particles of layer $1$ and $2$ respectively.
In this manner, one can figure out the microscopic origins of particle
and loop excitations. Substituting $\Sigma_{\mu\nu}^{3}$, $j_{\lambda}^{1}$
and $j_{\rho}^{2}$ into     Lagrangian (\ref{eq_Lagrangian_dualed})
and integrating out the Lagrange multiplier fields, we end up with
\begin{equation}
\begin{alignedat}{1}\mathcal{L}= & \frac{iN_{1}}{4\pi}\varepsilon^{\mu\nu\lambda\rho}B_{\mu\nu}^{1}\partial_{\lambda}A_{\rho}^{1}+\frac{iN_{2}}{4\pi}\varepsilon^{\mu\nu\lambda\rho}B_{\mu\nu}^{2}\partial_{\lambda}A_{\rho}^{2}\\
 & +\frac{iN_{3}}{4\pi}\varepsilon^{\mu\nu\lambda\rho}B_{\mu\nu}^{3}\partial_{\lambda}A_{\rho}^{3}+iN_{1}N_{2}N_{3}\Lambda\varepsilon^{\mu\nu\lambda\rho}A_{\mu}^{1}A_{\nu}^{2}B_{\lambda\rho}^{3}.
\end{alignedat}
\end{equation}
Finally, we effectively obtain the TQFT action
for a BR braiding: $S=\int\mathcal{L}dxdt\sim \int\sum_{i=1}^{3}\frac{N_{i}}{2\pi}B^{i}dA^{i}+A^{1}A^{2}B^{3}$.

In this theory, the particle excitations are $e_1,e_2,e_3$ carrying gauge charges of $\Z_{N_1},\Z_{N_2},\Z_{N_3}$ and minimally couple to $A^1,A^2,A^3$. The loop excitations are $m_1,m_2,m_3$ carrying gauge fluxes of $\Z_{N_1},\Z_{N_2},\Z_{N_3}$ and minimally couple to $B^1,B^2,B^3$.

\subsection{Particle-loop braiding and $BdA$ term\label{subsec_BdA}}

\begin{figure}
\includegraphics[scale=0.19]{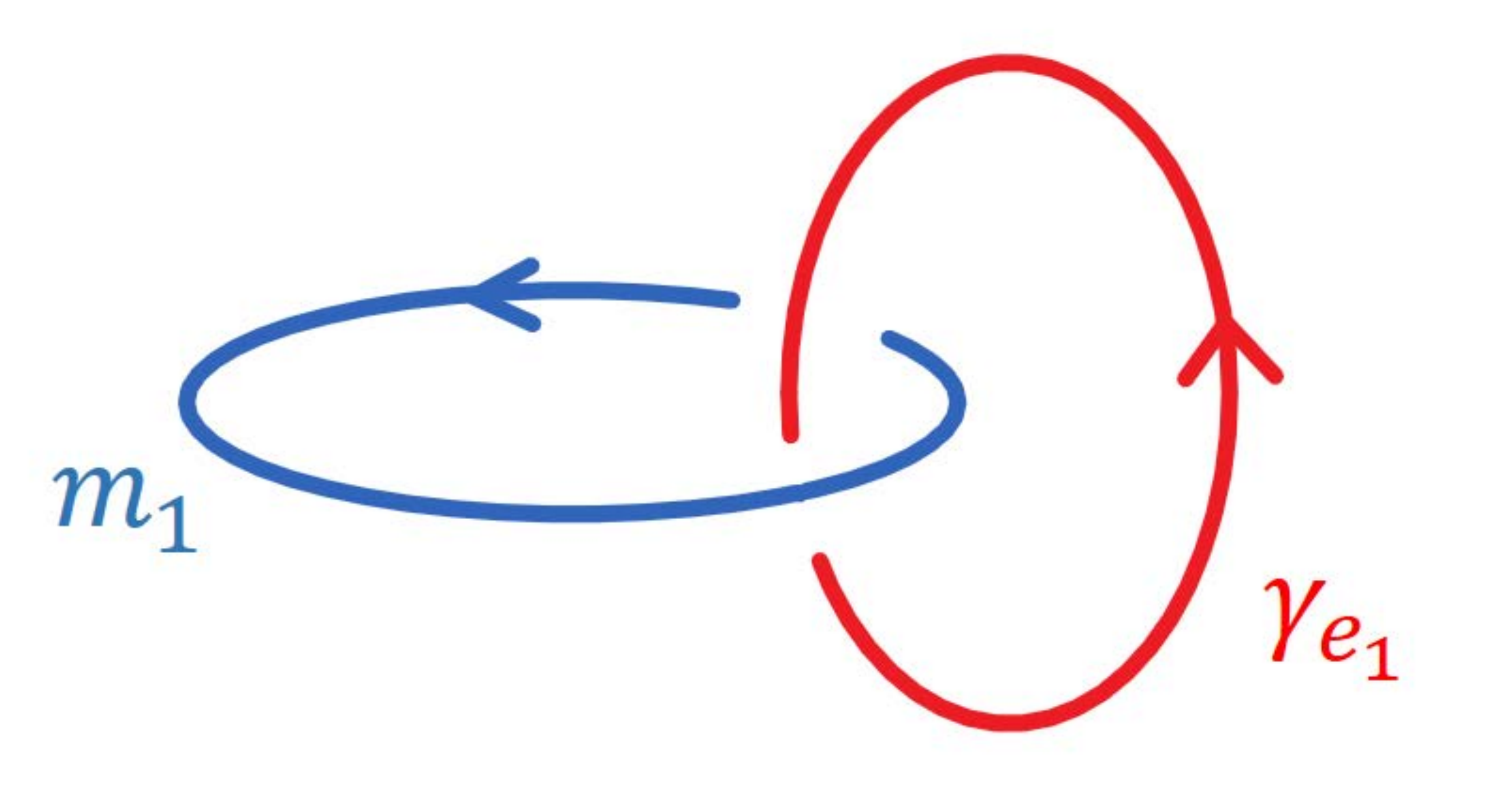}

\caption{Particle-loop braiding described by $B^{1}dA^{1}$ and $\mathsf{\Theta^{\text{H}}_{1}}$. $m_{1}$ is an elementary loop that carries unit gauge flux of  $\mathbb{Z}_{N_{1}}$ gauge group. $\gamma_{e_{1}}$
is the closed trajectory of an elementary particle  $e_{1}$ that carries unit gauge charge of   $\mathbb{Z}_{N_{1}}$ gauge group.
\label{figure_particle_loop_braiding}}

\end{figure}

The particle-loop braiding is essentially a quantized Aharonov-Bohm effect in a discrete gauge theory where local interactions are completely screened and long-range statistical interactions lead to nontrivial braiding phases. Given a $\Z_{N_1}$ gauge subgroup, an elementary particle $e_1$ is braided around a static elementary loop $m_1$. The trajectory $\gamma_{e_1}$ and $m_1$ as a whole form   a Hopf link, as shown in Fig.~\ref{figure_particle_loop_braiding}.  The corresponding braiding phase is given by $\mathsf{\Theta}^{\text{H}}_i=\frac{2\pi}{N_i}$ for unit Hopf linking number. The classification of equivalent trajectories of $e_1$ is essentially related to the mathematics of fundamental group and link homotopy, which was preliminarily introduced in the Supplemental Materials of Ref.~\cite{ye17b}. The corresponding TQFT is the following multi-component $BF$ action\footnote{In this paper, each summation is indicated by a $\sum$ symbol.} ($F=dA$, $\wedge$ is omitted):
\begin{equation}
S=\int\sum^n_{i=1}\frac{N_i}{2\pi}B^idA^i,\label{eq_action-Hopf link}
\end{equation}
where $N_i$ are positive integers that specify the discrete gauge group $G$. $\{A^i=\sum_\mu A^i_\mu dx_\mu\}$ and $\{B^i=\frac{1}{2!}\sum_{\mu\nu}B^i_{\mu\nu}dx_\mu dx_\nu\}$ are $1$-form and 2-form compact
$\mathbb{U}(1)$ gauge fields respectively. Although one may expect  a general matrix formalism $\sim K^{ij}B^idA^j$, the matrix $K$ can always be sent to a diagonal matrix with positive elements via two independent $\mathbb{GL}(n,\Z)$ transformations. The two transformations respectively act on $B$ fields and $A$ fields (A relevant discussion on  basis transformations can be found in Refs.~\cite{ye16a,ye16_set}.). The action (\ref{eq_action-Hopf link})
keeps invariant up to boundary terms under gauge transformations
\begin{equation}\label{eq_GT_BF_theory}
\begin{alignedat}{1}A^{i}\rightarrow & A^{i}+d\chi^{i},\\
B^{i}\rightarrow & B^{i}+dV^{i},
\end{alignedat}
\end{equation}where $\{\chi^i\}$ and $\{V^i\}$ are respectively $0$-form and $1$-form compact $\mathbb{U}(1)$
gauge parameters with $\int d\chi\in2\pi\mathbb{\mathbb{Z}}$ and $\int dV\in2\pi\mathbb{\mathbb{Z}}$. Once these integrals are nonzero, the corresponding gauge transformations are said to be ``large''. The coefficient quantization, i.e., $N_1$ is integral, is guaranteed by the invariance of the partition function $\mathcal{Z}=\int \mathscr{D}B\mathscr{D}Ae^{i\int \frac{N_1}{2\pi}BdA}$ under large gauge transformations on any compact oriented manifold.

\subsection{Multiloop braiding and $AAdA$, $AAAA$ terms}
\begin{figure}
\includegraphics[scale=0.15]{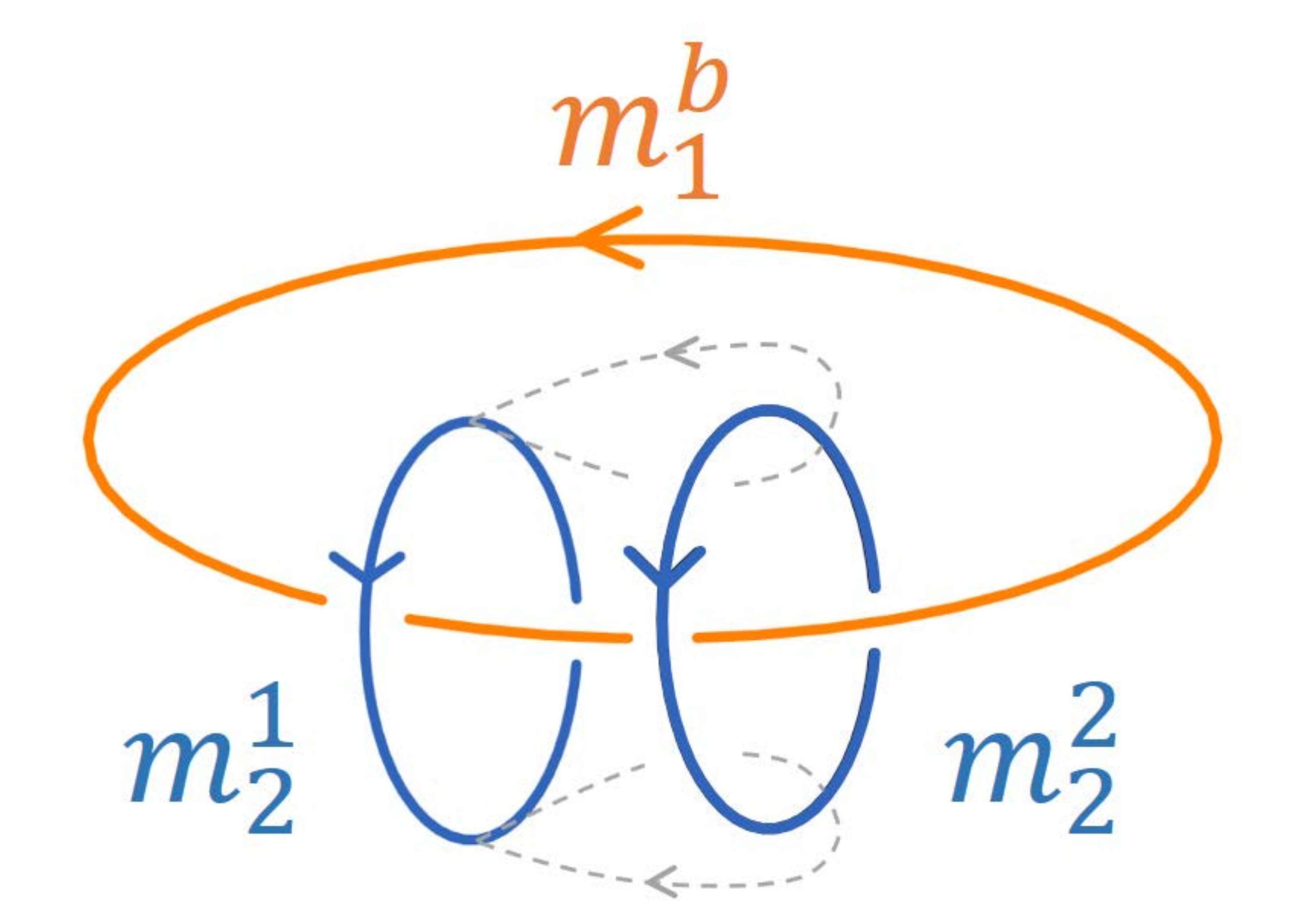}

\caption{Three-loop braiding process described by   $A^{1}A^{2}dA^{2}$
and $\mathsf{\Theta}_{2,2|1}^{\text{3L}}$.
In this case, the base loop $m_{1}^{b}$ carries the $\mathbb{Z}_{N_{1}}$ gauge
flux; Both $m_{2}^{1}$ and $m_{2}^{2}$   carry the $\mathbb{Z}_{N_{2}}$
gauge flux.
\label{figure_3_loop_braiding_zn1+zn2}}
\end{figure}

Next, we consider three-loop and four-loop braiding processes introduced in Sec.~\ref{sec_introduction}. The minimal number of gauge subgroups is two:  $G=\mathbb{Z}_{N_{1}}\times\mathbb{Z}_{N_{2}}$.   For this gauge group, we consider a three-loop braiding process (Fig.~\ref{figure_3_loop_braiding_zn1+zn2}) with braiding phase denoted as $\mathsf{\Theta}_{2,2|1}^{\text{3L}}$.
The TQFT action for this three-loop braiding process is
\begin{equation}
S=\int\sum_{i=1}^{2}\frac{N_{i}}{2\pi}B^{i}dA^{i}+\frac{q_{122}}{\left(2\pi\right)^{2}}A^{1}A^{2}dA^{2}\,.\label{eq_action_A1A2dA2+A2A1dA1}
\end{equation}
$S$ is invariant up to boundary terms under gauge transformations
\begin{equation}\label{eq_GT_BF_theory_3loop}
\begin{alignedat}{1}A^{i}\rightarrow & A^{i}+d\chi^{i},\\
B^{1}\rightarrow & B^{1}+dV^{1}+\frac{q_{122}}{2\pi N_{1}}d\chi^{2}A^{2},\\
B^{2}\rightarrow & B^{2}+dV^{2}-\frac{q_{122}}{2\pi N_{2}}d\chi^{1}A^{2}.
\end{alignedat}
\end{equation}
The coefficient $q_{122}$ is quantized and periodic:
 $q_{122}=\frac{kN_{1}N_{2}}{N_{12}}$,
where $k\in\mathbb{Z}_{N_{12}}$ and $N_{12}$ is the greatest common
divisor (GCD) of $N_{1}$ and $N_{2}$.

\begin{figure}
\includegraphics[scale=0.18]{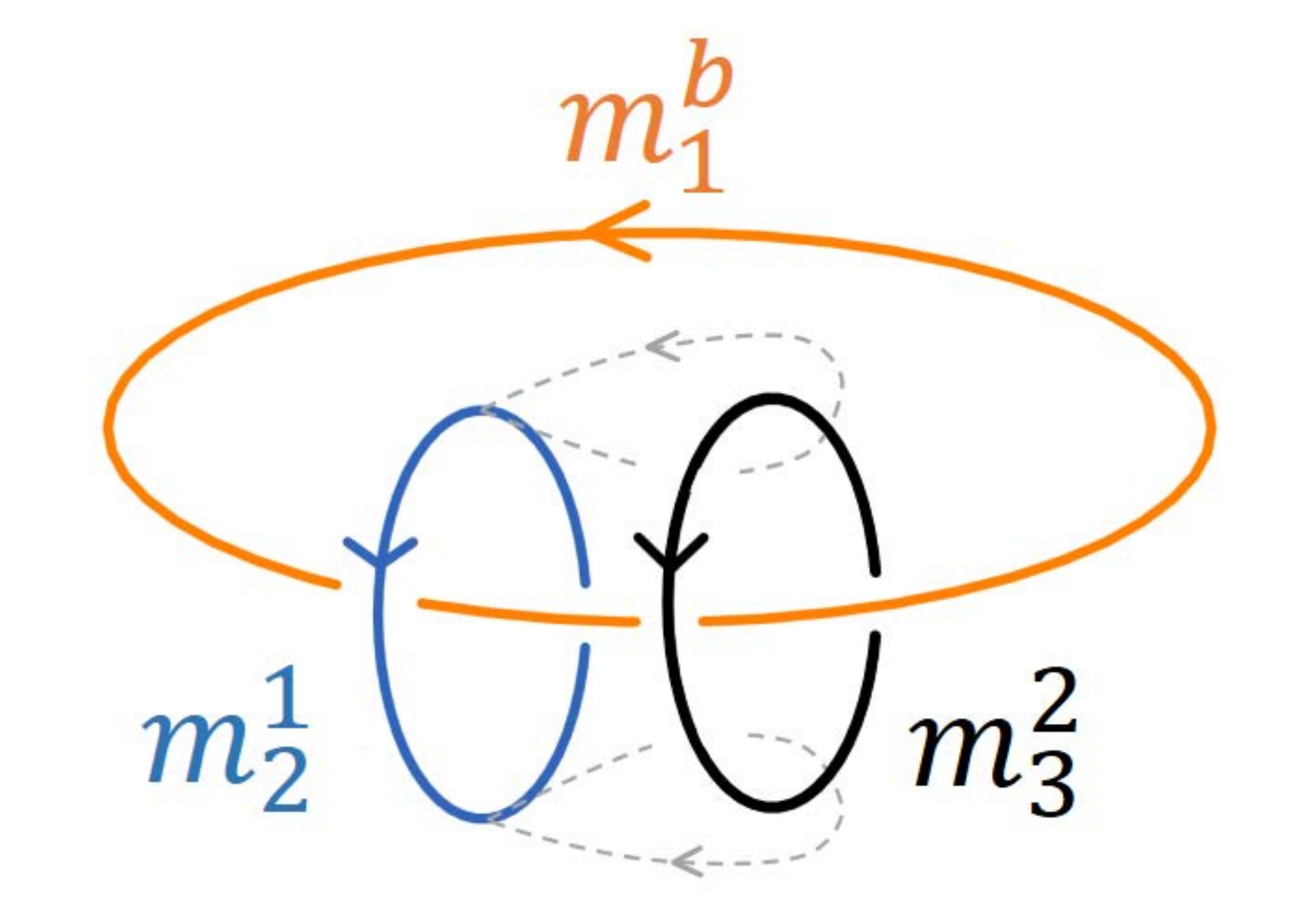}
\caption{Three-loop braiding process described by  $A^{1}A^{2}dA^{3}$
and $\mathsf{\Theta}_{2,3|1}^{\text{3L}}$.
In this case, the base loop $m_{1}^{b}$ carries the $\mathbb{Z}_{N_{3}}$ gauge
flux, $m_{2}^{1}$ carries the $\mathbb{Z}_{N_{2}}$ gauge flux, and
$m_{3}^{2}$ carries the $\mathbb{Z}_{N_{3}}$ gauge flux.\label{figure_3_loop_braiding}}

\end{figure}

When $G=\mathbb{Z}_{N_{1}}\times\mathbb{Z}_{N_{2}}\times\mathbb{Z}_{N_{3}}$,  we consider a three-loop
braiding process (Fig.~\ref{figure_3_loop_braiding}) associated with the braiding phase $\mathsf{\Theta}_{2,3|1}^{\text{3L}}$. The corresponding TQFT action
is
\begin{equation}
S=\int\sum_{i=1}^{3}\frac{N_{i}}{2\pi}B^{i}dA^{i}+\frac{q_{123}}{\left(2\pi\right)^{2}}A^{1}A^{2}dA^{3}\,.\label{eq:action-3-loop}
\end{equation}
The gauge transformations are defined as:
\begin{equation}\label{eq_GT_BF_theory_3loop_plus}
\begin{alignedat}{1}A^{i}\rightarrow & A^{i}+d\chi^{i},\\
B^{i}\rightarrow & B^{i}+dV^{i}+\frac{q_{123}}{2\pi N_{i}}\epsilon^{ij3}d\chi^{j}A^{3},
\end{alignedat}
\end{equation}
where $\epsilon$ is the Levi-Civita symbol with $\epsilon^{123}=-\epsilon^{213}=1$. The coefficient is $q_{123}=\frac{kN_{1}N_{2}}{N_{123}}$ where $k\in\mathbb{\mathbb{Z}}_{N_{123}}$ and $N_{123}$ is the GCD of $N_{1}$, $N_{2}$ and $N_{3}$.

When $G=\mathbb{Z}_{N_{1}}\times\mathbb{Z}_{N_{2}}\times\mathbb{Z}_{N_{3}}\times\mathbb{Z}_{N_{4}}$,
  four-loop braiding processes are realizable, in which four
loops carry the unit gauge fluxes of $\mathbb{Z}_{N_{1}},\mathbb{Z}_{N_{2}},\mathbb{Z}_{N_{3}},\mathbb{Z}_{N_{4}}$
gauge subgroups respectively. The corresponding TQFT action is
\begin{equation}
S=\int\sum_{i=1}^{4}\frac{N_{i}}{2\pi}B^{i}dA^{i}+\frac{q_{1234}}{\left(2\pi\right)^{3}}A^{1}A^{2}A^{3}A^{4}
\label{eq_action_4_loop}
\end{equation}
The action~(\ref{eq_action_4_loop}) is invariant up to boundary terms under gauge transformations
\begin{equation}\label{eq_GT_BF_theory_4loop}
\begin{alignedat}{1}A^{i}\rightarrow & A^{i}+d\chi^{i},\\
B^{i}\rightarrow & B^{i}+dV^{i}\\
 & -\frac{1}{2}\sum_{j,k,l}\frac{q_{1234}}{\left(2\pi\right)^{2}N_{i}}\epsilon^{ijkl}A^{j}A^{k}\chi^{l}\\
 & +\frac{1}{2}\sum_{j,k,l}\frac{q_{1234}}{\left(2\pi\right)^{2}N_{i}}\epsilon^{ijkl}A^{j}\chi^{k}d\chi^{l}\\
 & +\frac{1}{6}\sum_{j,k,l}\frac{q_{1234}}{\left(2\pi\right)^{2}N_{i}}\epsilon^{ijkl}\chi^{j}d\chi^{k}d\chi^{l},
\end{alignedat}
\end{equation}
where $\epsilon$ is the Levi-Civita symbol with $\epsilon^{1234}=-\epsilon^{1324}=1$. The coefficient is $q_{1234}=\frac{kN_{1}N_{2}N_{3}N_{4}}{N_{1234}}$ where $k\in\mathbb{Z}_{N_{1234}}$ and $N_{1234}$ is the GCD of $N_{1}$, $N_{2}$, $N_{3}$ and $N_{4}$. Till now, we have only reviewed gauge transformations in this review section. It is hard to visualize four-loop braiding in three-dimensional real space. We recommend  Fig.~$6$ of Ref.~\cite{2016arXiv161209298P}.

\subsection{Borromean Rings braiding and $AAB$ term}\label{subsec_br_braiding}
Last, we consider the particle-loop-loop braiding or Borromean Rings braiding with its braiding phase denoted as $\mathsf{\Theta}_{i,j|k}^{\text{BR}}$ \cite{ye17b}.  Likewise, the input data of BR braiding are Abelian gauge group $G=\prod^n_{i=1}\Z_{N_i}$ but with $n\geq 3$, such that all particles and loops can be labeled by gauge charges and gauge fluxes in a specific gauge subgroup $\Z_{N_i}$. This braiding   is  beyond Dijkgraaf-Witten gauge theory classification $\mathcal{H}^4(G,U(1))$.  The latter only includes   braiding phases of particle-loop braidings and multi-loop braidings. By further taking BR braiding into account,   we need to study   proper combinations of all braidings together to exhaust all topological orders.

The corresponding topological term of BR braidings is $A^{i}A^{j}B^{k}$.
\begin{figure}
\includegraphics[scale=0.18]{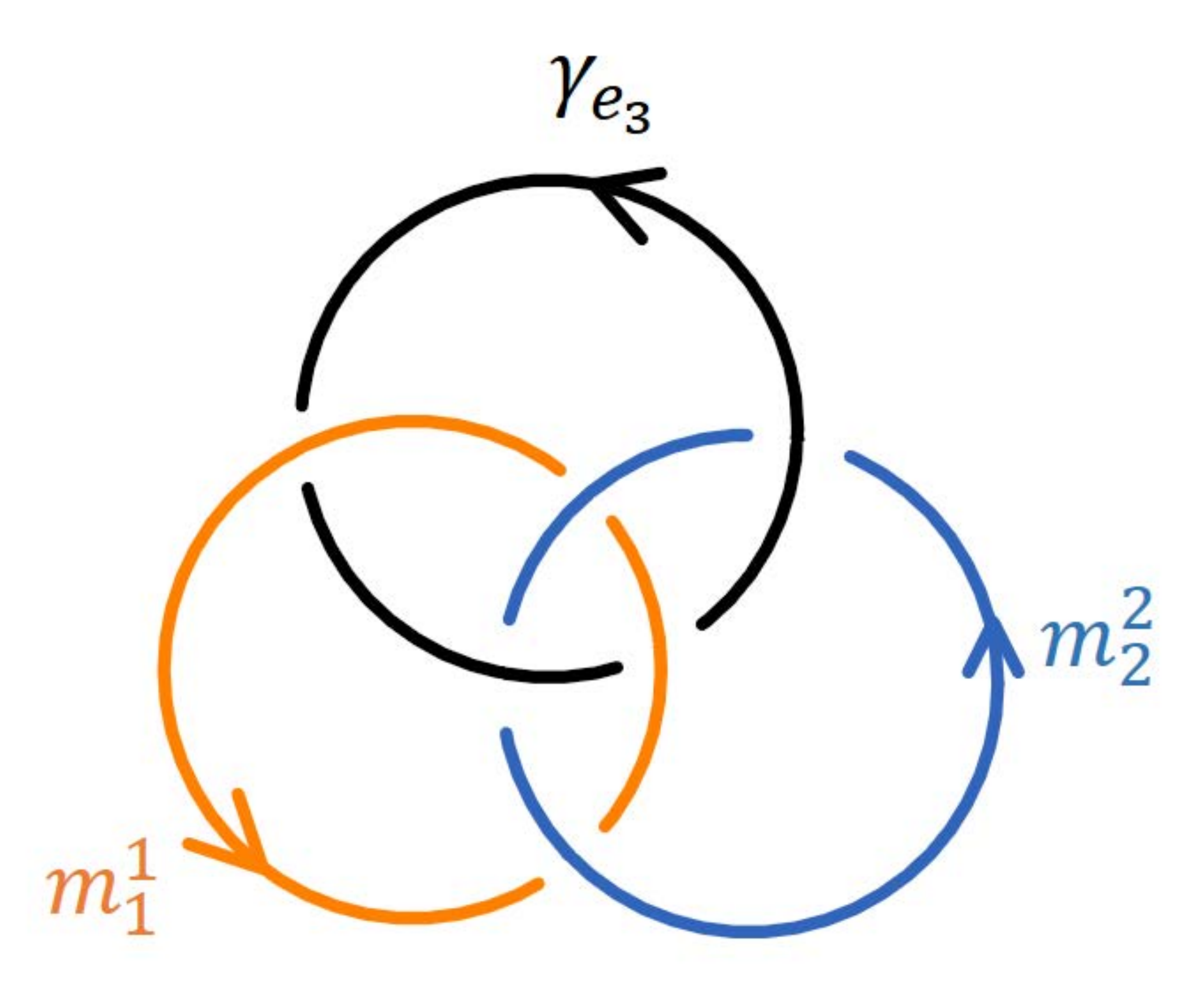}
\caption{Borromean Rings braiding described by $A^{1}A^{2}B^{3}$ and $\mathsf{\Theta}_{1,2|3}^{\text{BR}}$.
$m_{1}^{1}$ and $m_{2}^{2}$ respectively carry unit gauge flux
of $\mathbb{Z}_{N_{1}}$ and $\mathbb{Z}_{N_{2}}$ gauge subgroup.
$e_{3}$, carrying unit gauge charge of $\mathbb{Z}_{N_{3}}$ gauge
subgroup, moves around $m_{1}^{1}$ and $m_{2}^{2}$, such that the
trajectory of $e_{3}$ (denoted as $\gamma_{e_3}$) and other two loops form a Borromean Rings
link. \label{figure_br}}
\end{figure}
For example, consider
a BR braiding process shown in Fig.~\ref{figure_br}, the corresponding TQFT action is
\begin{equation}
S=\int\sum_{i=1}^{3}\frac{N_{i}}{2\pi}B^{i}dA^{i}+\frac{p_{12,3}}{\left(2\pi\right)^{2}}A^{1}A^{2}B^{3},\label{eq_action-BR braiding}
\end{equation}
where $p_{12,3}=\frac{l_{12,3}N_{1}N_{2}N_{3}}{N_{123}},l_{12,3}\in\mathbb{Z}_{N_{123}}$,
$N_{123}$ is the GCD of $N_{1}$, $N_{2}$
and $N_{3}$ . Ref.~\cite{ye17b} points out that $l_{12,3}=-l_{21,3}$ and $l_{ij,k}=0$ if any of the two indices are same. The quantization of $p_{12,3}$ is due to the large
gauge invariance. The
gauge transformations for (\ref{eq_action-BR braiding}) are
\begin{equation}\label{eq_GT_BF_theory_BRbraiding}
\begin{alignedat}{1}A^{1}\rightarrow & A^{1}+d\chi^{1},\\
A^{2}\rightarrow & A^{2}+d\chi^{2},\\
A^{3}\rightarrow & A^{3}+d\chi^{3}+X^{3},\\
B^{1}\rightarrow & B^{1}+dV^{1}+Y^{1},\\
B^{2}\rightarrow & B^{2}+dV^{2}+Y^{2},\\
B^{3}\rightarrow & B^{3}+dV^{3},
\end{alignedat}
\end{equation}
where
\begin{equation}
\begin{alignedat}{1}X^{3}= & -\frac{p_{12,3}}{2\pi N_{3}}\left(\chi^{1}A^{2}+\frac{1}{2}\chi^{1}d\chi^{2}\right)\\
 & +\frac{p_{12,3}}{2\pi N_{3}}\left(\chi^{2}A^{1}+\frac{1}{2}\chi^{2}d\chi^{1}\right),\\
Y^{1}= & -\frac{p_{12,3}}{2\pi N_{1}}\left(\chi^{2}B^{3}-A^{2}V^{3}+\chi^{2}dV^{3}\right),\\
Y^{2}= & \frac{p_{12,3}}{2\pi N_{2}}\left(\chi^{1}B^{3}-A^{1}V^{3}+\chi^{1}dV^{3}\right).
\end{alignedat}
\label{eq_GT-A1A2B3}
\end{equation}
$X^{3}$, $Y^{1}$ and $Y^{2}$ are so-called \emph{shift terms}. The Dirac
quantization of $A^{3}$, $B^{1}$ and $B^{2}$ requires that $\frac{1}{2\pi}\int dX^{3}\in\mathbb{Z}$,
$\frac{1}{2\pi}\int dY^{1}\in\mathbb{Z}$ and $\frac{1}{2\pi}\int dY^{2}\in\mathbb{Z}$.
The $\mathbb{Z}_{N_{i}}$ cyclic group structures are encoded in the cyclic
Wilson integrals of $A^{1}$, $A^{2}$ and $B^{3}$ respectively: $\oint A^{1}\in\frac{2\pi}{N_{1}}\mathbb{Z}_{N_{1}}$,
$\oint A^{2}\in\frac{2\pi}{N_{2}}\mathbb{Z}_{N_{2}}$ and $\oint B^{3}\in\frac{2\pi}{N_{3}}\mathbb{Z}_{N_{3}}$.

\section{Compatible braiding processes\label{section_compatible}}
Since we have reviewed the correspondence between topological terms, braiding processes and braiding phases, we may naively think that we can design a system which can support \emph{arbitrary} combinations of braiding processes. However, this is not true. In other words, there are illegitimate combinations in which the braiding processes are mutually incompatible. Such incompatibility reveals that some topological terms  are \emph{forbidden} to form a legitimate TQFT action.

In this section, we will compute the sets of compatible braiding phases
for different gauge groups, from which the compatible braiding processes
can be read out. For this purpose, we seek for a legitimate TQFT
action consisting of as many topological terms as possible for a given
gauge group.

In order to verify a TQFT action is legitimate or not, we need to take care of the following   aspects. First of all, a legitimate TQFT action should
be invariant up to boundary terms under proper gauge transformations.
Second, the gauge transformations are required to preserve the $\mathbb{Z}_{N_{i}}$
cyclic group structure. This requirement means that, if the $\mathbb{Z}_{N_{i}}$ cyclic group structure is encoded in $\oint A^{i}\in\frac{2\pi}{N_{i}}\mathbb{Z}_{N_{i}}$
$\left(\oint B^{i}\in\frac{2\pi}{N_{i}}\mathbb{Z}_{N_{i}}\right)$, $A^{i}$ $\left(B^{i}\right)$ must have the standard gauge transformation: $A^{i}\rightarrow A^{i}+d\chi^{i}$
$\left(B^{i}\rightarrow B^{i}+dV^{i}\right)$  such that the Wilson integrals of   $A^{i}$ $\left(B^{i}\right)$  are gauge-invariant. $\chi^{i}$ and
$V^{i}$ are $0$-form and $1$-form compact $\mathbb{U}(1)$ gauge parameters
with $\int d\chi^{i}\in2\pi\mathbb{Z}$ and $\int dV^{i}\in2\pi\mathbb{Z}$ respectively. From the perspective of microscopic origins (Sec.~\ref{section_microscopic_derivation}), if $A^i$ has the above standard gauge transformation, the $i$-th layer condensate must be a charge-$N_i$ boson condensate that higgses the Wilson integrals of $A^i$ down to $\Z_{N_i}$, and $B^i$ comes from the duality transformation shown in Sec.~\ref{section_microscopic_derivation} and Appendix \ref{appendix_microscopic_derivation}; likewise, if $B^i$ has the above standard gauge transformation, the $i$-th layer condensate must be a charge-$N_i$ vortexline condensate that higgses the Wilson integrals of $B^i$ down to $\Z_{N_i}$, and  $A^i$ comes from the duality transformation shown in Sec.~\ref{section_microscopic_derivation} and Appendix \ref{appendix_microscopic_derivation}. In summary, in a legitimate action, at least one of gauge fields  ($A^i$ and $B^i$), for a given $i$, should have the above standard gauge transformations and thus have the $\Z_{N_i}$ quantized Wilson integrals. For example, see Eq.~(\ref{eq_GT_BF_theory}), Eq.~(\ref{eq_GT_BF_theory_3loop}), Eq.~(\ref{eq_GT_BF_theory_3loop_plus}), Eq.~(\ref{eq_GT_BF_theory_4loop}), Eq.~(\ref{eq_GT_BF_theory_BRbraiding}).

Next, the gauge transformation of each gauge field should respect
the Dirac quantization. Last but not least, a legitimate TQFT action
should consist of topological terms with nontrivial coefficients.
If the coefficient of a topological term is identical to $0$, otherwise
the action cannot be gauge invariant under gauge transformations,
this topological term is actually incompatible with others in the action.

\subsection{$G=\mathbb{\mathbb{Z}}_{N_{1}}$ and $G=\mathbb{\mathbb{Z}}_{N_{1}}\times\mathbb{\mathbb{Z}}_{N_{2}}$\label{subsec_zn1zn2}}

When $G=\mathbb{Z}_{N_{1}}$, the elementary particle (loop) is $e_{1}$
$\left(m_{1}\right)$ carrying unit gauge charge (flux) of $\mathbb{Z}_{N_{1}}$gauge
group. The only nontrivial braiding process in this case is the particle-loop
braiding described by
\begin{equation}
S=\int\frac{N_{1}}{2\pi}B^{1}dA^{1}.\label{eq_action_partile_loop}
\end{equation}
The gauge transformations for (\ref{eq_action_partile_loop}) are
\begin{equation}
\begin{alignedat}{1}A^{1}\rightarrow & A^{1}+d\chi^{1},\\
B^{1}\rightarrow & B^{1}+dV^{1}.
\end{alignedat}
\label{eq_GT_particle_loop}
\end{equation}
The braiding phase of this particle-loop braiding is
\begin{equation}
\Theta_{1}^{\text{H}}=\frac{2\pi}{N_{1}}.
\end{equation}

When $G=\mathbb{Z}_{N_{1}}\times\mathbb{Z}_{N_{2}}$,
the elementary particles (loops) are denoted as $e_{1}$ and $e_{2}$
($m_{1}$ and $m_{2}$) carrying unit gauge charge (flux) of $\mathbb{Z}_{N_{1}}$
and $\mathbb{Z}_{N_{2}}$ gauge subgroups respectively. Beside particle-loop
braidings, three-loop braidings can be supported. There are $4$ kinds
of three-loop braiding phases in this case, described
by two linearly independent $AAdA$ terms:
\begin{enumerate}
\item $\Theta_{2,2|1}^{\text{3L}}$: $m_{l}^{b}$, $m_{j}^{1}$ and $m_{k}^{2}$
respectively carry unit gauge flux of $\mathbb{Z}_{N_{1}}$, $\mathbb{Z}_{N_{2}}$
and $\mathbb{Z}_{N_{2}}$ gauge subgroup;
\item $\Theta_{1,2|2}^{\text{3L}}$: $m_{l}^{b}$, $m_{j}^{1}$ and $m_{k}^{2}$
respectively carry unit gauge flux of $\mathbb{Z}_{N_{2}}$, $\mathbb{Z}_{N_{1}}$
and $\mathbb{Z}_{N_{2}}$ gauge subgroup;
\item $\Theta_{1,1|2}^{\text{3L}}$: $m_{l}^{b}$, $m_{j}^{1}$ and $m_{k}^{2}$
respectively carry unit gauge flux of $\mathbb{Z}_{N_{2}}$, $\mathbb{Z}_{N_{1}}$
and $\mathbb{Z}_{N_{1}}$ gauge subgroup;
\item $\Theta_{2,1|1}^{\text{3L}}$: $m_{l}^{b}$, $m_{j}^{1}$ and $m_{k}^{2}$
respectively carry unit gauge flux of $\mathbb{Z}_{N_{1}}$, $\mathbb{Z}_{N_{2}}$
and $\mathbb{Z}_{N_{1}}$ gauge subgroup.
\end{enumerate}
The first two braiding processes are described by $A^{1}A^{2}dA^{2}$
and the remainder by $A^{2}A^{1}dA^{1}$. All these three-loop braidings
and particle-loop braidings are compatible, described by a TQFT action
\begin{equation}
\begin{alignedat}{1}S= & \int\sum_{i=1}^{2}\frac{N_{i}}{2\pi}B^{i}dA^{i}\\
 & +\frac{q_{122}}{\left(2\pi\right)^{2}}A^{1}A^{2}dA^{2}+\frac{q_{211}}{\left(2\pi\right)^{2}}A^{2}A^{1}dA^{1}\,,
\end{alignedat}
\label{eq_S-Zn1+2-Hopf link+3L}
\end{equation}
where $q_{122}=\frac{kN_{1}N_{2}}{N_{12}}$, $k\in\mathbb{\mathbb{Z}}_{N_{12}}$ and $q_{211}=\frac{k^{\prime}N_{2}N_{1}}{N_{12}}$, $k^{\prime}\in\mathbb{Z}_{N_{12}}$. $N_{12}$ is the GCD of $N_{1}$ and $N_{2}$. The gauge transformations
for (\ref{eq_S-Zn1+2-Hopf link+3L}) are
\begin{equation}
\begin{alignedat}{1}A^{i}\rightarrow & A^{i}+d\chi^{i},\\
B^{1}\rightarrow & B^{1}+dV^{1}\\
 & +\frac{q_{122}}{\left(2\pi\right)N_{1}}d\chi^{2}A^{2}-\frac{q_{211}}{\left(2\pi\right)N_{1}}d\chi^{2}A^{1},\\
B^{2}\rightarrow & B^{2}+dV^{2}\\
 & -\frac{q_{122}}{\left(2\pi\right)N_{2}}d\chi^{1}A^{2}+\frac{q_{211}}{\left(2\pi\right)N_{2}}d\chi^{1}A^{1}.
\end{alignedat}
\label{eq_GT-Zn1+2-Hopf link+3L}
\end{equation}
The phases of particle-loop braidings and three-loop braidings
are
\begin{align}
\mathsf{\Theta}_{i}^{\text{H}}= & \frac{2\pi}{N_{i}},\\
\Theta_{2,2|1}^{\text{3L}}= & \frac{4\pi q_{122}}{N_{1}N_{2}},\\
\Theta_{1,2|2}^{\text{3L}}= & -\frac{2\pi q_{122}}{N_{1}N_{2}},\\
\Theta_{1,1|2}^{\text{3L}}= & \frac{4\pi q_{211}}{N_{1}N_{2}},\\
\Theta_{2,1|1}^{\text{3L}}= & -\frac{2\pi q_{211}}{N_{1}N_{2}}.
\end{align}

Legitimate TQFT actions for $G=\mathbb{\mathbb{Z}}_{N_{1}}$ and $G=\mathbb{\mathbb{Z}}_{N_{1}}\times\mathbb{\mathbb{Z}}_{N_{2}}$
are summarized in Table~\ref{table_zn1zn2}. The sets of compatible braiding phases for $G=\mathbb{Z}_{N_{1}}$
and $G=\prod_{i=1}^{2}\mathbb{Z}_{N_{i}}$ are respectively
\begin{equation}
\left\{ \Theta_{1}^{\text{H}}=\frac{2\pi}{N_{1}}\right\} _{\mathbb{Z}_{N_{1}}}
\end{equation}
and
\begin{equation}
\left\{ \Theta_{i}^{\text{H}};\underline{\Theta_{2,2|1}^{\text{3L}},\Theta_{1,2|2}^{\text{3L}}},\underline{\Theta_{1,1|2}^{\text{3L}},\Theta_{2,1|1}^{\text{3L}}}\right\} _{\prod_{i=1}^{2}\mathbb{Z}_{N_{i}}}
\end{equation}
where the underlines denote the linear dependence of two braiding phases, for example, $\underline{\Theta_{2,2|1}^{\text{3L}},\Theta_{1,2|2}^{\text{3L}}}$ indicates that $\Theta_{2,2|1}^{\text{3L}}=-2\cdot\Theta_{1,2|2}^{\text{3L}}$.

\begin{table*}
\caption{\textbf{Compatible braiding phases and TQFTs when $G=\mathbb{\mathbb{\mathbb{Z}}}_{N_{1}}$
and $G=\mathbb{\mathbb{\mathbb{Z}}}_{N_{1}}\times\mathbb{\mathbb{\mathbb{Z}}}_{N_{2}}$}.
Each row represents a class of topological order(s). The \nth{1} row
stands for a topological order whose braiding data is given by $\left\{ \mathsf{\Theta}_{i}^{\text{H}}\right\} _{\mathbb{Z}_{N_{1}}}$
with $i=1,2,3$. The \nth{2} row represents topological orders which
are characterized by $\left\{ \mathsf{\Theta}_{i}^{\text{H}},\text{any combinations of compatible $\mathsf{\Theta}^{\text{3L}}$'s}\right\}$. The coefficient of $AAdA$ term is given by $q_{ijj}=\frac{nN_{i}N_{j}}{N_{ij}},n\in\mathbb{\mathbb{Z}}_{N_{ij}}$,
which is determined by the large gauge invariance. $\chi$ and $V$
are $0$-form and $1$-form compact $\mathbb{U}(1)$ gauge parameters
with $\int d\chi\in2\pi\mathbb{Z}$ and $\int dV\in2\pi\mathbb{Z}$.
\label{table_zn1zn2}}
\begin{tabular*}{\textwidth}{@{\extracolsep{\fill}}ccc}
\hline

\hline

\hline
Compatible braiding phases & TQFT actions & Gauge transformations\tabularnewline
\hline
$\Theta_{1}^{\text{H}}=\frac{2\pi}{N_{1}}\mathbb{\mathbb{\mathbb{Z}}}_{N_{1}}$ & $\int\frac{N_{1}}{2\pi}B^{1}dA^{1}$ & $\begin{alignedat}{1}A^{1}\rightarrow & A^{1}+d\chi^{1}\\
B^{1}\rightarrow & B^{1}+dV^{1}
\end{alignedat}
$\tabularnewline
\hline
$\begin{alignedat}{1}\Theta_{i}^{\text{H}}= & \frac{2\pi}{N_{i}}\\
\Theta_{22|1}^{\text{3L}}= & -2\cdot\Theta_{12|2}^{\text{3L}}=\frac{4\pi q_{122}}{N_{1}N_{2}}\\
\Theta_{11|2}^{\text{3L}}= & -2\cdot\Theta_{21|1}^{\text{3L}}=\frac{4\pi q_{211}}{N_{1}N_{2}}
\end{alignedat}
$ & $\begin{alignedat}{1} & \int\sum_{i=1}^{2}\frac{N_{i}}{2\pi}B^{i}dA^{i}\\
+ & \frac{q_{122}}{\left(2\pi\right)^{2}}A^{1}A^{2}dA^{2}+\frac{q_{211}}{\left(2\pi\right)^{2}}A^{2}A^{1}dA^{1}
\end{alignedat}
$ & $\begin{alignedat}{1}A^{i}\rightarrow & A^{i}+d\chi^{i}\\
B^{1}\rightarrow & B^{1}+dV^{1}\\
 & +\frac{q_{122}}{\left(2\pi\right)N_{1}}d\chi^{2}A^{2}-\frac{q_{211}}{\left(2\pi\right)N_{1}}d\chi^{2}A^{1}\\
B^{2}\rightarrow & B^{2}+dV^{2}\\
 & -\frac{q_{122}}{\left(2\pi\right)N_{2}}d\chi^{1}A^{2}+\frac{q_{211}}{\left(2\pi\right)N_{2}}d\chi^{1}A^{1}
\end{alignedat}
$\tabularnewline
\hline

\hline

\hline
\end{tabular*}
\end{table*}

\subsection{$G=\mathbb{Z}_{N_{1}}\times\mathbb{Z}_{N_{2}}\times\mathbb{Z}_{N_{3}}$}\label{subsec_zn1zn2zn3}

When $G=\mathbb{Z}_{N_{1}}\times\mathbb{Z}_{N_{2}}\times\mathbb{Z}_{N_{3}}$,
the elementary particles (loops) are $e_{1}$, $e_{2}$ and $e_{3}$
($m_{1}$, $m_{2}$ and $m_{3}$) carrying unit gauge charge (flux)
of $\mathbb{Z}_{N_{1}}$, $\mathbb{Z}_{N_{2}}$ and $\mathbb{Z}_{N_{3}}$
gauge subgroups respectively. Besides particle-loop braidings and
three-loop braidings discussed in Sec.~\ref{subsec_zn1zn2}, BR braidings
described by $AAB$ terms and three-loop braidings described by $A^{m}A^{n}dA^{o}$
terms ($m,n,o$ are mutually different) are realizable. For the reason
why BR braiding is absent when $G=\prod_{i=1}^{2}\mathbb{Z}_{N_{i}}$,
one can refer to Sec.~\ref{subsec_A1A2B2_not_compatible}.

The compatibility issues of $G=\prod_{i=1}^{3}\mathbb{Z}_{N_{i}}$
are discussed as follows. First of all, if we neglect BR braidings, all particle-loop braidings and three-loop
braidings are compatible, we can write down a legitimate TQFT action
for all root braiding processes except BR braidings:
\begin{equation}
\begin{alignedat}{1} & \int\sum_{i=1}^{3}\frac{N_{i}}{2\pi}B^{i}dA^{i}\\
+ & \frac{q_{122}}{\left(2\pi\right)^{2}}A^{1}A^{2}dA^{2}+\frac{q_{211}}{\left(2\pi\right)^{2}}A^{2}A^{1}dA^{1}\\
+ & \frac{q_{133}}{\left(2\pi\right)^{2}}A^{1}A^{3}dA^{3}+\frac{q_{311}}{\left(2\pi\right)^{2}}A^{3}A^{1}dA^{1}\\
+ & \frac{q_{233}}{\left(2\pi\right)^{2}}A^{2}A^{3}dA^{3}+\frac{q_{322}}{\left(2\pi\right)^{2}}A^{3}A^{2}dA^{2}\\
+ & \frac{q_{123}}{\left(2\pi\right)^{2}}A^{1}A^{2}dA^{3}+\frac{q_{231}}{\left(2\pi\right)^{2}}A^{2}A^{3}dA^{1},
\end{alignedat}
\end{equation}
where $q_{ijk}=\frac{lN_{i}N_{j}}{N_{ij}}$, $l\in\mathbb{Z}_{N_{ijk}}$, $N_{ij}$ ($N_{ijk}$) is the GCD of $N_{i}$, $N_{j}$ (and $N_{k}$). Next, we consider a BR braiding, say, $\mathsf{\Theta}_{1,2|3}^{\text{BR}}$
described by an $A^{1}A^{2}B^{3}$ term. All particle-loop braidings
are compatible with this BR braiding. However, among three-loop braiding phases,
only $\mathsf{\Theta}_{2,2|1}^{\text{3L}}$, $\mathsf{\Theta}_{1,2|2}^{\text{3L}}$,
$\mathsf{\Theta}_{1,1|2}^{\text{3L}}$ and $\mathsf{\Theta}_{2,1|1}^{\text{3L}}$
are compatible with $\mathsf{\Theta}_{1,2|3}^{\text{BR}}$. Other
$\mathsf{\Theta}_{i,j|k}^{\text{3L}}$'s with one index equal to $3$
are not compatible with $\mathsf{\Theta}_{1,2|3}^{\text{BR}}$. The reasons
for this incompatibility between three-loop braidings and BR braidings
are discussed in Sec.~\ref{subsec_A1A3dA3+A1A2B3_not_compatible},~\ref{subsec_A1A2dA3+A1A2B3_not_compatible} and~\ref{subsec_A2A3dA1+A1A2B3_not_compatible}.
The total TQFT action for $\mathsf{\Theta}_{1,2|3}^{\text{BR}}$ and
its compatible braiding processes is
\begin{equation}
\begin{alignedat}{1}\int & \sum_{i=1}^{3}\frac{N_{i}}{2\pi}B^{i}dA^{i}+\frac{p_{12,3}}{\left(2\pi\right)^{2}}A^{1}A^{2}B^{3}\\
 & +\frac{q_{122}}{\left(2\pi\right)^{2}}A^{1}A^{2}dA^{2}+\frac{q_{211}}{\left(2\pi\right)^{2}}A^{2}A^{1}dA^{1},
\end{alignedat}
\end{equation}
where $p_{12,3}=\frac{lN_{1}N_{2}N_{3}}{N_{123}},l\in\mathbb{Z}_{N_{123}}$ and $q_{ijj}=\frac{l^{\prime}N_{i}N_{j}}{N_{ij}},l^{\prime}\in\mathbb{Z}_{N_{ij}}$. Furthermore, if we consider two different BR braidings, we find that
it is impossible to write down a legitimate TQFT action which contains two different $AAB$ terms. The reason is explained in Sec.~\ref{subsec_A1A2B3+A2A3B1_not_compatible}. Nevertheless, if the gauge
group is $G=\prod_{i=1}^{n}\mathbb{Z}_{N_{i}}$ with $n\geq4$, legitimate
TQFT actions for two different BR braidings are possible, which is detailed in Sec.~\ref{subsec_zn1zn2zn3zn4}.

By checking all combinations of braiding processes with the criteria of compatibility, we summarized all legitimate TQFT actions for $G=\mathbb{Z}_{N_{1}}\times\mathbb{Z}_{N_{2}}\times\mathbb{Z}_{N_{3}}$
and corresponding gauge transformations in Table~\ref{table_zn1zn2zn3}. The sets of compatible braiding phases for $G=\prod_{i=1}^{3}\mathbb{Z}_{N_{i}}$
can be summarized as
\begin{equation}
\left\{ \Theta_{i}^{\text{H}};\underline{\Theta_{j,j|i}^{\text{3L}},\Theta_{i,j|j}^{\text{3L}}},\underline{\Theta_{2,3|1}^{\text{3L}},\Theta_{1,3|2}^{\text{3L}}},\underline{\Theta_{3,1|2}^{\text{3L}},\Theta_{2,1|3}^{\text{3L}}}\right\} _{\prod_{i=1}^{3}\mathbb{Z}_{N_{i}}},
\end{equation}
\begin{equation}
\left\{ \Theta_{i}^{\text{H}};\underline{\Theta_{3,3|2}^{\text{3L}},\Theta_{2,3|3}^{\text{3L}}},\underline{\Theta_{2,2|3}^{\text{3L}},\Theta_{3,2|2}^{\text{3L}}};\mathsf{\Theta}_{2,3|1}^{\text{BR}}\right\} _{\prod_{i=1}^{3}\mathbb{Z}_{N_{i}}},
\end{equation}
\begin{equation}
\left\{ \Theta_{i}^{\text{H}};\underline{\Theta_{1,1|3}^{\text{3L}},\Theta_{3,1|1}^{\text{3L}}},\underline{\Theta_{3,3|1}^{\text{3L}},\Theta_{1,3|3}^{\text{3L}}};\mathsf{\Theta}_{3,1|2}^{\text{BR}}\right\} _{\prod_{i=1}^{3}\mathbb{Z}_{N_{i}}},
\end{equation}
and
\begin{equation}
\left\{ \Theta_{i}^{\text{H}};\underline{\Theta_{2,2|1}^{\text{3L}},\Theta_{1,2|2}^{\text{3L}}},\underline{\Theta_{1,1|2}^{\text{3L}},\Theta_{2,1|1}^{\text{3L}}};\mathsf{\Theta}_{1,2|3}^{\text{BR}}\right\} _{\prod_{i=1}^{3}\mathbb{Z}_{N_{i}}},
\end{equation}
where the underlines denote the linear dependence between two compatible
braiding phases: $\Theta_{3,3|2}^{\text{3L}}=-2\cdot\Theta_{2,3|3}^{\text{3L}}$,
$\Theta_{2,3|1}^{\text{3L}}=-\Theta_{1,3|2}^{\text{3L}}$, etc.

\begin{table*}
\caption{\textbf{Compatible braiding phases and TQFTs when $G=\mathbb{\mathbb{\mathbb{Z}}}_{N_{1}}\times\mathbb{\mathbb{\mathbb{Z}}}_{N_{2}}\times\mathbb{\mathbb{\mathbb{Z}}}_{N_{3}}$}.
Each row represents a class of topological orders which are characterized by braiding data $\left\{ \mathsf{\Theta}_{i}^{\text{H}};\text{combinations of compatible braiding phases}\right\}$. Take the \nth{2} row as an example: the sets of braiding phases $\left\{ \mathsf{\Theta}_{i}^{\text{H}};\mathsf{\Theta}_{3,3|2}^{\text{3L}}\right\} $,
$\left\{ \mathsf{\Theta}_{i}^{\text{H}};\mathsf{\Theta}_{2,3|1}^{\text{BR}}\right\} $,$\left\{ \mathsf{\Theta}_{i}^{\text{H}};\mathsf{\Theta}_{3,3|2}^{\text{3L}};\mathsf{\Theta}_{2,3|1}^{\text{BR}}\right\} $,
$\left\{ \Theta_{i}^{\text{H}};\Theta_{3,3|2}^{\text{3L}},\Theta_{2,3|3}^{\text{3L}},\Theta_{2,2|3}^{\text{3L}},\Theta_{3,2|2}^{\text{3L}};\mathsf{\Theta}_{2,3|1}^{\text{BR}}\right\} $,
etc., respectively characterize a topological order. The coefficients are given by $q_{ijk}=\frac{mN_{i}N_{j}}{N_{ij}}$, $m\in\mathbb{\mathbb{Z}}_{N_{ijk}}$ and $p_{ij,k}=\frac{l_{ij,k}N_{i}N_{j}N_{k}}{N_{ijk}}$, $l_{ij,k}\in\mathbb{\mathbb{Z}}_{N_{ijk}}$,
as a result of large gauge invariance. $l_{ij,k}=-l_{ji,k}$ and $l_{ij,k}=0$
if any of the two indices are the same. The Levi-Civita
symbol is defined by $\epsilon^{a_{1}a_{2}\cdots a_{n}}=\prod_{1\protect\leq i<j\protect\leq n}\text{sgn}\left(a_{j}-a_{i}\right)$ where $\text{sgn}\left(x\right)=1$, $0$, or $-1$ if $x>0$, $x=0$,
or $x<0$.
\label{table_zn1zn2zn3}}

\begin{tabular*}{\textwidth}{@{\extracolsep{\fill}}ccc}
\hline

\hline

\hline
Compatible braiding phases & TQFT actions & Gauge transformations\tabularnewline
\hline
$\begin{alignedat}{1}\Theta_{i}^{\text{H}}= & \frac{2\pi}{N_{i}}\\
\Theta_{j,j|i}^{\text{3L}}= & -2\cdot\Theta_{i,j|j}^{\text{3L}}=\frac{4\pi q_{ijj}}{N_{i}N_{j}}\\
\Theta_{2,3|1}^{\text{3L}}= & -\Theta_{1,3|2}^{\text{3L}}=\frac{2\pi q_{123}}{N_{1}N_{2}}\\
\Theta_{3,1|2}^{\text{3L}}= & -\Theta_{2,1|3}^{\text{3L}}=\frac{2\pi q_{231}}{N_{2}N_{3}}
\end{alignedat}
$ & $\begin{alignedat}{1} & \int\sum_{i=1}^{3}\frac{N_{i}}{2\pi}B^{i}dA^{i}\\
+ & \sum_{i,j}\frac{q_{ijj}}{\left(2\pi\right)^{2}}A^{i}A^{j}dA^{j}\\
+ & \frac{q_{123}}{\left(2\pi\right)^{2}}A^{1}A^{2}dA^{3}+\frac{q_{231}}{\left(2\pi\right)^{2}}A^{2}A^{3}dA^{1}
\end{alignedat}
$ & $\begin{alignedat}{1}A^{i}\rightarrow & A^{i}+d\chi^{i}\\
B^{i}\rightarrow & B^{i}+dV^{i}\\
 & +\sum_{j}\left(\frac{q_{ijj}}{2\pi N_{i}}d\chi^{j}A^{j}-\frac{q_{jii}}{2\pi N_{i}}d\chi^{j}A^{i}\right)\\
 & +\frac{1}{2\pi N_{i}}\left(q_{123}\epsilon^{ij3}d\chi^{j}A^{3}+q_{231}\epsilon^{ij1}d\chi^{j}A^{1}\right)
\end{alignedat}
$\tabularnewline
\hline
$\begin{alignedat}{1}\Theta_{i}^{\text{H}}= & \frac{2\pi}{N_{i}}\\
\Theta_{3,3|2}^{\text{3L}}= & -2\cdot\Theta_{2,3|3}^{\text{3L}}=\frac{4\pi q_{233}}{N_{2}N_{3}}\\
\Theta_{2,2|3}^{\text{3L}}= & -2\cdot\Theta_{3,2|2}^{\text{3L}}=\frac{4\pi q_{322}}{N_{3}N_{2}}\\
\Theta_{2,3|1}^{\text{BR}}= & \frac{2\pi p_{23,1}}{N_{1}N_{2}N_{3}}
\end{alignedat}
$ & $\begin{alignedat}{1}& \int \sum_{i=1}^{3}\frac{N_{i}}{2\pi}B^{i}dA^{i}\\
+ & \frac{q_{233}}{\left(2\pi\right)^{2}}A^{2}A^{3}dA^{3}+\frac{q_{322}}{\left(2\pi\right)^{2}}A^{3}A^{2}dA^{2}\\
+&
 \frac{p_{23,1}}{\left(2\pi\right)^{2}}A^{2}A^{3}B^{1}
\end{alignedat}
$ & $\begin{alignedat}{1}A^{i}\rightarrow & A^{i}+d\chi^{i}\\
 & -\epsilon^{i23}\frac{p_{23,1}}{2\pi N_{i}}\left(\chi^{2}A^{3}+\frac{1}{2}\chi^{2}d\chi^{3}\right)\\
 & +\epsilon^{i23}\frac{p_{23,1}}{2\pi N_{i}}\left(\chi^{3}A^{2}+\frac{1}{2}\chi^{3}d\chi^{2}\right)\\
B^{i}\rightarrow & B^{i}+dV^{i}\\
 & +\left(\frac{q_{ijj}}{2\pi N_{i}}d\chi^{j}A^{j}-\frac{q_{jii}}{2\pi N_{i}}d\chi^{j}A^{i}\right)\\
 & -\epsilon^{ij1}\frac{p_{23,1}}{2\pi N_{i}}\left(\chi^{j}B^{1}-A^{j}V^{1}+\chi^{j}dV^{1}\right)
\end{alignedat}
$\tabularnewline
\hline
$\begin{alignedat}{1}\Theta_{i}^{\text{H}}= & \frac{2\pi}{N_{i}}\\
\Theta_{3,3|1}^{\text{3L}}= & -2\cdot\Theta_{1,3|3}^{\text{3L}}=\frac{4\pi q_{133}}{N_{1}N_{3}}\\
\Theta_{1,1|3}^{\text{3L}}= & -2\cdot\Theta_{3,1|1}^{\text{3L}}=\frac{4\pi q_{311}}{N_{3}N_{1}}\\
\Theta_{3,1|2}^{\text{BR}}= & \frac{2\pi p_{31,2}}{N_{1}N_{2}N_{3}}
\end{alignedat}
$ & $\begin{alignedat}{1} & \int\sum_{i=1}^{3}\frac{N_{i}}{2\pi}B^{i}dA^{i}\\
+ & \frac{q_{133}}{\left(2\pi\right)^{2}}A^{1}A^{3}dA^{3}+\frac{q_{311}}{\left(2\pi\right)^{2}}A^{3}A^{1}dA^{1}\\
+ & \frac{p_{31,2}}{\left(2\pi\right)^{2}}A^{3}A^{1}B^{2}
\end{alignedat}
$ & $\begin{alignedat}{1}A^{i}\rightarrow & A^{i}+d\chi^{i}\\
 & -\epsilon^{1i3}\frac{p_{31,2}}{2\pi N_{i}}\left(\chi^{3}A^{1}+\frac{1}{2}\chi^{3}d\chi^{1}\right)\\
 & +\epsilon^{1i3}\frac{p_{31,2}}{2\pi N_{i}}\left(\chi^{1}A^{3}+\frac{1}{2}\chi^{1}d\chi^{3}\right)\\
B^{i}\rightarrow & B^{i}+dV^{i}\\
 & +\left(\frac{q_{ijj}}{2\pi N_{i}}d\chi^{j}A^{j}-\frac{q_{jii}}{2\pi N_{i}}d\chi^{j}A^{i}\right)\\
 & -\epsilon^{ij2}\frac{p_{31,2}}{2\pi N_{i}}\left(\chi^{j}B^{3}-A^{j}V^{3}+\chi^{j}dV^{3}\right)
\end{alignedat}
$\tabularnewline
\hline
$\begin{alignedat}{1}\Theta_{i}^{\text{H}}= & \frac{2\pi}{N_{i}}\\
\Theta_{2,2|1}^{\text{3L}}= & -2\cdot\Theta_{1,2|2}^{\text{3L}}=\frac{4\pi q_{122}}{N_{1}N_{2}}\\
\Theta_{1,1|2}^{\text{3L}}= & -2\cdot\Theta_{2,1|1}^{\text{3L}}=\frac{4\pi q_{211}}{N_{1}N_{2}}\\
\Theta_{1,2|3}^{\text{BR}}= & \frac{2\pi p_{12,3}}{N_{1}N_{2}N_{3}}
\end{alignedat}
$ & $\begin{alignedat}{1} & \int\sum_{i=1}^{3}\frac{N_{i}}{2\pi}B^{i}dA^{i}\\
+ & \frac{q_{122}}{\left(2\pi\right)^{2}}A^{1}A^{2}dA^{2}+\frac{q_{211}}{\left(2\pi\right)^{2}}A^{2}A^{1}dA^{1}\\
+ & \frac{p_{12,3}}{\left(2\pi\right)^{2}}A^{1}A^{2}B^{3}
\end{alignedat}
$ & $\begin{alignedat}{1}A^{i}\rightarrow & A^{i}+d\chi^{i}\\
 & -\epsilon^{i12}\frac{p_{12,3}}{2\pi N_{i}}\left(\chi^{1}A^{2}+\frac{1}{2}\chi^{1}d\chi^{2}\right)\\
 & +\epsilon^{i12}\frac{p_{12,3}}{2\pi N_{i}}\left(\chi^{2}A^{1}+\frac{1}{2}\chi^{2}d\chi^{1}\right)\\
B^{i}\rightarrow & B^{i}+dV^{i}\\
 & +\left(\frac{q_{ijj}}{2\pi N_{i}}d\chi^{j}A^{j}-\frac{q_{jii}}{2\pi N_{i}}d\chi^{j}A^{i}\right)\\
 & -\epsilon^{ij3}\frac{p_{12,3}}{2\pi N_{i}}\left(\chi^{j}B^{3}-A^{j}V^{3}+\chi^{j}dV^{3}\right)
\end{alignedat}
$\tabularnewline
\hline

\hline

\hline
\end{tabular*}
\end{table*}

\subsection{$G=\mathbb{Z}_{N_{1}}\times\mathbb{Z}_{N_{2}}\times\mathbb{Z}_{N_{3}}\times\mathbb{Z}_{N_{4}}$}\label{subsec_zn1zn2zn3zn4}
Similar to the case in Sec.~\ref{subsec_zn1zn2zn3}, particle-loop braidings
and multi-loop braidings are compatible with each other when $G=\prod_{i=1}^{4}\mathbb{Z}_{N_{i}}$. The TQFT action for them is
\begin{equation}
\begin{alignedat}{1} & \int\sum_{i=1}^{4}\frac{N_{i}}{2\pi}B^{i}dA^{i}\\
+ & \sum_{i<j}\left[\frac{q_{ijj}}{\left(2\pi\right)^{2}}A^{i}A^{j}dA^{j}+\frac{q_{jii}}{\left(2\pi\right)^{2}}A^{j}A^{i}dA^{i}\right]\\
+ & \sum_{i<j<k}\left[\frac{q_{ijk}}{\left(2\pi\right)^{2}}A^{i}A^{j}dA^{k}+\frac{q_{jki}}{\left(2\pi\right)^{2}}A^{j}A^{k}dA^{i}\right]\\
+ & \frac{q_{1234}}{\left(2\pi\right)^{3}}A^{1}A^{2}A^{3}A^{4}.
\end{alignedat}
\label{eq_action_zn1-4_particleloop+multiloop}
\end{equation}
The set of compatible braiding phases
for (\ref{eq_action_zn1-4_particleloop+multiloop}) is
\begin{equation}
\begin{alignedat}{1}\left\{ \Theta_{i}^{\text{H}};\underline{\Theta_{j,j|i}^{\text{3L}},\Theta_{i,j|j}^{\text{3L}}},\right. & \underline{\Theta_{i,i|j}^{\text{3L}},\Theta_{j,i|i}^{\text{3L}}},\\
\underline{\Theta_{j,k|i}^{\text{3L}},\Theta_{i,k|j}^{\text{3L}}}, & \left.\underline{\Theta_{k,i|j}^{\text{3L}},\Theta_{j,i|k}^{\text{3L}}};\Theta_{1,2,3,4}^{\text{4L}}\right\}_{\prod_{i=1}^{4}\mathbb{Z}_{i}}
\end{alignedat}
\end{equation}
where $i<j<k$ and $\left\{ i,j,k\right\} \subset\left\{ 1,2,3,4\right\} $.

When $G=\prod_{i=1}^{4}\mathbb{Z}_{i}$, if we take BR braidings into account, we need to treat them carefully.
If these BR braidings involve only \emph{one} kind of elementary particles,
say, $e_{1}$, corresponding to $A^{2}A^{3}B^{1}$, $A^{2}A^{4}B^{1}$
and $A^{3}A^{4}B^{1}$ terms, the legitimate TQFT action is
\begin{equation}
\begin{alignedat}{1} & \int\sum_{i=1}^{4}\frac{N_{i}}{2\pi}B^{i}dA^{i}+\sum_{i,j\neq1}\frac{q_{ijj}}{\left(2\pi\right)^{2}}A^{i}A^{j}dA^{j}\\
+ & \frac{q_{234}}{\left(2\pi\right)^{2}}A^{2}A^{3}dA^{4}+\frac{q_{342}}{\left(2\pi\right)^{2}}A^{3}A^{4}dA^{2}\\
+ & \frac{p_{23,1}}{\left(2\pi\right)^{2}}A^{2}A^{3}B^{1}+\frac{p_{24,1}}{\left(2\pi\right)^{2}}A^{2}A^{4}B^{1}+\frac{p_{34,1}}{\left(2\pi\right)^{2}}A^{3}A^{4}B^{1},
\end{alignedat}
\end{equation}
corresponding to the set of compatible braiding phases
\begin{equation}
\begin{alignedat}{1}\left\{ \Theta_{i}^{\text{H}};\underline{\Theta_{s,s|r}^{\text{3L}},\Theta_{r,s|s}^{\text{3L}}},\right. & \underline{\Theta_{3,4|2}^{\text{3L}},\Theta_{2,4|3}^{\text{3L}}},\\
\underline{\Theta_{4,2|3}^{\text{3L}},\Theta_{3,2|4}^{\text{3L}}}; & \left.\mathsf{\Theta}_{4,2|3}^{\text{BR}};\mathsf{\Theta}_{2,4|1}^{\text{BR}};\mathsf{\Theta}_{3,4|1}^{\text{BR}}\right\}_{\prod_{i=1}^{4}\mathbb{Z}_{i}}
\end{alignedat}
\end{equation}
with $r\neq1$ and $s\neq1$. In other words, any multi-loop braiding which involves elementary
loop $m_{1}$ is incompatible with BR braidings which involve elementary
particle $e_{1}$.

If we consider BR braidings which involve \emph{two} kinds of elementary
particles, e.g., $e_{1}$ and $e_{2}$, the legitimate TQFT action
is
\begin{equation}
\begin{alignedat}{1} & \int\sum_{i=1}^{4}\frac{N_{i}}{2\pi}B^{i}dA^{i}\\
+ & \frac{q_{344}}{\left(2\pi\right)^{2}}A^{3}A^{4}dA^{4}+\frac{q_{433}}{\left(2\pi\right)^{2}}A^{4}A^{3}dA^{3}\\
+ & \frac{p_{34,1}}{\left(2\pi\right)^{2}}A^{3}A^{4}B^{1}+\frac{p_{34,2}}{\left(2\pi\right)^{2}}A^{3}A^{4}B^{2}
\end{alignedat}
\end{equation}
which excludes multi-loop braidings which involve elementary loops $m_{1}$
and $m_{2}$. The compatible braiding phases form a set:
\begin{equation}
\left\{ \Theta_{i}^{\text{H}};\underline{\Theta_{4,4|3}^{\text{3L}},\Theta_{3,4|4}^{\text{3L}}},\underline{\Theta_{3,3|4}^{\text{3L}},\Theta_{4,3|3}^{\text{3L}}};\mathsf{\Theta}_{3,4|1}^{\text{BR}};\mathsf{\Theta}_{3,4|2}^{\text{BR}}\right\} _{\prod_{i=1}^{4}\mathbb{Z}_{N_{i}}}.
\end{equation}
Notice that when we write down a BR braiding which involves
one kind of elementary particles, we have 4 choices since there are
4 distinct gauge subgroups. If we consider two BR braidings with two kinds of elementary particles, we have 6 different combinations of these two different elementary particles. Therefore, there are total
$1+4+6=11$ legitimate TQFT actions, i.e., 11 different sets of compatible braiding phases, when $G=\prod_{i=1}^{4}\mathbb{Z}_{N_{i}}$.
Due to the limiting space of page, Table~\ref{table_zn1zn2zn3zn4}
only list 3 examples of TQFT actions and their gauge transformations.
By properly reassigning the indices of the TQFT actions listed in Table~\ref{table_zn1zn2zn3zn4}, we can construct all legitimate TQFT actions and corresponding gauge transformations. One can find details in Appendix~\ref{sec_all_actions_zn1zn2zn3zn4} and Table~\ref{table_ALL_zn1zn2zn3zn4} therein.

\subsection{Results for general gauge groups}\label{sec_general_gauge_groups}

In fact, our previous discussions can be easily generalized to the
case where the number of $\mathbb{Z}_{N_{i}}$ subgroups are arbitrary.
As we can see in the previous main text, compatibility is always guaranteed
for particle-loop braidings and multi-loop braidings, or, particle-loop
braidings and BR braidings. The general rule of incompatibility is
that if two braiding processes (multi-loop braiding and BR braiding,
or two different BR braidings ) involve particle and loop that carry
gauge charge and gauge flux of the \emph{same} $\mathbb{Z}_{N_{i}}$
gauge subgroup, these two braiding processes are incompatible with
each other. In the language of TQFT, a $\mathbb{Z}_{N_{i}}$ gauge
subgroup is associated with a topological term $B^{i}dA^{i}$. In
a legitimate TQFT action, given the $BF$ term $B^{i}dA^{i}$, only
$A^{i}$ or $B^{i}$, not both, can appear in the twisted terms (i.e.,
$AAdA$, $AAAA$ and $AAB$). Otherwise, the TQFT action would be
illegitimate, e.g., violating the gauge invariance.

Viewed from the microscopic origin of topological terms, the rule
of incompatibility is natural. Each layer of condensate (corresponding
to each $\mathbb{Z}_{N_{i}}$ gauge subgroup) is either in the charge
condensation or in the vortexline condensation, never both. The $\mathbb{Z}_{N_{i}}$
charge condensation implies particles that carry gauge charge $e_{i}$
(corresponding to $A^{i}$), while the $\mathbb{Z}_{N_{i}}$ string
condensation results in loops that carry gauge flux $m_{i}$ (corresponding
to $B^{i}$). If a multi-loop braiding or BR braiding involves the
$\mathbb{Z}_{N_{i}}$ particle, it is incompatible with other braiding process that involves the $\mathbb{Z}_{N_{i}}$
loop, and vice versa.

\section{Incompatible braiding processes\label{section_incompatible}}
In Sec.~\ref{section_compatible}, we discuss the compatible braiding
processes in different cases. We find that, when the gauge
group is given, the set of compatible braiding processes is only a
subset of $\left\{ \text{all possible root braiding processes}\right\}$. This means that some braiding
processes are mutually incompatible. For example, when $G=\prod_{i=1}^{3}\mathbb{Z}_{i}$ (see Sec.~\ref{subsec_zn1zn2zn3}),
a set of compatible braiding phases is $\left\{ \Theta_{i}^{\text{H}};\Theta_{2,2|1}^{\text{3L}},\Theta_{1,2|2}^{\text{3L}},\Theta_{1,1|2}^{\text{3L}},\Theta_{2,1|1}^{\text{3L}};\mathsf{\Theta}_{1,2|3}^{\text{BR}}\right\} _{\prod_{i=1}^{3}\mathbb{Z}_{N_{i}}}$. We can see that, among all possible root braiding phases, $\mathsf{\Theta}_{i,j|k}^{\text{3L}}$ with one index equal to 3,
$\mathsf{\Theta}_{2,3|1}^{\text{BR}}$ and $\mathsf{\Theta}_{3,1|2}^{\text{BR}}$
are excluded from this set, because they are not compatible with $\mathsf{\Theta}_{1,2|3}^{\text{BR}}$.
In this section, we will demonstrate several examples of such incompatibility between braiding processes via the TQFT perspective. The basic idea is: if an action consists of topological terms corresponding to incompatible braiding processes, it would never be a legitimate TQFT theory.
\begin{table*}[tbp]
\caption{\textbf{Compatible braiding phases and TQFTs when $G=\mathbb{Z}_{N_{1}}\times\mathbb{Z}_{N_{2}}\times\mathbb{Z}_{N_{3}}\times\mathbb{Z}_{N_{4}}$}. Each row represents a class of topological orders which are characterized by braiding data $\left\{ \mathsf{\Theta}_{i}^{\text{H}};\text{combinations of compatible braiding phases}\right\}$. Only 3 of 11 (see the main text) legitimate TQFT actions are listed here. General expressions of legitimate TQFT actions and gauge transformations are given in Table~\ref{table_ALL_zn1zn2zn3zn4}. By properly assigning the indices in the general expressions, one can obtain all legitimate TQFT actions for $G=\prod_{i=1}^{4}\mathbb{Z}_{N_{i}}$.
Coefficients are given by $q_{ijk}=\frac{mN_{i}N_{j}}{N_{ij}}$, $m\in\mathbb{\mathbb{Z}}_{N_{ijk}}$;
$q_{ijkl}=\frac{nN_{i}N_{j}N_{k}N_{l}}{N_{ijkl}}$, $n\in\mathbb{\mathbb{Z}}_{N_{ijkl}}$
and $p_{ij,k}=\frac{l_{ij,k}N_{i}N_{j}N_{k}}{N_{ijk}}$, $l_{ij,k}\in\mathbb{\mathbb{Z}}_{N_{ijk}}$
as a result of large gauge invariance. $l_{ij,k}=-l_{ji,k}$ and $l_{ij,k}=0$
if any of the two indices are the same. The Levi-Civita
symbol is defined by $\epsilon^{a_{1}a_{2}\cdots a_{n}}=\prod_{1\protect\leq i<j\protect\leq n}\text{sgn}\left(a_{j}-a_{i}\right)$ where $\text{sgn}\left(x\right)=1$, $0$, or $-1$ if $x>0$, $x=0$,
or $x<0$.
\label{table_zn1zn2zn3zn4}}
\begin{tabular*}{\textwidth}{@{\extracolsep{\fill}}ccc}
\hline

\hline

\hline
Compatible braiding phases & TQFT actions & Gauge transformations\tabularnewline
\hline
$\begin{alignedat}{1}\Theta_{i}^{\text{H}}= & \frac{2\pi}{N_{i}}\\
\Theta_{j,j|i}^{\text{3L}}= & -2\cdot\Theta_{i,j|j}^{\text{3L}}=\frac{4\pi q_{ijj}}{N_{i}N_{j}}\\
\Theta_{i,i|j}^{\text{3L}}= & -2\cdot\Theta_{j,i|i}^{\text{3L}}=\frac{4\pi q_{jii}}{N_{j}N_{i}}\\
\Theta_{j,k|i}^{\text{3L}}= & -\Theta_{i,k|j}^{\text{3L}}=\frac{2\pi q_{ijk}}{N_{i}N_{j}}\\
\Theta_{k,i|j}^{\text{3L}}= & -\Theta_{j,i|k}^{\text{3L}}=\frac{2\pi q_{jki}}{N_{j}N_{k}}\\
\Theta_{1,2,3,4}^{\text{4L}}= & \frac{2\pi q_{1234}}{N_{1}N_{2}N_{3}N_{4}}
\end{alignedat}
$ & $\begin{alignedat}{1} &  \int\sum_{i=1}^{4}\frac{N_{i}}{2\pi}B^{i}dA^{i}\\
+ & \sum_{i<j}\frac{q_{ijj}}{\left(2\pi\right)^{2}}A^{i}A^{j}dA^{j}\\
+ & \sum_{i<j}\frac{q_{jii}}{\left(2\pi\right)^{2}}A^{j}A^{i}dA^{i}\\
+ & \sum_{i<j<k}\frac{q_{ijk}}{\left(2\pi\right)^{2}}A^{i}A^{j}dA^{k}\\
+ & \sum_{i<j<k}\frac{q_{jki}}{\left(2\pi\right)^{2}}A^{j}A^{k}dA^{i}\\
+ & \frac{q_{1234}}{\left(2\pi\right)^{3}}A^{1}A^{2}A^{3}A^{4}
\end{alignedat}
$ & $\begin{alignedat}{1}A^{i}\rightarrow & A^{i}+d\chi^{i}\\
B^{i}\rightarrow & B^{i}+dV^{i}\\
 & +\sum_{j}\left(\frac{q_{ijj}}{2\pi N_{i}}d\chi^{j}A^{j}-\frac{q_{jii}}{2\pi N_{i}}d\chi^{j}A^{i}\right)\\
 & +\sum_{m<n<l}\frac{q_{mnl}}{2\pi N_{i}}\left(\delta_{i,m}d\chi^{n}A^{l}-\delta_{i,n}d\chi^{m}A^{l}\right)\\
 & +\sum_{m<n<l}\frac{q_{nml}}{2\pi N_{i}}\left(\delta_{i,n}d\chi^{l}A^{n}-\delta_{i,l}d\chi^{n}A^{m}\right)\\
 & -\frac{1}{2}\sum_{j,k,l}\frac{q_{1234}}{\left(2\pi\right)^{2}N_{i}}\epsilon^{ijkl}A^{j}A^{k}\chi^{l}\\
 & +\frac{1}{2}\sum_{j,k,l}\frac{q_{1234}}{\left(2\pi\right)^{2}N_{i}}\epsilon^{ijkl}A^{j}\chi^{k}d\chi^{l}\\
 & +\frac{1}{6}\sum_{j,k,l}\frac{q_{1234}}{\left(2\pi\right)^{2}N_{i}}\epsilon^{ijkl}\chi^{j}d\chi^{k}d\chi^{l}
\end{alignedat}
$\tabularnewline
\hline
$\begin{alignedat}{1}\Theta_{i}^{\text{H}}= & \frac{2\pi}{N_{i}}\\
\Theta_{j,j|i}^{\text{3L}}= & -2\cdot\Theta_{i,j|j}^{\text{3L}}=\frac{4\pi q_{ijj}}{N_{i}N_{j}}\\
\Theta_{2,4|1}^{\text{3L}}= & -\Theta_{1,4|2}^{\text{3L}}=\frac{2\pi q_{124}}{N_{1}N_{2}}\\
\Theta_{4,1|2}^{\text{3L}}= & -\Theta_{2,1|4}^{\text{3L}}=\frac{2\pi q_{241}}{N_{2}N_{4}}\\
\Theta_{1,2|3}^{\text{BR}}= & \frac{2\pi p_{12,3}}{N_{1}N_{2}N_{3}}\\
\Theta_{4,1|3}^{\text{BR}}= & \frac{2\pi p_{41,3}}{N_{4}N_{1}N_{3}}\\
\Theta_{4,2|3}^{\text{BR}}= & \frac{2\pi p_{42,3}}{N_{4}N_{2}N_{3}}
\end{alignedat}
$ & $\begin{alignedat}{1} & \int\sum_{i=1}^{4}\frac{N_{i}}{2\pi}B^{i}dA^{i}\\
+ & \sum_{i\neq3,j\neq3}\frac{q_{ijj}}{\left(2\pi\right)^{2}}A^{i}A^{j}dA^{j}\\
+ & \frac{q_{124}}{\left(2\pi\right)^{2}}A^{1}A^{2}dA^{4}+\frac{q_{241}}{\left(2\pi\right)^{2}}A^{2}A^{4}dA^{1}\\
+ & \frac{p_{12,3}}{\left(2\pi\right)^{2}}A^{1}A^{2}B^{3}\\
+ & \frac{p_{41,3}}{\left(2\pi\right)^{2}}A^{4}A^{1}B^{3}\\
+ & \frac{p_{42,3}}{\left(2\pi\right)^{2}}A^{4}A^{2}B^{3}
\end{alignedat}
$ & $\begin{alignedat}{1}A^{i}\rightarrow & A^{i}+d\chi^{i}\\
 & -\sum_{a,b}\frac{p_{ab,3}}{2\pi N_{3}}\delta_{i,3}\left(\chi^{a}A^{b}+\frac{1}{2}\chi^{a}d\chi^{b}\right)\\
B^{i}\rightarrow & B^{i}+dV^{i}\\
 & +\sum_{i\neq3,j\neq3}\left[\frac{q_{ijj}}{2\pi N_{i}}d\chi^{j}A^{j}-\frac{q_{jii}}{2\pi N_{i}}d\chi^{j}A^{i}\right]\\
 & +\sum_{i\neq3,j\neq3}\left(\frac{q_{124}}{2\pi N_{i}}\epsilon^{ij4}d\chi^{j}A^{4}+\frac{q_{241}}{2\pi N_{i}}\epsilon^{ij1}d\chi^{j}A^{1}\right)\\
 & -\sum_{i\neq3,j\neq3}\frac{p_{ij,3}}{2\pi N_{i}}\left(\chi^{j}B^{3}-A^{j}V^{3}+\chi^{j}dV^{3}\right)
\end{alignedat}
$\tabularnewline
\hline
$\begin{alignedat}{1}\Theta_{i}^{\text{H}}= & \frac{2\pi}{N_{i}}\\
\Theta_{2,2|1}^{\text{3L}}= & -2\cdot\Theta_{1,2|2}^{\text{3L}}=\frac{4\pi q_{122}}{N_{1}N_{2}}\\
\Theta_{1,1|2}^{\text{3L}}= & -2\cdot\Theta_{2,1|1}^{\text{3L}}=\frac{4\pi q_{211}}{N_{2}N_{1}}\\
\Theta_{1,2|3}^{\text{BR}}= & \frac{2\pi p_{12,3}}{N_{1}N_{2}N_{3}}\\
\Theta_{1,2|4}^{\text{BR}}= & \frac{2\pi p_{12,4}}{N_{1}N_{2}N_{4}}
\end{alignedat}
$ & $\begin{alignedat}{1} & \int\sum_{i=1}^{4}\frac{N_{i}}{2\pi}B^{i}dA^{i}\\
+ & \frac{q_{122}}{\left(2\pi\right)^{2}}A^{1}A^{2}dA^{2}+\frac{q_{211}}{\left(2\pi\right)^{2}}A^{2}A^{1}dA^{1}\\
+ & \frac{p_{12,3}}{\left(2\pi\right)^{2}}A^{1}A^{2}B^{3}+\frac{p_{12,4}}{\left(2\pi\right)^{2}}A^{1}A^{2}B^{4}
\end{alignedat}
$ & $\begin{alignedat}{1}A^{i}\rightarrow & A^{i}+d\chi^{i}\\
 & -\left(\frac{p_{12,i}}{2\pi N_{i}}\delta_{i,3}+\frac{p_{12,i}}{2\pi N_{i}}\delta_{i,4}\right)\left(\chi^{1}A^{2}+\frac{1}{2}\chi^{1}d\chi^{2}\right)\\
 & +\left(\frac{p_{12,i}}{2\pi N_{i}}\delta_{i,3}+\frac{p_{12,i}}{2\pi N_{i}}\delta_{i,4}\right)\left(\chi^{2}A^{1}+\frac{1}{2}\chi^{2}d\chi^{1}\right)\\
B^{i}\rightarrow & B^{i}+dV^{i}\\
 & +\sum_{j\neq3,4}\left(\frac{q_{ijj}}{2\pi N_{i}}d\chi^{j}A^{j}-\frac{q_{jii}}{2\pi N_{i}}d\chi^{j}A^{i}\right)\\
 & -\sum_{j}\frac{p_{12,j}}{2\pi N_{i}}\delta_{i,1}\left(\chi^{2}B^{j}-A^{2}V^{j}+\chi^{2}dV^{j}\right)\\
 & +\sum_{j}\frac{p_{12,j}}{2\pi N_{i}}\delta_{i,2}\left(\chi^{1}B^{j}-A^{1}V^{j}+\chi^{1}dV^{3}\right)
\end{alignedat}
$\tabularnewline
\hline

\hline

\hline
\end{tabular*}
\end{table*}

\subsection{Absence of BR braidings when $G=\prod_{i=1}^{2}\mathbb{Z}_{i}$  \label{subsec_A1A2B2_not_compatible}}

In Sec.~\ref{subsec_br_braiding}, we review the BR braiding in which a particle carrying unit $\mathbb{Z}_{N_{3}}$
gauge charge moves around two loops which carry unit $\mathbb{Z}_{N_{1}}$
and $\mathbb{Z}_{N_{2}}$ gauge flux respectively. Such a braiding process is
only possible when there are \emph{three} distinct $\mathbb{Z}_{N_{i}}$ gauge subgroups.
It is natural to ask: why it is impossible when there are only \emph{two }$\mathbb{Z}_{N_{i}}$
gauge subgroups, i.e., $G=\prod_{i=1}^{2}\mathbb{Z}_{N_{i}}$? Without
loss of generality, we can consider the following braiding process: a
particle $e_{2}$ carrying unit $\mathbb{Z}_{N_{2}}$ gauge charge moves around
two loops $m_{1}$ and $m_{2}$ which carry unit $\mathbb{Z}_{N_{1}}$ and $\mathbb{Z}_{N_{2}}$
gauge flux respectively such that the trajectory of $e_{1}$ and two loops
form a Borromean Rings link. One may naively think that, this braiding
process, along with its braiding phase $\mathsf{\Theta}_{12,2}^{\text{BR}}$,
is described by the topological term $A^{1}A^{2}B^{2}$, analogous
to the case in Sec.~\ref{subsec_br_braiding}. The corresponding TQFT action should be
\begin{equation}
S=\int\sum_{i=1}^{2}\frac{N_{i}}{2\pi}B^{i}dA^{i}+\frac{p}{\left(2\pi\right)^{2}}A^{1}A^{2}B^{2}.
\label{eq_action-A1A2B2}
\end{equation}
However, the action~(\ref{eq_action-A1A2B2}) is not a legitimate TQFT action because
we cannot find gauge transformations  which respects gauge invariance
and $\mathbb{Z}_{N}$ cyclic group structure simultaneously.
We are going to illustrate this in details.

In the action (\ref{eq_action-A1A2B2}), $B^{1}$ serves as a Lagrange
multiplier imposing a local constraint $dA^{1}=0$. This means that the gauge
transformation of $A^{1}$ is $A^{1}\rightarrow A^{1}+d\chi^{1}$. Thus, the $\mathbb{Z}_{N_{1}}$ cyclic group structure is encoded in $\oint A^{i}\in\frac{2\pi}{N_{i}}\mathbb{Z}_{N_{i}}$.
The $\mathbb{Z}_{N_{2}}$ cyclic group structure requires
that at least one of $A^{2}\rightarrow A^{2}+d\chi^{2}$ and $B^{2}\rightarrow B^{2}+dV^{2}$
holds.

First, we assume that the gauge transformations are
\begin{equation}
\begin{alignedat}{1}A^{i}\rightarrow & A^{i}+d\chi^{i},\\
B^{i}\rightarrow & B^{i}+dV^{i}+Y^{i},
\end{alignedat}
\label{eq_GT-A1A2B2-A1+A2-nromal}
\end{equation}
 where $Y^{i}$ is a shift term with $\oint Y^{i}\notin\frac{2\pi}{N_{i}}\mathbb{\mathbb{\mathbb{Z}}}_{N_{i}}$ hence the $\mathbb{Z}_{N_{2}}$ cyclic group
structure is encoded in $\oint A^{2}\in\frac{2\pi}{N_{2}}\mathbb{Z}_{N_{2}}$.
Under the gauge transformations~(\ref{eq_GT-A1A2B2-A1+A2-nromal}), the variation of action~(\ref{eq_action-A1A2B2}) (boundary terms are neglected) is
\begin{equation}
\begin{alignedat}{1}\Delta S= & \int\frac{N_{1}}{2\pi}Y^{1}dA^{1}+\frac{N_{2}}{2\pi}Y^{2}dA^{2}\\
 & +\frac{p}{\left(2\pi\right)^{2}}\left[\left(d\chi^{1}A^{2}B^{2}+A^{1}d\chi^{2}B^{2}+d\chi^{1}d\chi^{2}B^{2}\right)\right.\\
 & +\left(A^{1}A^{2}dV^{2}+d\chi^{1}A^{2}dV^{2}+A^{1}d\chi^{2}dV^{2}\right)\\
 & +\left.\left(A^{1}A^{2}Y^{2}+d\chi^{1}A^{2}Y^{2}+A^{1}d\chi^{2}Y^{2}+d\chi^{1}d\chi^{2}Y^{2}\right)\right]
\end{alignedat}
\label{eq_Delta-S-A1A2B2}
\end{equation}
which should be an integral of total derivative terms in order to be gauge invariant. Focus on the $d\chi^{1}d\chi^{2}B^{2}$ term which is not a total derivative term: we want to eliminate it
by subtraction or by absorbing it into a total derivative term. If we want to
eliminate $d\chi^{1}d\chi^{2}B^{2}$ by subtraction, the only way is to require
\begin{equation}
\begin{alignedat}{1} & \frac{p}{\left(2\pi\right)^{2}}d\chi^{1}d\chi^{2}B^{2}+\frac{p}{\left(2\pi\right)^{2}}d\chi^{1}d\chi^{2}Y^{2}\\
= & \frac{p}{\left(2\pi\right)^{2}}d\chi^{1}d\chi^{2}\left(B^{2}-B^{2}+\cdots\right)
\end{alignedat}
\label{eq_A1A2B2-eliminate-1}
\end{equation}
since $d\chi^{1}d\chi^{2}Y^{2}$ is the only term containing $d\chi^{1}d\chi^{2}$ in (\ref{eq_Delta-S-A1A2B2}).
However, Eq.~(\ref{eq_A1A2B2-eliminate-1}) means that $Y^{2}=-B^{2}+\cdots$, thus the gauge transformation of $B^{2}$ is
\begin{equation}\label{eq_A1A2B2-GT-of-B2}
  B^{2}\rightarrow dV^{2}+\cdots,
\end{equation}
which is not a proper gauge transformation for $B^{2}$. If
we want to absorb $d\chi^{1}d\chi^{2}B^{2}$ to a total derivative term,
we could make use of
\begin{equation}
d\left(\chi^{1}d\chi^{2}B^{2}\right)=d\chi^{1}d\chi^{2}B^{2}-\chi^{1}d\chi^{2}dB^{2}.
\end{equation}
However, this attempt  fails since there is no $-\chi^{1}d\chi^{2}dB^{2}$ term
in $\Delta S$. Therefore, (\ref{eq_GT-A1A2B2-A1+A2-nromal}) is not the proper gauge transformations for $S$.

Next, we assume that gauge transformations are
\begin{equation}\label{eq_GT-A1A2B2-A1+B2-normal}
\begin{alignedat}{1}A^{1}\rightarrow & A^{1}+d\chi^{1},\\
A^{2}\rightarrow & A^{2}+d\chi^{2}+X^{2},\\
B^{1}\rightarrow & B^{1}+dV^{1}+Y^{1},\\
B^{2}\rightarrow & B^{2}+dV^{2},
\end{alignedat}
\end{equation}
where $X^{2}$ and $Y^{1}$ are shift terms with $\oint X^{2}\notin\frac{2\pi}{N_{2}}\mathbb{Z}_{N_{2}}$ and $\oint Y^{1}\notin\frac{2\pi}{N_{1}}\mathbb{Z}_{N_{1}}$. The $\mathbb{Z}_{N_{2}}$ cyclic group
structure is encoded in $\oint B^{2}\in\frac{2\pi}{N_{2}}\mathbb{Z}_{N_{2}}$. Under~(\ref{eq_GT-A1A2B2-A1+B2-normal}),  the variation of action~(\ref{eq_action-A1A2B2}) is
\begin{equation}
\begin{alignedat}{1}\Delta S= & \int\frac{N_{1}}{2\pi}Y^{1}dA^{1}+\frac{N_{2}}{2\pi}B^{2}dX^{2}\\
 & +\frac{p}{\left(2\pi\right)^{2}}\left(d\chi^{1}A^{2}B^{2}+A^{1}d\chi^{2}B^{2}+d\chi^{1}d\chi^{2}B^{2}\right)\\
 & +\frac{p}{\left(2\pi\right)^{2}}\left(A^{1}X^{2}B^{2}+d\chi^{1}X^{2}B^{2}\right)\\
 & +\frac{p}{\left(2\pi\right)^{2}}\left(A^{1}A^{2}dV^{2}+d\chi^{1}A^{2}dV^{2}+A^{1}d\chi^{2}dV^{2}\right)\\
 & +\frac{p}{\left(2\pi\right)^{2}}\left(A^{1}X^{2}dV^{2}+d\chi^{1}X^{2}dV^{2}\right)
\end{alignedat}
\end{equation}
which should be an integral of total derivative terms.
Focus on the $d\chi^{1}A^{2}dV^{2}$ term: we want to eliminate it by subtraction
or by absorbing it into a total derivative term. If we want to eliminate
$d\chi^{1}A^{2}dV^{2}$ by subtraction, the only way is to require,
\begin{equation}
\begin{alignedat}{1} & \frac{p}{\left(2\pi\right)^{2}}d\chi^{1}A^{2}dV^{2}+\frac{p}{\left(2\pi\right)^{2}}d\chi^{1}X^{2}dV^{2}\\
= & \frac{p}{\left(2\pi\right)^{2}}d\chi^{1}\left(A^{2}-A^{2}+\cdots\right)dV^{2},
\end{alignedat}
\end{equation}
which means that $X^{2}=-A^{2}+\cdots$ hence
\begin{equation}
  A^{2}\rightarrow d\chi^{2}+\cdots,
\end{equation}
an ill-defined gauge transformation for $A^{2}$. If
we want to absorb $d\chi^{1}A^{2}dV^{2}$ into a total derivative term,
we need a $\chi^{1}dA^{2}dV^{2}$ term since
\begin{equation}
d\left(\chi^{1}A^{2}dV^{2}\right)=d\chi^{1}A^{2}dV^{2}+\chi^{1}dA^{2}dV^{2},
\end{equation}
but there is no a $\chi^{1}dA^{2}dV^{2}$ term in $\Delta S$,
so this attempt fails.
Therefore, Eq.~(\ref{eq_GT-A1A2B2-A1+B2-normal}) is not the proper gauge transformations for the action~(\ref{eq_action-A1A2B2}) either.

Since neither (\ref{eq_GT-A1A2B2-A1+A2-nromal}) nor (\ref{eq_GT-A1A2B2-A1+B2-normal}) serves as proper gauge transformations, it is impossible to construct gauge
transformations respecting $\mathbb{Z}_{N_{2}}$
cyclic group structure for the action (\ref{eq_action-A1A2B2}). This means that the action (\ref{eq_action-A1A2B2}) is not a legitimate TQFT action. Therefore, $\mathsf{\Theta}_{1,2|2}^{\text{BR}}$ is not even a well-defined braiding phase, not to mention its compatibility with other braiding phases. In other words, when $G=\prod_{i=1}^{2}\mathbb{Z}_{N_{i}}$, BR braidings cannot be realized.

\subsection{Incompatibility: $\mathsf{\Theta}_{1,2|3}^{\text{BR}}$ and $\mathsf{\Theta}_{2,3|1}^{\text{BR}}$\label{subsec_A1A2B3+A2A3B1_not_compatible}}
In previous sections we have known that the $A^{1}A^{2}B^{3}$ term
describes the BR braiding with a phase denoted as $\mathsf{\Theta}_{1,2|3}^{\text{BR}}$.
One may expect that, in the \emph{same} system, there could exists another
BR braiding process described by an $A^{2}A^{3}B^{1}$
term and its braiding phase $\mathsf{\Theta}_{2,3|1}^{\text{BR}}$.
Assumed that these two BR braidings could be supported in the same system,
the TQFT action should be
\begin{equation}
S=\int\sum_{i=1}^{3}\frac{N_{i}}{2\pi}B^{i}dA^{i}+\frac{p_{12,3}}{\left(2\pi\right)^{2}}A^{1}A^{2}B^{3}+\frac{p_{23,1}}{\left(2\pi\right)^{2}}A^{2}A^{3}B^{1}.\label{eq_action_A1A2B3+A2A3B1}
\end{equation}
In the action (\ref{eq_action_A1A2B3+A2A3B1}), $B^{2}$ is a Lagrange multiplier imposing $dA^{2}=0$, i.e., $A^{2}\rightarrow A^{2}+d\chi^{2}$.
Therefore the $\mathbb{Z}_{N_{2}}$ cyclic group structure is encoded in
$\oint A^{2}\in\frac{2\pi}{N_{2}}\mathbb{Z}_{N_{2}}$. In order to preserve the $\mathbb{\mathbb{\mathbb{Z}}}_{N_{1}}$ and $\mathbb{\mathbb{\mathbb{Z}}}_{N_{3}}$
cyclic group structure, at least one of $A^{i}\rightarrow A^{i}+d\chi^{i}$
and $B^{i}\rightarrow B^{i}+dV^{i}$ ($i=1,3$) is required. We find that, if
the gauge transformations preserve $\mathbb{\mathbb{\mathbb{Z}}}_{N_{1}}$
and $\mathbb{\mathbb{\mathbb{Z}}}_{N_{3}}$ cyclic group structure, the action~(\ref{eq_action_A1A2B3+A2A3B1})
would never be gauge invariant, i.e., the braiding processes it describes
are gauge-dependent. The reason is that no matter how we modify the
gauge transformations, there are ``stubborn terms'' in $\Delta S$
which cannot be eliminated by subtraction nor be absorbed into a total
derivative term, making $\Delta S$ could never be $0\mod2\pi$. The details of calculation can be found in Appendix~\ref{subsec_Derivation_A1A2B3+A2A3B1}.
Therefore, $A^{1}A^{2}B^{3}$ and is incompatible with $A^{2}A^{3}B^{1}$,
i.e., $\mathsf{\Theta}_{2,3|1}^{\text{BR}}$ is incompatible with
$\mathsf{\Theta}_{1,2|3}^{\text{BR}}$.

In the same manner, we can derive that any two of $\left( \mathsf{\Theta}_{2,3|1}^{\text{BR}},\mathsf{\Theta}_{3,1|2}^{\text{BR}},\mathsf{\Theta}_{1,2|3}^{\text{BR}}\right) $ are mutually incompatible.

\subsection{Incompatibility: $\mathsf{\Theta}_{3,3|1}^{\text{3L}}$ and $\mathsf{\Theta}_{1,2|3}^{\text{BR}}$\label{subsec_A1A3dA3+A1A2B3_not_compatible}}

In this and following subsections, we investigate the incompatibility between three-loop braidings and BR braidings.

As the first example, we consider these two braiding processes: a three-loop braiding with
phase $\mathsf{\Theta}_{3,3|1}^{\text{3L}}$, corresponding to an
$A^{1}A^{3}dA^{3}$ term, and a BR braiding with phase $\mathsf{\Theta}_{1,2|3}^{\text{BR}}$, corresponding to an $A^{1}A^{2}B^{3}$ term. If theses two braiding processes are compatible, the total TQFT action should be:
\begin{equation}
S=\int\sum_{i=1}^{3}\frac{N_{i}}{2\pi}B^{i}dA^{i}+\frac{q_{133}}{\left(2\pi\right)^{2}}A^{1}A^{3}dA^{3}+\frac{p_{12,3}}{\left(2\pi\right)^{2}}A^{1}A^{2}B^{3}.
\label{eq_action-A1A3dA3+A1A2B3}
\end{equation}
In the action~(\ref{eq_action-A1A3dA3+A1A2B3}), $B^{1}$ and $B^{2}$ serve as Lagrange multipliers, imposing $dA^{1}=dA^{2}=0$, i.e., $A^{1,2}\rightarrow A^{1,2}+dg^{1,2}$.
The $\mathbb{Z}_{N_{1}}$ and $\mathbb{Z}_{N_{2}}$ cyclic group structure
are encoded in $\oint A^{1}\in\frac{2\pi}{N_{1}}\mathbb{Z}_{N_{1}}$
and $\oint A^{2}\in\frac{2\pi}{N_{2}}\mathbb{Z}_{N_{2}}$ respectively.
For the $\mathbb{Z}_{N_{3}}$ cyclic group structure, we have two choices:
we can encoded it in $\oint A^{3}\in\frac{2\pi}{N_{3}}\mathbb{Z}_{N_{3}}$, corresponding to gauge transformations
\begin{equation}
\begin{alignedat}{1}A^{i}\rightarrow & A^{i}+d\chi^{i},\\
B^{i}\rightarrow & B^{i}+dV^{i}+Y^{i};
\end{alignedat}
\label{eq_GT-A1A3dA3+A1A2B3-A3-normal-1}
\end{equation}
or $\oint B^{3}\in\frac{2\pi}{N_{3}}\mathbb{Z}_{N_{3}}$, corresponding
to gauge transformations
\begin{equation}
\begin{alignedat}{1}A^{1,2}\rightarrow & A^{1,2}+d\chi^{1,2},\\
A^{3}\rightarrow & A^{3}+d\chi^{3}+X^{3},\\
B^{1,2}\rightarrow & B^{1,2}+dV^{1,2}+Y^{1,2},\\
B^{3}\rightarrow & B^{3}+dV^{3}.
\end{alignedat}
\label{eq_GT-A1A3dA3+A1A2B3-B3-normal-1}
\end{equation}
However, similar to the case in Sec.~\ref{subsec_A1A2B2_not_compatible} and Sec.~\ref{subsec_A1A2B3+A2A3B1_not_compatible},
neither (\ref{eq_GT-A1A3dA3+A1A2B3-A3-normal-1}) nor (\ref{eq_GT-A1A3dA3+A1A2B3-B3-normal-1})
could make the action (\ref{eq_action-A1A3dA3+A1A2B3}) gauge invariant
up to boundary terms, due to the existence of ``stubborn terms''. Details of derivations can be found in Appendix~\ref{subsec_derivation_incompatibility_A1A3dA3+A1A2B3}.
If the action is forced to be gauge invariant up to boundary terms,
the $\mathbb{Z}_{N_{3}}$ cyclic group structure would be violated. This dilemma
reveals that (\ref{eq_action-A1A3dA3+A1A2B3}) is not a legitimate
TQFT action. So we conclude that $A^{1}A^{3}dA^{3}$ is incompatible
with $A^{1}A^{2}B^{3}$, i.e., $\mathsf{\Theta}_{1,3|3}^{\text{3L}}$
is incompatible with $\mathsf{\Theta}_{1,2|3}^{\text{BR}}$.

\subsection{Incompatibility: $\mathsf{\Theta}_{2,3|1}^{\text{3L}}$ and $\mathsf{\Theta}_{1,2|3}^{\text{BR}}$ \label{subsec_A1A2dA3+A1A2B3_not_compatible}}

In this subsection, we consider these two braiding processes: a three-loop braiding corresponding to $\mathsf{\Theta}_{2,3|1}^{\text{3L}}$ and an $A^{1}A^{2}dA^{3}$ term, and a BR braiding corresponding to $\mathsf{\Theta}_{1,2|3}^{\text{BR}}$ and an $A^{1}A^{2}B^{3}$
term. If $\mathsf{\Theta}_{2,3|1}^{\text{3L}}$ is compatible with $\mathsf{\Theta}_{1,2|3}^{\text{BR}}$, the total TQFT action should be:
\begin{equation}
S=\int\sum_{i=1}^{3}\frac{N_{i}}{2\pi}B^{i}dA^{i}+\frac{q_{123}}{\left(2\pi\right)^{2}}A^{1}A^{2}dA^{3}+\frac{p_{12,3}}{\left(2\pi\right)^{2}}A^{1}A^{2}B^{3},\label{eq_action-A1A2dA3+A1A2B3}
\end{equation}
where the coefficients are $q_{123}=\frac{k_{1}N_{1}N_{2}}{N_{12}}$, $k_{1}\in\mathbb{Z}_{N_{123}}$ and $p_{12,3}=\frac{k_{2}N_{1}N_{2}N_{3}}{N_{123}}$, $k_{2}\in\mathbb{Z}_{N_{123}}$, determined by the large gauge invariance. In (\ref{eq_action-A1A2dA3+A1A2B3}),
$B^{1}$ and $B^{2}$ serve as Lagrange multipliers, imposing $dA^{1}=dA^{2}=0$,
i.e., $A^{1,2}\rightarrow A^{1,2}+d\chi^{1,2}$. We can find a set of gauge
transformations that seems to respect both $\mathbb{\mathbb{\mathbb{Z}}}_{N_{3}}$
cyclic group structure and gauge invariance (see Appendix~\ref{subsec_Derivation-GT-A1A2dA3+A1A2B3} for
derivation):
\begin{widetext}
\begin{equation}
\begin{alignedat}{1}A^{1}\rightarrow & A^{1}+d\chi^{1},\\
A^{2}\rightarrow & A^{2}+d\chi^{2},\\
A^{3}\rightarrow & A^{3}+d\chi^{3}+X^{3},\\
B^{1}\rightarrow & B^{1}+dV^{1}+Y^{1},\\
B^{2}\rightarrow & B^{2}+dV^{2}+Y^{2},\\
B^{3}\rightarrow & B^{3}+dV^{3},
\end{alignedat}
\label{eq_GT-A1A2dA3+A1A2B3-B3 normal}
\end{equation}
where
\begin{equation}
\begin{alignedat}{1}X^{3}= & -\frac{p_{12,3}}{\left(2\pi\right)N_{3}}\left(\chi^{1}A^{2}+\frac{1}{2}\chi^{1}d\chi^{2}\right)+\frac{p_{12,3}}{\left(2\pi\right)N_{3}}\left(\chi^{2}A^{1}+\frac{1}{2}\chi^{2}d\chi^{1}\right),\\
Y^{1}= & \frac{q_{123}}{2\pi N_{1}}d\chi^{2}A^{3}-\frac{p}{\left(2\pi\right)N_{1}}\left(\chi^{2}B^{3}-A^{2}V^{3}+\chi^{2}dV^{3}\right)\\
 & +\frac{q_{123}}{\left(2\pi\right)^{2}}\cdot\frac{p_{12,3}}{\left(2\pi\right)N_{3}}\cdot\frac{2\pi}{N_{1}}\cdot\left[\chi^{1}A^{2}d\chi^{2}-\left(A^{1}A^{2}\chi^{2}+A^{1}d\chi^{2}\chi^{2}+d\chi^{1}A^{2}\chi^{2}+d\chi^{1}d\chi^{2}\chi^{2}\right)\right],\\
Y^{2}= & -\frac{q_{123}}{2\pi N_{2}}d\chi^{1}A^{3}+\frac{p_{12,3}}{\left(2\pi\right)N_{2}}\left(\chi^{1}B^{3}-A^{1}V^{3}+\chi^{1}dV^{3}\right)\\
 & +\frac{q_{123}}{\left(2\pi\right)^{2}}\cdot\frac{p_{12,3}}{\left(2\pi\right)N_{3}}\cdot\frac{2\pi}{N_{2}}\cdot\left[-d\left(A^{1}\chi^{1}\right)\chi^{2}+\left(A^{1}A^{2}\chi^{1}+A^{1}d\chi^{2}\chi^{1}+d\chi^{1}A^{2}\chi^{1}+d\chi^{1}d\chi^{2}\chi^{1}\right)\right].
\end{alignedat}
\label{eq_GT-shift-A1A2dA3+A1A2B3-B3 normal}
\end{equation}
\end{widetext}

Nevertheless, the action~(\ref{eq_action-A1A2dA3+A1A2B3}) is still not a legitimate
TQFT theory: the coefficient of $A^{1}A^{2}dA^{3}$ in~(\ref{eq_action-A1A2dA3+A1A2B3}) is
actually identical to $0$. To see this, we first integrate out $A^{3}$ in $S$, i.e.,
sum over all possible nontrivial compactness contributions from $dA^{3}$, leading to a constraint
\begin{equation}
\int\frac{q_{123}}{\left(2\pi\right)^{2}}A^{1}A^{2}+\frac{N_{3}}{2\pi}B^{3}\in\mathbb{Z}.\label{eq_constraint-A1A2dA3+A1A2B3-sum dA3}
\end{equation}
Since the $\mathbb{Z}_{N_{3}}$ cyclic group structure is encoded in $\oint\frac{N_{3}}{2\pi}B^{3}\in\mathbb{Z}$,
Eq.~(\ref{eq_constraint-A1A2dA3+A1A2B3-sum dA3}) requires that
\begin{equation}
\frac{q_{123}}{\left(2\pi\right)^{2}}\int A^{1}A^{2}\in\mathbb{Z},
\end{equation}
i.e., (consider a 2-manifold $\mathcal{M}^{2}=S^{1}\times S^{1}$)
\begin{equation}
\begin{alignedat}{1}\frac{q_{123}}{\left(2\pi\right)^{2}}\int_{\mathcal{M}^{2}} A^{1}A^{2} & =\frac{k_{1}N_{1}N_{2}}{\left(2\pi\right)^{2}N_{12}}\int_{\mathcal{S}^{1}}A^{1}\int_{\mathcal{S}^{1}}A^{2}\\
 & =\frac{k_{1}N_{1}N_{2}}{\left(2\pi\right)^{2}N_{12}}\cdot\frac{2\pi m_{1}}{N_{1}}\cdot\frac{2\pi m_{2}}{N_{2}}\\
 & =\frac{k_{1}m_{1}m_{2}}{N_{12}}\in\mathbb{Z},
\end{alignedat}
\label{eq_coeff-constriant-A1A2dA3+A1A2B3}
\end{equation}
where $k_{1}\in\mathbb{Z}_{N_{123}}$, $m_{1}\in\mathbb{Z}$, $m_{2}\in\mathbb{Z}$.
Since $m_{1}$ and $m_{2}$ can be arbitrary integers, the sufficient
condition for Eq.~(\ref{eq_coeff-constriant-A1A2dA3+A1A2B3}) is
\begin{equation}
k_{1}=lN_{12},l\in\mathbb{Z}.
\end{equation}
On the other hand, $N_{12}$ is the integral
multiple of $N_{123}$, thus
\begin{equation}
k_{1}=lN_{12}=l\left(nN_{123}\right)=\left(ln\right)N_{123},
\end{equation}
where $n=\frac{N_{12}}{N_{123}}$. Notice that $k_{1}\in\mathbb{Z}_{N_{123}}$,
this means that
\begin{equation}
k_{1}=\left(ln\right)N_{123}\simeq0.
\end{equation}
Therefore the coefficient of $A^{1}A^{2}dA^{3}$ term is required to be identical
to $0$:
\begin{equation}
q_{123}=\frac{k_{1}N_{1}N_{2}}{N_{12}}\simeq0.
\end{equation}
So far, we have argued that, in order to preserve the $\mathbb{Z}_{N_{3}}$ fusion
structure with arbitrary values of $N_{1}$, $N_{2}$ and $N_{3}$, the coefficient of $A^{1}A^{2}dA^{3}$ term in (\ref{eq_action-A1A2dA3+A1A2B3})
is required to be trivial. If the coefficient of $A^{1}A^{2}dA^{3}$ is
nontrivial, the action (\ref{eq_action-A1A2dA3+A1A2B3}) cannot be
gauge invariant up to boundary terms while the $\mathbb{Z}_{N_{3}}$ cyclic group
structure is preserved. The restriction on the coefficient of $A^{1}A^{2}dA^{3}$
indicates the incompatibility between $A^{1}A^{2}dA^{3}$ and $A^{1}A^{2}B^{3}$,
i.e., the incompatibility between $\mathsf{\Theta}_{1,2|3}^{\text{3L}}$
and $\mathsf{\Theta}_{1,2|3}^{\text{BR}}$.

\subsection{Incompatibility: $\mathsf{\Theta}_{3,1|2}^{\text{3L}}$ and $\mathsf{\Theta}_{1,2|3}^{\text{BR}}$\label{subsec_A2A3dA1+A1A2B3_not_compatible}}

As the last example of the incompatibility between three-loop braidings
and BR braidings, we consider a three-loop braiding corresponding to $\mathsf{\Theta}_{3,1|2}^{\text{3L}}$ as well as an $A^{2}A^{3}dA^{1}$ term, and a BR braiding corresponding to $\mathsf{\Theta}_{1,2|3}^{\text{BR}}$ as well as an $A^{1}A^{2}B^{3}$
term. The total action should be:
\begin{equation}
\begin{alignedat}{1}S= & \int\sum_{i=1}^{3}\frac{N_{i}}{2\pi}B^{i}dA^{i}\\
 & +\frac{q_{231}}{\left(2\pi\right)^{2}}A^{2}A^{3}dA^{1}+\frac{p_{12,3}}{\left(2\pi\right)^{2}}A^{1}A^{2}B^{3}
\end{alignedat}
\label{eq_action-A2A3dA1+A1A2B3}
\end{equation}
where the coefficients are
$q_{231}=\frac{kN_{2}N_{3}}{N_{23}}$, $k\in\mathbb{Z}_{N_{123}}$ and $p_{12,3}=\frac{lN_{1}N_{2}N_{3}}{N_{123}}$, $l\in\mathbb{Z}_{N_{123}}$, determined by the large gauge invariance. In the action~(\ref{eq_action-A2A3dA1+A1A2B3}),
$B^{1}$ and $B^{2}$ serve as Lagrange multipliers, imposing $dA^{1}=dA^{2}=0$,
i.e., $A^{1,2}\rightarrow A^{1,2}+dg^{1,2}$.

If the $\mathbb{Z}_{N_{3}}$ cyclic group structure is
encoded in $\oint A^{3}\in\frac{2\pi}{N_{3}}\mathbb{Z}_{N_{3}}$,
i.e., $A^{3}\rightarrow A^{3}+d\chi^{3}$, we cannot find a set of gauge transformations with $A^{i}\rightarrow A^{i}+d\chi^{i}$ ($i=1,2,3$) under which the action (\ref{eq_action-A2A3dA1+A1A2B3})
is invariant up to boundary terms. The reason is same as those
in Sec.~\ref{subsec_A1A2B2_not_compatible} and Sec.~\ref{subsec_A1A3dA3+A1A2B3_not_compatible}.

If we encode the $\mathbb{\mathbb{Z}}_{N_{3}}$ cyclic group structure in $\oint B^{3}\in\frac{2\pi}{N_{3}}\mathbb{\mathbb{\mathbb{Z}}}_{N_{3}}$,
i.e., $B^{3}\rightarrow B^{3}+dV^{3}$, we can find a set of gauge transformations under which (\ref{eq_action-A2A3dA1+A1A2B3}) is invariant up to boundary
terms:
\begin{equation}
\begin{alignedat}{1}A^{1}\rightarrow & A^{1}+dg^{1},\\
A^{2}\rightarrow & A^{2}+dg^{2},\\
A^{3}\rightarrow & A^{3}+d\chi^{3}+\frac{p_{12,3}}{\left(2\pi\right)N_{3}}\left(-\chi^{1}A^{2}+A^{1}\chi^{2}-\chi^{1}d\chi^{2}\right),\\
B^{1}\rightarrow & B^{1}+dV^{1}-\frac{p_{12,3}}{\left(2\pi\right)N_{1}}\left(\chi^{2}B^{3}-A^{2}V^{3}+\chi^{2}dV^{3}\right)\\
 & -\frac{q_{231}}{\left(2\pi\right)N_{1}}\left(d\chi^{2}A^{3}+A^{2}d\chi^{3}+A^{2}X^{3}+d\chi^{2}X^{3}\right),\\
B^{2}\rightarrow & B^{2}+dV^{2}+\frac{p_{12,3}}{\left(2\pi\right)N_{2}}\left(\chi^{1}B^{3}-A^{1}V^{3}+\chi^{1}dV^{3}\right),\\
B^{3}\rightarrow & B^{3}+dV^{3}.
\end{alignedat}
\label{eq_GT-A2A3dA1+A1A2B3-B3 normal}
\end{equation}
At first glance, one may think that (\ref{eq_GT-A2A3dA1+A1A2B3-B3 normal}) is a proper set of gauge transformations for (\ref{eq_action-A2A3dA1+A1A2B3}) thus the action~(\ref{eq_action-A2A3dA1+A1A2B3})
is a legitimate TQFT action. However, we argue that this is not true. We provide the following two arguments.

In the first argument, we consider the gauge transformations at limiting cases. Let us set $p_{12,3}=0$, i.e., turn off the $A^{1}A^{2}B^{3}$ term, the action~(\ref{eq_action-A2A3dA1+A1A2B3}) becomes
\begin{equation}
S=\int\sum_{i=1}^{3}\frac{N_{i}}{2\pi}B^{i}dA^{i}+\frac{q_{231}}{\left(2\pi\right)^{2}}A^{2}A^{3}dA^{1}
\label{eq_action-A2A3dA1}
\end{equation}
and the corresponding gauge transformations should be
\begin{equation}
\begin{alignedat}{1}A^{i}\rightarrow & A^{i}+d\chi^{i},\\
B^{1}\rightarrow & B^{1}+dV^{1},\\
B^{2}\rightarrow & B^{2}+dV^{2}+\frac{q_{231}}{2\pi N_{2}}d\chi^{3}A^{1},\\
B^{3}\rightarrow & B^{3}+dV^{3}-\frac{q_{231}}{2\pi N_{3}}d\chi^{2}A^{1}.
\end{alignedat}
\label{eq_GT-A2A3dA1}
\end{equation}
According to Ref~\cite{PhysRevB.99.235137}, gauge transformations~(\ref{eq_GT-A2A3dA1})
is motivated by a microscopic derivation of the action~(\ref{eq_action-A2A3dA1}).
However, if we set $p_{12,3}=0$ in~(\ref{eq_GT-A2A3dA1+A1A2B3-B3 normal}),
we cannot correctly recover the gauge transformations to~(\ref{eq_GT-A2A3dA1}). Therefore, we claim that (\ref{eq_GT-A2A3dA1+A1A2B3-B3 normal}) is not proper gauge transformations for the action (\ref{eq_action-A2A3dA1+A1A2B3}).

In fact, if we expect that by setting $q_{231}=0$ or $p_{12,3}=0$ the gauge transformations
for the action (\ref{eq_action-A2A3dA1+A1A2B3}) would recover to
(\ref{eq_GT-A1A2B3}) or (\ref{eq_GT-A2A3dA1}), the gauge transformations for the action (\ref{eq_action-A2A3dA1+A1A2B3})
should take the form of
\begin{widetext}
\begin{equation}
\begin{alignedat}{1}A^{1}\rightarrow & A^{1}+d\chi^{1},\\
A^{2}\rightarrow & A^{2}+d\chi^{2},\\
A^{3}\rightarrow & A^{3}+d\chi^{3}+\frac{p_{12,3}}{\left(2\pi\right)^{2}}\frac{2\pi}{N_{3}}\left(-\chi^{1}A^{2}+A^{1}\chi^{2}-\chi^{1}d\chi^{2}\right)+f_{A}^{3}\left(p_{12,3},q_{231}\right)\\
B^{1}\rightarrow & B^{1}+dV^{1}-\frac{p_{12,3}}{\left(2\pi\right)N_{1}}\left(\chi^{2}B^{3}-A^{2}V^{3}+\chi^{2}dV^{3}\right)+f_{B}^{1}\left(p_{12,3},q_{231}\right)\\
B^{2}\rightarrow & B^{2}+dV^{2}+\frac{q_{231}}{2\pi N_{2}}d\chi^{3}A^{1}+\frac{p_{12,3}}{\left(2\pi\right)N_{2}}\left(\chi^{1}B^{3}-A^{1}V^{3}+\chi^{1}dV^{3}\right)+f_{B}^{2}\left(p_{12,3},q_{231}\right)\\
B^{3}\rightarrow & B^{3}+dV^{3}-\frac{q_{231}}{2\pi N_{3}}d\chi^{2}A^{1}+f_{B}^{3}\left(p_{12,3},q_{231}\right),
\end{alignedat}
\label{eq_GT-modifeied-A2A3dA1+A1A2B3}
\end{equation}
\end{widetext}
where the gauge parameters $f_{A}^{3}\left(p_{12,3},q_{231}\right)$ and $f_{B}^{i}\left(p_{12,3},q_{231}\right)$ with $i=1,2,3$ are identical to $0$ once $p_{12,3}=0$ or $q_{231}=0$. We can see that, the $\mathbb{\mathbb{\mathbb{Z}}}_{N_{3}}$ cyclic group structure is no longer preserved under gauge transformation~(\ref{eq_GT-modifeied-A2A3dA1+A1A2B3}).

The second argument exploits the identity
\begin{equation}
\label{eq_identity_dA1A2A3}
A^{2}A^{3}dA^{1}=d\left(A^{1}A^{2}A^{3}\right)-A^{3}A^{1}dA^{2}-A^{1}A^{2}dA^{3}.
\end{equation}
Using~(\ref{eq_identity_dA1A2A3}) we can rewrite the action~(\ref{eq_action-A2A3dA1+A1A2B3}) as
\begin{equation}
\begin{alignedat}{1}S= & \int\sum_{i=1}^{3}\frac{N_{i}}{2\pi}B^{i}dA^{i}+\frac{p_{12,3}}{\left(2\pi\right)^{2}}A^{1}A^{2}B^{3}\\
 & +\frac{q_{231}}{\left(2\pi\right)^{2}}\left[d\left(A^{1}A^{2}A^{3}\right)-A^{3}A^{1}dA^{2}-A^{1}A^{2}dA^{3}\right].
\end{alignedat}
\end{equation}
Similar to the case in Sec.~\ref{subsec_A1A2dA3+A1A2B3_not_compatible},
we can integrate out $A^{3}$ and obtain a constraint
\begin{equation}
\int\frac{N_{3}}{2\pi}B^{3}-\frac{q_{231}}{\left(2\pi\right)^{2}}A^{1}A^{2}\in\mathbb{Z}.\label{eq_int-dA3-constraint-A2A3dA1+A1A2B3}
\end{equation}
Since in the gauge transformations (\ref{eq_GT-A1A2dA3+A1A2B3-B3 normal})
the $\mathbb{Z}_{N_{3}}$ cyclic group structure is encoded in
$\oint B^{3}=\frac{2\pi}{N_{3}}\mathbb{\mathbb{Z}}_{N_{3}}$, we require that
\begin{equation}
\int\frac{q_{231}}{\left(2\pi\right)^{2}}A^{1}A^{2}\in\mathbb{Z}.
\end{equation}
In other words, we require that (consider $\mathcal{M}^{2}=S^{1}\times S^{1}$)
\begin{equation}
\begin{alignedat}{1}\frac{q_{231}}{\left(2\pi\right)^{2}}\int_{\mathcal{M}^{2}} A^{1}A^{2} & =\frac{kN_{2}N_{3}}{\left(2\pi\right)^{2}N_{23}}\int_{\mathcal{S}^{1}}A^{1}\int_{\mathcal{S}^{1}}A^{2}\\
 & =\frac{kN_{2}N_{3}}{\left(2\pi\right)^{2}N_{23}}\cdot\frac{2\pi m_{1}}{N_{1}}\cdot\frac{2\pi m_{2}}{N_{2}}\\
 & =\frac{kN_{3}m_{1}m_{2}}{N_{1}N_{23}}\in\mathbb{\mathbb{\mathbb{Z}}},
\end{alignedat}
\label{eq_coef-constraint-A2A3dA1+A1A2B3}
\end{equation}
where $k\in\mathbb{Z}_{N_{123}}$. Since $m_{1}$ and $m_{2}$ can
be arbitrary integers, we need
\begin{equation}
\frac{kN_{3}}{N_{1}N_{23}}\in\mathbb{\mathbb{\mathbb{Z}}}.
\end{equation}
Since $\frac{N_{3}}{N_{23}}\in\mathbb{Z}$, the sufficient condition for Eq. (\ref{eq_coef-constraint-A2A3dA1+A1A2B3}) to hold is
\begin{equation}
k=mN_{1},m\in\mathbb{Z}.
\end{equation}
Notice that $\frac{N_{1}}{N_{123}}\in\mathbb{Z}$ and $k\in\mathbb{Z}_{N_{123}}$, therefore
\begin{equation}
k=mN_{1}=mnN_{123}\simeq0,
\end{equation}
where $n=\frac{N_{1}}{N_{123}}\in\mathbb{Z}$. We see that the coefficient of $A^{2}A^{3}dA^{1}$ is required to be identical to $0$:
\begin{equation}
q_{231}=\frac{kN_{2}N_{3}}{N_{23}}\simeq0,
\end{equation}
to ensure that the $\mathbb{Z}_{N_{3}}$ cyclic group structure is preserved no matter what the values of $N_{1}$, $N_{2}$ and $N_{3}$ are.
This restriction on the coefficient of $A^{2}A^{3}dA^{1}$ reveals
that, $A^{2}A^{3}dA^{1}$ is incompatible with $A^{1}A^{2}B^{3}$,
hence $\Theta_{31,2}^{\text{3L}}$ and $\Theta_{12,3}^{\text{BR}}$
are mutually incompatible.

\subsection{Incompatibility: $\mathsf{\Theta}_{1,2,3,4}^{\text{4L}}$ and $\mathsf{\Theta}_{1,2|4}^{\text{BR}}$\label{subsec_A1A2A3A4+A1A2B4_not_compatible}}

Since we have demonstrated the incompatibility between BR braidings and three-loop braidings, one may wonder whether BR braiding is compatible
with four-loop braiding or not. The conclusion is that, when $G=\prod_{i=1}^{4}\mathbb{Z}_{N_{i}}$,
no BR braiding is compatible with four-loop braiding, i.e.,
each $\mathsf{\Theta}_{i,j|k}^{\text{BR}}$ is incompatible with $\mathsf{\Theta}_{1,2,3,4}^{\text{4L}}$.
The proof of this claim, which follows the same idea discussed in
previous sections, is detailed in Appendix~\ref{subsec_derivation_incompatibility-A1A2A3A4_A1A2B4}.
Notice that, if $G=\prod_{i=1}^{5}\mathbb{Z}_{N_{i}}$, $\mathsf{\Theta}_{1,2,3,4}^{\text{4L}}$
is compatible with $\mathsf{\Theta}_{i,j|5}^{\text{BR}}$, as long as $i\neq5$, $j\neq5$ and $i\neq j$. Such compatibility due to an extra gauge
subgroup can be checked by applying the criteria for legitimacy of
TQFT actions.

\section{Compatible braidings in (4+1)D spacetime}\label{section_4d}
In the above discussions, we have obtained compatible braidings in $(3+1)$D spacetime. In the following,  we concisely discuss compatible braidings in (4+1)D spacetime. More systematic and quantitative studies  will appear in  Ref.~\cite{ZhangYeUnpublished} separately.
\subsection{Excitations and braiding processes in $\left(4+1\right)$D }
We limit the gauge group to be $G=\prod_{i=1}^{n}\mathbb{Z}_{N_{i}}$
where $n$ is the total number of cyclic subgroups. In $\left(4+1\right)$D
spacetime, besides particles and loops, there is a kind exotic topological
excitations, dubbed \emph{membranes}, which are closed 2D surface objects
in the continuum limit. The membrane excitations in $3$-dimensional
space is topologically trivial because they are impenetrable hence identified as
particle excitations. However, in $4$-dimensional space, the interior
of a membrane excitation becomes accessible due to the extra dimension.
Therefore, nontrivial braiding processes involving particles, loops
and membranes are possible in $\left(4+1\right)$D spacetime.

In field theory, a $\mathbb{Z}_{N_{i}}$ gauge theory is realized
by a $BF$ term in the continuum limit. In $\left(3+1\right)$D spacetime,
the $\mathbb{Z}_{N_{i}}$ cyclic group structure is encoded in the $B^{i}dA^{i}$
term, the only possible $BF$ term in $\left(3+1\right)$D spacetime
which corresponds to the particle-loop braiding (see Sec.~\ref{subsec_BdA}). In $\left(4+1\right)$D
spacetime, there are two kinds of $BF$ terms with distinct physical
meanings: $C^{i}dA^{i}$ and $\widetilde{B}^{i}dB^{i}$, where the $3$-form
$C^{i}$, $1$-form $A^{i}$, $2$-form $\widetilde{B}^{i}$ and $2$-form
$B^{i}$ are respectively compact $\mathbb{U}(1)$ gauge fields.
Therefore, we can encode the $\mathbb{Z}_{N_{i}}$ cyclic group structure
either in $C^{i}dA^{i}$, corresponding to a \emph{particle-membrane
braiding}; or in $\widetilde{B}^{i}dB^{i}$, corresponding to a \emph{loop-loop
braiding}. If we consider a $\prod_{i=1}^{n}\mathbb{Z}_{N_{i}}$ topological
order in $\left(4+1\right)$D spacetime, we can even encode some
$\mathbb{Z}_{N_{i}}$ gauge subgroups in $C^{i}dA^{i}$ terms while
the others in $\widetilde{B}^{i}dB^{i}$ terms, which would lead to
a more complicated situation in which different gauge fields can couple together resulting in many nontrivial braiding processes.

Since there are three kinds of topological excitations in $\left(4+1\right)$D
spacetime, we can classify the braiding processes and identify the
corresponding topological terms:  {$\left(1\right)$particle-membrane
braiding}, $CdA$; $\left(2\right)${particle-membrane-membrane braiding}:
$CAA$; $\left(3\right)${loop-loop braiding}, $\widetilde{B}dB$;
$\left(4\right)${loop-membrane braidings}, $BBA$, $BAdA$, $AAAB$
and $AAdB$; $\left(5\right)${multi-membrane braidings}, $AAAAA$,
$AAAdA$ and $AdAdA$.

\subsection{An overview of compatible braiding processes in $\left(4+1\right)$D}

Following the discussions in $(3+1)$D spacetime, we can investigate the compatibility
of braiding processes in $\left(4+1\right)$D spacetime.   In other words, we check the legitimacy of TQFT actions consisting of topological terms, then distinguish compatible/incompatible braiding processes in $\left(4+1\right)$D spacetime.

First, if all $\mathbb{Z}_{N_{i}}$
cyclic group structures are encoded in particle-membrane braidings, i.e.,
$C^{i}dA^{i}$ terms, the compatible braiding processes are multi-membrane
braidings and particle-membrane-membrane braiding. Here are some examples.
When $G=\prod_{i=1}^{5}\mathbb{Z}_{N_{i}}$, one of legitimate TQFT
actions is
\begin{equation}
\begin{alignedat}{1}S\sim & \int\sum_{i=1}^{5}\frac{N_{i}}{2\pi}C^{i}dA^{i}+A^{1}A^{2}A^{3}A^{4}A^{5}\\
 & +\sum_{i,j,k,l=1}^{5}A^{i}A^{j}A^{k}dA^{l}+\sum_{i,j,k=1}^{5}A^{i}dA^{j}dA^{k}
\end{alignedat}
\label{eq_action_4+1D_CdA+all A}
\end{equation}
where the quantized coefficients of topological terms are neglected. The action (\ref{eq_action_4+1D_CdA+all A}) indicates the compatibility of particle-membrane braidings
and multi-membrane braidings when $G=\prod_{i=1}^{5}\mathbb{Z}_{N_{i}}$.
We believe that TQFT action (\ref{eq_action_4+1D_CdA+all A}) describes
braiding processes within the Dijkgraaf-Witten model.
There are also braiding processes beyond the Dijkgraaf-Witten model.
Consider $G=\prod_{i=1}^{3}\mathbb{Z}_{N_{i}}$, one of legitimate TQFT actions (coefficients
neglected) is
\begin{equation}
S\sim\int\sum_{i=1}^{3}\frac{N_{i}}{2\pi}C^{i}dA^{i}+C^{3}A^{1}A^{2}
\end{equation}
as an analog of the BR braiding in $\left(3+1\right)$D, $S_{\text{BR}}\sim\int\sum_{i=1}^{3}B^{i}dA^{i}+A^{1}A^{2}B^{3}$. We have known that, in $\left(3+1\right)$D,
a BR braiding is incompatible with some of the multi-loop braidings (Sec.~\ref{subsec_A1A3dA3+A1A2B3_not_compatible}, \ref{subsec_A1A2dA3+A1A2B3_not_compatible} and \ref{subsec_A2A3dA1+A1A2B3_not_compatible}). Similar incompatibility occurs in $\left(4+1\right)$D.
When $G=\prod_{i=1}^{3}\mathbb{Z}_{N_{i}}$, a multi-membrane braiding
described by the $A^{1}A^{2}A^{3}dA^{1}$ term is incompatible with
the loop-membrane-membrane braiding described by the $C^{3}A^{1}A^{2}$
term. Using the language of TQFT, we claim that a legitimate TQFT
action in $\left(4+1\right)$D can only include one of $C^{3}A^{1}A^{2}$
term and $A^{1}A^{2}A^{3}dA^{1}$ term. When $G=\prod_{i=1}^{3}\mathbb{Z}_{N_{i}}$,
two of the legitimate TQFT actions are
\begin{equation}
\begin{alignedat}{1}S\sim & \int\sum_{i=1}^{3}\frac{N_{i}}{2\pi}C^{i}dA^{i}+\sum_{i,j,k=1}^{2}A^{i}dA^{j}dA^{k}\\
 & +C^{3}A^{1}A^{2}
\end{alignedat}
\end{equation}
and
\begin{equation}
\begin{alignedat}{1}S\sim  & \int\sum_{i=1}^{3}\frac{N_{i}}{2\pi}C^{i}dA^{i}+\sum_{i=1}^{3}A^{1}A^{2}A^{3}dA^{i}\\
 & +\sum_{i,j,k=1}^{3}A^{i}dA^{j}dA^{k}.
\end{alignedat}
\end{equation}

Next, we consider the case in which all $\mathbb{Z}_{N_{i}}$ cyclic group
structures are encoded in $\widetilde{B}^{i}dB^{i}$ terms, i.e.,
loop-loop braidings. In this case, the only topological excitations
are loops, thus only loop-loop braidings can be supported. We can
simply write down the legitimate TQFT action:
\begin{equation}
S=\int\sum_{i=1}^{n}\frac{N_{i}}{2\pi}\widetilde{B}^{i}dB^{i}.
\end{equation}

Last, we tackle the more complicated yet more interesting situation: some $\mathbb{Z}_{N}$ cyclic group structures (e.g., $\mathbb{Z}_{N_{i}}$ gauge subgroups) are encoded
in $\frac{N_{i}}{2\pi}C^{i}dA^{i}$ terms while the others (e.g., $\mathbb{Z}_{K_{i}}$ gauge subgroups) in $\frac{K_{i}}{2\pi}\widetilde{B}^{i}dB^{i}$
terms. Different assignments for $N_{i}$'s and $K_{i}$'s correspond to different topological orders in $(4+1)$D. In this case, there are particle, loop and membrane excitations in the system. All nontrivial braiding processes in $(4+1)$D should be taken into consideration. In the followings we will demonstrate several examples of legitimate TQFT actions from which we can directly read out compatible braiding processes in $\left(4+1\right)$D.

Consider a $\prod_{i=1}^{3}\mathbb{Z}_{N_{i}}$ case in which the $\mathbb{Z}_{N_{1}}$and $\mathbb{Z}_{N_{2}}$ cyclic group structures are encoded in $CdA$ while the $\mathbb{Z}_{K_{1}}$ cyclic group structure in $\widetilde{B}dB$, i.e., the gauge group is $G=\mathbb{Z}_{N_{1}}\times\mathbb{Z}_{N_{2}}\times\mathbb{Z}_{K_{1}}$, the legitimate TQFT actions is
\begin{equation}
\begin{alignedat}{1}S\sim & \int \frac{N_{1}}{2\pi}C^{1}dA^{1}+\frac{N_{2}}{2\pi}C^{2}dA^{2}+\frac{K_{1}}{2\pi}\widetilde{B}^{1}dB^{1}\\
 & +\sum_{i,j,k=1}^{2}A^{i}dA^{j}dA^{k}+\sum_{i=1}^{2}B^{1}B^{1}A^{i}\\
 & +\sum_{j,k=1}^{2}B^{1}A^{j}dA^{k}+\sum_{i,j=1}^{2}A^{i}A^{j}dB^{1}.
\end{alignedat}
\end{equation}
Consider again a $\prod_{i=1}^{3}\mathbb{Z}_{N_{i}}$ case but this time $G=\mathbb{Z}_{N_1}\times\mathbb{Z}_{K_{1}}\times\mathbb{Z}_{K_{2}}$, the corresponding legitimate TQFT action is
\begin{equation}
\begin{alignedat}{1} S\sim& \int\frac{N_{1}}{2\pi}C^{1}dA^{1}+\frac{K_{1}}{2\pi}\widetilde{B}^{1}dB^{1}+\frac{K_{2}}{2\pi}\widetilde{B}^{2}dB^{2}\\
 & +A^{1}dA^{1}dA^{1}\\
 & +\sum_{i=1}^{2}B^{i}B^{j}A^{1}\\
 & +\sum_{i=1}^{2}B^{i}A^{1}dA^{1}.
\end{alignedat}
\end{equation}

Consider a $\prod_{i=1}^{4}\mathbb{Z}_{N_{i}}$ case in which $G=\mathbb{Z}_{N_1}\times\mathbb{Z}_{N_{2}}\times\mathbb{Z}_{N_{3}}\times\mathbb{Z}_{K_{1}}$, the legitimate TQFT actions is
\begin{equation}
\begin{alignedat}{1}S\sim  & \int \sum_{i=1}^{3}\frac{N_{i}}{2\pi}C^{i}dA^{i}+\frac{K_{1}}{2\pi}\widetilde{B}^{1}dB^{1}\\
 & +\sum_{i,j,k=1}^{3}A^{i}dA^{j}dA^{k}+C^{3}A^{1}A^{2}\\
 & +\sum_{i=1}^{2}B^{1}B^{1}A^{i}+\sum_{i,j=1}^{2}B^{1}A^{i}dA^{j}.
\end{alignedat}
\end{equation}

Consider a $\prod_{i=1}^{5}\mathbb{Z}_{N_{i}}$ case in which $G=\mathbb{Z}_{N_1}\times\mathbb{Z}_{N_{2}}\times\mathbb{Z}_{N_{3}}\times\mathbb{Z}_{K_{1}}\times\mathbb{Z}_{K_{2}}$, the legitimate TQFT actions
is
\begin{equation}
\begin{alignedat}{1}S\sim  & \int\sum_{i=1}^{3}\frac{N_{i}}{2\pi}C^{i}dA^{i}+\sum_{i=1}^{2}\frac{K_{i}}{2\pi}\widetilde{B}^{i}dB^{i}\\
 & +\sum_{i,j,k=1}^{3}A^{i}dA^{j}dA^{k}\\
 & +\sum_{i,j=1}^{2}\sum_{k=1}^{3}B^{i}B^{j}A^{k}\\
 & +\sum_{i=1}^{2}\sum_{j,k=1}^{2}B^{i}A^{j}dA^{k}\\
 & +C^{3}A^{1}A^{2}.
\end{alignedat}
\end{equation}

We should point out that the $BC$ term is also a topological term
in $\left(4+1\right)$D spacetime, but we do not regard that it corresponds
to any braiding process in the discussion above. Naively, because
the $2$-form $B$ is related to the loop current $\Sigma$ via $\Sigma=\frac{1}{2\pi}*dB$
and the $3$-form $C$ is related to the particle current $J$ via
$J=\frac{1}{2\pi}*dC$, one may think that $BC$ is related to
the particle-loop braiding in $\left(4+1\right)$D. But we argue that
particle-loop braiding is actually trivial in $\left(4+1\right)$D spacetime
thus the $BC$ term does not describe the particle-loop braiding.
To show the triviality of particle-loop braiding in $\left(4+1\right)$D spacetime,
we can first consider the particle-loop braiding in $\left(3+1\right)$D spacetime
in which the Hopf link formed by trajectory $\gamma_{e}$ of the particle
$e$ and the loop $m$ (e.g., FIG.~\ref{figure_particle_loop_braiding}) \emph{cannot} be unlinked in
$3$D space. However, in $4$D space, due to an extra dimension, we
can smoothly deform $\gamma_{e}$ such that it is unlinked with the
loop $m$. This argument is analog to the fact that the nontrivial
particle-particle braiding in $\left(2+1\right)$D spacetime becomes
trivial in $\left(3+1\right)$D spacetime. In this manner, we claim
that the particle-loop braiding in $\left(4+1\right)$D spacetime
is trivial.

A question naturally arises: what is the physical effect of the $BC$
term? We noticed that there are topological terms which can generate
trivial fermions from a bosonic model: $\frac{k}{4\pi}AdA$ with an
odd integer $k$ in $\left(2+1\right)$D and $BB$ in $\left(3+1\right)$D.
We conjecture that the effect of $BC$ term is similar: the emergence
of trivial fermions is possible in a bosonic model described by the
action $S\sim \int CdA+\widetilde{B}dB+BC$.

\section{Conclusion}\label{section_conclusion}

In this paper, when a gauge group is given, we investigate the compatibility
between all three classes of root braiding processes in $(3+1)$D spacetime, i.e., particle-loop braidings, multi-loop braidings and Borromean Rings braidings. We find that not all root braiding processes are mutually compatible, which is not so obvious on a superficial level. By incompatibility we mean that  two braiding processes cannot be supported in the \emph{same} system, i.e., there is no legitimate topological order in $(3+1)$-dimensional spacetime characterized by incompatible braiding phases. In the language of TQFT, a TQFT action describing incompatible braiding processes is not gauge invariant while preserving the cyclic group structure of each gauge subgroup. Our
conclusions are drawn as follows.

When the gauge group is $G=\mathbb{Z}_{N_{1}}$, only one class of
root braiding processes, i.e., particle-loop braiding, is realizable.
Therefore, there is no incompatibility when $G=\mathbb{Z}_{N_{1}}$.
The set of compatible braiding phase is simply $\left\{ \mathsf{\Theta}_{1}^{\text{H}}\right\} _{\mathbb{Z}_{N_{1}}}$. The corresponding TQFT action and gauge transformations are listed in Table~\ref{table_zn1zn2}.

When $G=\prod_{i=1}^{2}\mathbb{Z}_{N_{i}}$, besides particle-loop
braidings, three-loop braidings are realizable. In this case, all
braiding processes are compatible with each other. The set of compatible
braiding phase is the collection of all possible braiding phases:
$\left\{ \Theta_{i}^{\text{H}};\underline{\Theta_{2,2|1}^{\text{3L}},\Theta_{1,2|2}^{\text{3L}}},\underline{\Theta_{1,1|2}^{\text{3L}},\Theta_{2,1|1}^{\text{3L}}}\right\} _{\prod_{i=1}^{2}\mathbb{Z}_{N_{i}}}$. Table~\ref{table_zn1zn2} lists the TQFT action corresponding to compatible braiding phases along with its gauge transformations.

As for $G=\prod_{i=1}^{3}\mathbb{Z}_{N_{i}}$, the possible braiding
processes are particle-loop braidings, three-loop braidings and BR
braidings. We find that, particle-loop braidings are compatible with
three-loop braidings or BR braidings respectively. However, incompatibility occurs
between three-loop braidings and BR braidings, even between different BR braidings. In general notations, a BR braiding phase $\mathsf{\Theta}_{i,j|k}^{\text{BR}}$
is incompatible with three-loop braiding phases $\mathsf{\Theta}_{n,o|m}^{\text{3L}}$
in which one of the indices $\left\{ m,n,o\right\} $ is equal to
$k$. Moreover, in the case of $G=\prod_{i=1}^{3}\mathbb{Z}_{N_{i}}$,
$\mathsf{\Theta}_{i,j|k}^{\text{BR}}$ is incompatible with $\mathsf{\Theta}_{r,s|t}^{\text{BR}}$
($k\neq t$) if $k=r$ or $k=s$. From the viewpoint of TQFT, such
incompatibility can be explained by that there are no proper gauge
transformations for the TQFT action which consists of topological terms corresponding to incompatible braiding processes. Compatible braiding phases and corresponding TQFT actions of $G=\prod_{i=1}^{3}\mathbb{Z}_{N_{i}}$ are summarized in Table~\ref{table_zn1zn2zn3}.

For the case of $G=\prod_{i=1}^{4}\mathbb{Z}_{N_{i}}$, besides the
braidings mentioned above, four-loop braiding is realizable, classified
as one of multi-loop braidings. Similar to the case of $G=\prod_{i=1}^{3}\mathbb{Z}_{N_{i}}$,
when $G=\prod_{i=1}^{4}\mathbb{Z}_{N_{i}}$, particle-loop braidings
are compatible with multi-loop braidings or BR braidings respectively. However, the
incompatibility between BR braidings and multi-loop braidings still
exists. $\mathsf{\Theta}_{i,j|k}^{\text{BR}}$ is only compatible
with those $\mathsf{\Theta}_{s,t|r}^{\text{3L}}$ in which none of
the indices $\left\{ s,t,r\right\} $ is equal to $k$. No $\mathsf{\Theta}_{i,j|k}^{\text{BR}}$
is compatible with $\mathsf{\Theta}_{1,2,3,4}^{\text{4L}}$. Different from the case of $G=\prod_{i=1}^{3}\mathbb{Z}_{N_{i}}$, when $G=\prod_{i=1}^{4}\mathbb{Z}_{N_{i}}$,
different BR braidings may be compatible: for example, $\left\{ \mathsf{\Theta}_{1,2|4}^{\text{BR}},\mathsf{\Theta}_{1,3|4}^{\text{BR}},\mathsf{\Theta}_{2,3|4}^{\text{BR}}\right\} $
and $\left\{ \mathsf{\Theta}_{1,2|3}^{\text{BR}},\mathsf{\Theta}_{1,2|4}^{\text{BR}}\right\} $is respectively a subset of the sets of compatible braiding phases. Table~\ref{table_zn1zn2zn3zn4} and Table~\ref{table_ALL_zn1zn2zn3zn4} list all possible sets of compatible braiding phases, legitimate TQFT actions and corresponding gauge transformations at the case of $G=\prod_{i=1}^{4}\mathbb{Z}_{N_{i}}$.

Motivated by the compatibility of braiding processes in $(3+1)$D spacetime, we initiate  an attempt to the physics of  braiding processes in $(4+1)$D spacetime and their compatibility. Several results are given in Sec.~\ref{section_4d}. A comprehensive understanding of compatible braiding processes and TQFTs in $(4+1)$D topological orders will be discussed in Ref.~\cite{ZhangYeUnpublished}.

There are several interesting future investigations based on the compatibility analysis of braiding phases. For example, it is interesting to study the connection between  incompatibility and quantum anomaly \cite{2016arXiv161008645Y}. Meanwhile, we have not considered $BB$ term \cite{string7,horowitz89,bti2,Kapustin2014,kb2015,walker_wang_model_2012} which can switch self-statistics (either bosonic or fermionic) of particles. In this sense, adding $BB$ term may be useful when considering compatible braiding phases of either fermionic systems or bosonic systems with fermionic particle excitations that carry nontrivial gauge charges. Our work focuses only on topological orders where all particles are bosonic, so it is unnecessary to consider $BB$. In addition,  $BB$ can   drastically change the gauge group. In other words, the gauge group $G$ is simultaneously determined by the coefficients of $BF$ term and $BB$ term. Nevertheless, this complexity leads to superfluous  difficulty in determining gauge group and does not provide new braidings among topological excitations. In addition, from experiences in two-dimensional topological orders, exhausting all non-Abelian braidings with non-Abelian gauge group are very challenging if still applying the present field theory method. Algebraic tools from mathematics may be a much better way. It is definitely interesting to explore how to exhaust all legitimate topological orders with a non-Abelian gauge group.

\acknowledgements
The authors  would like to  thank AtMa P.O. Chan, Shinsei Ryu,  Zheng-Cheng Gu, Juven Wang, Meng Cheng, Eduardo Fradkin, Huan He, Apoorv Tiwari, and Xiao-Gang Wen for  earlier discussions and/or collaborations on topological phases of matter in higher dimensions. This work was supported in part by the Sun Yat-sen University startup grant, Guangdong Basic and Applied Basic Research Foundation under Grant No.~2020B1515120100, National Natural Science Foundation of China (NSFC) Grant (No.~11847608 \& No.~12074438).

\appendix
\onecolumngrid
\clearpage
\section{Microscopic derivation of the TQFT action $S\sim\int \sum_{i=1}^{3}B^{i}dA^{i}+A^{1}A^{2}B^{3}$}\label{appendix_microscopic_derivation}

  We can derive the $\int BdA+AAB$ action from a $3$-layer $3$D superfluid
where the layer $1$ and $2$ are in charge-$N_{1}$ and charge-$N_{2}$
particle condensations while the layer $3$ is in chareg-$N_{3}$
vortexline condensation. To illustrate the vortexline condensation picture,
we start with a normal superfluid state:
\begin{equation}
\mathcal{L}_{3}=\frac{\rho_{3}}{2}\left(\partial_{\mu}\theta^{3}\right)^{2}.
\end{equation}
Separate the phase angle into smooth part and singular part $\theta^{3}=\theta_{s}^{3}+\theta_{v}^{3}$
and introduce a Hubbard-Stratonovich auxiliary vector field $J_{\mu}^{3}$,
the above $\mathcal{L}_{3}$ can be expressed as
\begin{equation}
\mathcal{L}_{3}=\frac{1}{2\rho_{3}}\left(J_{\mu}^{3}\right)^{2}+iJ_{\mu}^{3}\left(\partial_{\mu}\theta_{s}^{3}+\partial_{\mu}\theta_{v}^{3}\right).
\end{equation}
Integrating out the smooth part $\theta_{s}^{3}$ leads to a constraint
$\partial_{\mu}J_{\mu}^{3}=0$, which can be solved by introducing
a $2$-form noncompact $\mathbb{U}\left(1\right)$ gauge filed $B_{\mu\nu}^{3}$
:
\begin{equation}
J_{\mu}^{3}=\frac{1}{4\pi}\varepsilon^{\mu\nu\lambda\rho}\partial_{\nu}B_{\mu\nu}^{3}.
\end{equation}
Thus $\mathcal{L}_{3}$ is dualed to the following gauge theory
\begin{equation}
\begin{alignedat}{1}\mathcal{L}_{3}= & \frac{1}{2\rho_{3}}\left(\frac{1}{4\pi}\varepsilon^{\mu\nu\lambda\rho}\partial_{\nu}B_{\mu\nu}^{3}\right)^{2}+i\frac{1}{4\pi}\varepsilon^{\mu\nu\lambda\rho}\partial_{\nu}B_{\mu\nu}^{3}\partial_{\mu}\theta_{v}^{3}\\
= & \frac{1}{2\rho_{3}}\left(\frac{1}{4\pi}\varepsilon^{\mu\nu\lambda\rho}\partial_{\nu}B_{\mu\nu}^{3}\right)^{2}+\frac{i}{2}B_{\mu\nu}^{3}\cdot\frac{1}{2\pi}\varepsilon^{\mu\nu\lambda\rho}\partial_{\lambda}\partial_{\rho}\theta_{v}^{3}.
\end{alignedat}
\end{equation}
The string current operator is defined by
\begin{equation}
\Sigma_{\mu\nu}^{3}=\frac{1}{2\pi}\varepsilon^{\mu\nu\lambda\rho}\partial_{\lambda}\partial_{\rho}\theta_{v}^{3}.
\end{equation}

Now consider a $3$-layer $3$D superfluid where the layer $1$ and
$2$ are in particle condensation while the layer $3$ is in string
condensation. The Lagrangian is
\begin{equation}
\begin{alignedat}{1}\mathcal{L}= & \frac{\rho_{3}}{2}\left(\partial_{[\mu}\Theta_{\nu]}-N_{3}B_{\mu\nu}^{3}\right)^{2}+\frac{\rho_{1}}{2}\left(\partial_{\mu}\theta^{1}-N_{1}A_{\mu}^{1}\right)^{2}+\frac{\rho_{2}}{2}\left(\partial_{\mu}\theta^{2}-N_{2}A_{\mu}^{2}\right)^{2}\\
 & +i\Lambda\varepsilon^{\mu\nu\lambda\rho}\left(\partial_{[\mu}\Theta_{\nu]}-N_{3}B_{\mu\nu}^{3}\right)\left(\partial_{\lambda}\theta^{1}-N_{1}A_{\lambda}^{1}\right)\left(\partial_{\rho}\theta^{2}-N_{2}A_{\rho}^{2}\right)+\mathcal{L}_{\text{Maxwell}}\\
= & \frac{\rho_{3}}{2}\left(\partial_{[\mu}\Theta_{\nu]}-N_{3}B_{\mu\nu}^{3}\right)^{2}+\frac{\rho_{1}}{2}\left(\partial_{\mu}\theta^{1}-N_{1}A_{\mu}^{1}\right)^{2}+\frac{\rho_{2}}{2}\left(\partial_{\mu}\theta^{2}-N_{2}A_{\mu}^{2}\right)^{2}+\mathcal{L}_{\text{Maxwell}}\\
 & +i\Lambda\varepsilon^{\mu\nu\lambda\rho}\partial_{[\mu}\Theta_{\nu]}\partial_{\lambda}\theta^{1}\partial_{\rho}\theta^{2}-i\Lambda\varepsilon^{\mu\nu\lambda\rho}\partial_{[\mu}\Theta_{\nu]}\partial_{\lambda}\theta^{1}N_{2}A_{\rho}^{2}-i\Lambda\varepsilon^{\mu\nu\lambda\rho}\partial_{[\mu}\Theta_{\nu]}\partial_{\rho}\theta^{2}N_{1}A_{\lambda}^{1}+i\Lambda\varepsilon^{\mu\nu\lambda\rho}\partial_{[\mu}\Theta_{\nu]}N_{1}A_{\lambda}^{1}N_{2}A_{\rho}^{2}\\
 & -i\Lambda\varepsilon^{\mu\nu\lambda\rho}N_{3}B_{\mu\nu}^{3}\partial_{\lambda}\theta^{1}\partial_{\rho}\theta^{2}+i\Lambda\varepsilon^{\mu\nu\lambda\rho}N_{3}B_{\mu\nu}^{3}\partial_{\lambda}\theta^{1}N_{2}A_{\rho}^{2}+i\Lambda\varepsilon^{\mu\nu\lambda\rho}N_{3}B_{\mu\nu}^{3}\partial_{\rho}\theta^{2}N_{1}A_{\lambda}^{1}-i\Lambda\varepsilon^{\mu\nu\lambda\rho}N_{1}A_{\lambda}^{1}N_{2}A_{\rho}^{2}N_{3}B_{\mu\nu}^{3}\\
= & \frac{\rho_{3}}{2}\left(\partial_{[\mu}\Theta_{\nu]}-N_{3}B_{\mu\nu}^{3}\right)^{2}+\frac{\rho_{1}}{2}\left(\partial_{\mu}\theta^{1}-N_{1}A_{\mu}^{1}\right)^{2}+\frac{\rho_{2}}{2}\left(\partial_{\mu}\theta^{2}-N_{2}A_{\mu}^{2}\right)^{2}+\mathcal{L}_{\text{Maxwell}}+\text{boundary terms}\\
 & +i\Lambda\varepsilon^{\mu\nu\lambda\rho}\left[2\Theta_{\mu}\partial_{\nu}\left(\partial_{\lambda}\theta^{1}N_{2}A_{\rho}^{2}\right)\right]+i\Lambda\varepsilon^{\mu\nu\lambda\rho}\left[2\Theta_{\mu}\partial_{\nu}\left(\partial_{\rho}\theta^{2}N_{1}A_{\lambda}^{1}\right)\right]+i\Lambda\varepsilon^{\mu\nu\lambda\rho}\frac{1}{2}\left[\theta^{1}\partial_{\lambda}\left(\partial_{\rho}\theta^{2}N_{3}B_{\mu\nu}^{3}\right)+\theta^{2}\partial_{\rho}\left(\partial_{\lambda}\theta^{1}N_{3}B_{\mu\nu}^{3}\right)\right]\\
 & -i\Lambda\varepsilon^{\mu\nu\lambda\rho}\left[2\Theta_{\mu}\partial_{\nu}\left(N_{1}N_{2}A_{\lambda}^{1}A_{\rho}^{2}\right)\right]+i\Lambda\varepsilon^{\mu\nu\lambda\rho}\left[-\theta^{1}\partial_{\lambda}\left(N_{2}A_{\rho}^{2}N_{3}B_{\mu\nu}^{3}\right)\right]+i\Lambda\varepsilon^{\mu\nu\lambda\rho}\left[-\theta^{2}\partial_{\rho}\left(N_{3}B_{\mu\nu}^{3}N_{1}A_{\lambda}^{1}\right)\right]\\
 & -i\Lambda\varepsilon^{\mu\nu\lambda\rho}A_{\lambda}^{1}A_{\rho}^{2}N_{3}B_{\mu\nu}^{3},
\end{alignedat}
\end{equation}
where $\partial_{[\mu}\Theta_{\nu]}=\partial_{\mu}\Theta_{\nu}-\partial_{\nu}\Theta_{\mu}$. Introduce Hubbard-Stratonovich fields $\Sigma_{\mu\nu}^{3}$, $j^{1}$
and $j^{2}$:
\begin{equation}
\begin{alignedat}{1}\mathcal{L}= & \frac{1}{8\rho_{3}}\left(\Sigma_{\mu\nu}^{3}\right)^{2}+i\frac{1}{2}\Sigma_{\mu\nu}^{3}\left(\partial_{[\mu}\Theta_{\nu]}-N_{3}B_{\mu\nu}^{3}\right)+\frac{1}{2\rho_{1}}\left(j^{1}\right)^{2}+ij_{\lambda}^{1}\left(\partial_{\lambda}\theta^{1}-N_{1}A_{\lambda}^{1}\right)+\frac{1}{2\rho_{2}}\left(j^{2}\right)^{2}+ij_{\rho}^{2}\left(\partial_{\rho}\theta^{2}-N_{2}A_{\rho}^{2}\right)\\
 & +i\Lambda\varepsilon^{\mu\nu\lambda\rho}\left[2\Theta_{\mu}\partial_{\nu}\left(\partial_{\lambda}\theta^{1}N_{2}A_{\rho}^{2}\right)\right]+i\Lambda\varepsilon^{\mu\nu\lambda\rho}\left[2\Theta_{\mu}\partial_{\nu}\left(\partial_{\rho}\theta^{2}N_{1}A_{\lambda}^{1}\right)\right]+i\Lambda\varepsilon^{\mu\nu\lambda\rho}\frac{1}{2}\left[\theta^{1}\partial_{\lambda}\left(\partial_{\rho}\theta^{2}N_{3}B_{\mu\nu}^{3}\right)+\theta^{2}\partial_{\rho}\left(\partial_{\lambda}\theta^{1}N_{3}B_{\mu\nu}^{3}\right)\right]\\
 & -i\Lambda\varepsilon^{\mu\nu\lambda\rho}\left[2\Theta_{\mu}\partial_{\nu}\left(N_{1}N_{2}A_{\lambda}^{1}A_{\rho}^{2}\right)\right]+i\Lambda\varepsilon^{\mu\nu\lambda\rho}\left[-\theta^{1}\partial_{\lambda}\left(N_{2}A_{\rho}^{2}N_{3}B_{\mu\nu}^{3}\right)\right]+i\Lambda\varepsilon^{\mu\nu\lambda\rho}\left[-\theta^{2}\partial_{\rho}\left(N_{3}B_{\mu\nu}^{3}N_{1}A_{\lambda}^{1}\right)\right]\\
 & -iN_{1}N_{2}N_{3}\Lambda\varepsilon^{\mu\nu\lambda\rho}A_{\lambda}^{1}A_{\rho}^{2}B_{\mu\nu}^{3}\\
 & +\mathcal{L}_{\text{Maxwell}}+\text{boundary terms}
\end{alignedat}
\end{equation}
Further introduce Lagrange multiplier fields $\xi^{I}$ and $\eta^{I}$
to decouple terms like $i\Lambda\varepsilon^{\mu\nu\lambda\rho}\left[-\theta^{1}\partial_{\lambda}\left(\partial_{\rho}\theta^{2}N_{3}B_{\mu\nu}^{3}\right)\right]$:
\begin{equation}
\begin{alignedat}{1}\mathcal{L}= & \frac{1}{8\rho_{3}}\left(\Sigma_{\mu\nu}^{3}\right)^{2}-i\Theta_{\mu}\partial_{\nu}\Sigma_{\mu\nu}^{3}-i\frac{1}{2}N_{3}B_{\mu\nu}^{3}\Sigma_{\mu\nu}^{3}\\
 & +\frac{1}{2\rho_{1}}\left(j^{1}\right)^{2}-i\theta^{1}\partial_{\lambda}j_{\lambda}^{1}-iN_{1}A_{\lambda}^{1}j_{\lambda}^{1}+\frac{1}{2\rho_{2}}\left(j^{2}\right)^{2}-i\theta^{2}\partial_{\rho}j_{\rho}^{2}-iN_{2}A_{\rho}^{2}j_{\rho}^{2}\\
 & +i\Lambda\varepsilon^{\mu\nu\lambda\rho}\left[2\Theta_{\mu}\partial_{\nu}\left(N_{1}N_{2}A_{\lambda}^{1}A_{\rho}^{2}\right)+\theta^{1}\partial_{\lambda}\left(N_{2}A_{\rho}^{2}N_{3}B_{\mu\nu}^{3}\right)+\theta^{2}\partial_{\rho}\left(N_{3}B_{\mu\nu}^{3}N_{1}A_{\lambda}^{1}\right)\right]\\
 & -iN_{1}N_{2}N_{3}\Lambda\varepsilon^{\mu\nu\lambda\rho}A_{\lambda}^{1}A_{\rho}^{2}B_{\mu\nu}^{3}\\
 & +i\eta_{\lambda}^{1}\left[\xi_{\lambda}^{1}-\Lambda\varepsilon^{\mu\nu\lambda\rho}\cdot\frac{1}{2}\partial_{\rho}\theta^{2}N_{3}B_{\mu\nu}^{3}\right]+i\theta^{1}\partial_{\lambda}\xi_{\lambda}^{1}\\
 & +i\eta_{\rho}^{2}\left[\xi_{\rho}^{2}-\Lambda\varepsilon^{\mu\nu\lambda\rho}\cdot\frac{1}{2}\partial_{\lambda}\theta^{1}N_{3}B_{\mu\nu}^{3}\right]+i\theta^{2}\partial_{\rho}\xi_{\rho}^{2}\\
 & +i\eta_{\mu\nu}^{3}\left[\xi_{\mu\nu}^{3}-\Lambda\varepsilon^{\mu\nu\lambda\rho}\left(2\partial_{\lambda}\theta^{1}N_{2}A_{\rho}^{2}+2\partial_{\rho}\theta^{2}N_{1}A_{\lambda}^{1}\right)\right]+i\Theta_{\mu}\partial_{\nu}\xi_{\mu\nu}^{3}+\mathcal{L}_{\text{Maxwell}}+\text{boundary terms}\\
= & \frac{1}{8\rho_{3}}\left(\Sigma_{\mu\nu}^{3}\right)^{2}+\frac{1}{2\rho_{1}}\left(j^{1}\right)^{2}+\frac{1}{2\rho_{2}}\left(j^{2}\right)^{2}-i\frac{1}{2}N_{3}B_{\mu\nu}^{3}\Sigma_{\mu\nu}^{3}-iN_{1}A_{\lambda}^{1}j_{\lambda}^{1}-iN_{2}A_{\rho}^{2}j_{\rho}^{2}\\
 & -i\Theta_{\mu}\partial_{\nu}\left[\Sigma_{\mu\nu}^{3}-\xi_{\mu\nu}^{3}+\Lambda\varepsilon^{\mu\nu\lambda\rho}\cdot2N_{1}N_{2}A_{\lambda}^{1}A_{\rho}^{2}\right]+i\eta_{\mu\nu}^{3}\xi_{\mu\nu}^{3}\\
 & -i\theta^{1}\partial_{\lambda}\left[j_{\lambda}^{1}-\xi_{\lambda}^{1}+\Lambda\varepsilon^{\mu\nu\lambda\rho}\left(N_{2}A_{\rho}^{2}N_{3}B_{\mu\nu}^{3}-\frac{1}{2}\eta_{\rho}^{2}N_{3}B_{\mu\nu}^{3}-2\eta_{\mu\nu}^{3}N_{2}A_{\rho}^{2}\right)\right]+i\eta_{\lambda}^{1}\xi_{\lambda}^{1}\\
 & -i\theta^{2}\partial_{\rho}\left[j_{\rho}^{2}-\xi_{\rho}^{2}+\Lambda\varepsilon^{\mu\nu\lambda\rho}\left(N_{3}B_{\mu\nu}^{3}N_{1}A_{\lambda}^{1}-\frac{1}{2}\eta_{\lambda}^{1}N_{3}B_{\mu\nu}^{3}-2\eta_{\mu\nu}^{3}N_{1}A_{\lambda}^{1}\right)\right]+i\eta_{\rho}^{1}\xi_{\rho}^{1}\\
 & -iN_{1}N_{2}N_{3}\Lambda\varepsilon^{\mu\nu\lambda\rho}A_{\lambda}^{1}A_{\rho}^{2}B_{\mu\nu}^{3}+\mathcal{L}_{\text{Maxwell}}+\text{boundary terms}
\end{alignedat}
\end{equation}
Integrating out $\Theta_{\mu}$, $\theta^{1}$ and $\theta^{2}$ yield
constraints
\begin{equation}
\partial_{\nu}\left[\Sigma_{\mu\nu}^{3}-\xi_{\mu\nu}^{3}+\Lambda\varepsilon^{\mu\nu\lambda\rho}\cdot2N_{1}N_{2}A_{\lambda}^{1}A_{\rho}^{2}\right]=0,
\end{equation}
\begin{equation}
\partial_{\lambda}\left[j_{\lambda}^{1}-\xi_{\lambda}^{1}+\Lambda\varepsilon^{\mu\nu\lambda\rho}\left(N_{2}A_{\rho}^{2}N_{3}B_{\mu\nu}^{3}-\frac{1}{2}\eta_{\rho}^{2}N_{3}B_{\mu\nu}^{3}-2\eta_{\mu\nu}^{3}N_{2}A_{\rho}^{2}\right)\right]=0
\end{equation}
 and
\begin{equation}
\partial_{\rho}\left[j_{\rho}^{2}-\xi_{\rho}^{2}+\Lambda\varepsilon^{\mu\nu\lambda\rho}\left(N_{3}B_{\mu\nu}^{3}N_{1}A_{\lambda}^{1}-\frac{1}{2}\eta_{\lambda}^{1}N_{3}B_{\mu\nu}^{3}-2\eta_{\mu\nu}^{3}N_{1}A_{\lambda}^{1}\right)\right]=0.
\end{equation}
These constraints can be solved by
\begin{equation}
\Sigma_{\mu\nu}^{3}=\frac{1}{2\pi}\varepsilon^{\mu\nu\lambda\rho}\partial_{\lambda}A_{\rho}^{3}+\xi_{\mu\nu}^{3}-\Lambda\varepsilon^{\mu\nu\lambda\rho}\cdot2N_{1}N_{2}A_{\lambda}^{1}A_{\rho}^{2},
\end{equation}
\begin{equation}
j_{\lambda}^{1}=\frac{1}{4\pi}\varepsilon^{\lambda\rho\mu\nu}\partial_{\rho}B_{\mu\nu}^{1}+\xi_{\lambda}^{1}-\Lambda\varepsilon^{\mu\nu\lambda\rho}\left(N_{2}A_{\rho}^{2}N_{3}B_{\mu\nu}^{3}-\frac{1}{2}\eta_{\rho}^{2}N_{3}B_{\mu\nu}^{3}-2\eta_{\mu\nu}^{3}N_{2}A_{\rho}^{2}\right),
\end{equation}
\begin{equation}
j_{\rho}^{2}=\frac{1}{4\pi}\varepsilon^{\rho\lambda\mu\nu}\partial_{\lambda}B_{\mu\nu}^{2}+\xi_{\rho}^{2}-\Lambda\varepsilon^{\mu\nu\lambda\rho}\left(N_{3}B_{\mu\nu}^{3}N_{1}A_{\lambda}^{1}-\frac{1}{2}\eta_{\lambda}^{1}N_{3}B_{\mu\nu}^{3}-2\eta_{\mu\nu}^{3}N_{1}A_{\lambda}^{1}\right).
\end{equation}
Then we obtain
\begin{equation}
\begin{alignedat}{1}\mathcal{L}= & -i\frac{1}{2}N_{3}B_{\mu\nu}^{3}\left[\frac{1}{2\pi}\varepsilon^{\mu\nu\lambda\rho}\partial_{\lambda}A_{\rho}^{3}+\xi_{\mu\nu}^{3}-\Lambda\varepsilon^{\mu\nu\lambda\rho}\cdot2N_{1}N_{2}A_{\lambda}^{1}A_{\rho}^{2}\right]\\
 & -iN_{1}A_{\lambda}^{1}\left[\frac{1}{4\pi}\varepsilon^{\lambda\rho\mu\nu}\partial_{\rho}B_{\mu\nu}^{1}+\xi_{\lambda}^{1}-\Lambda\varepsilon^{\mu\nu\lambda\rho}\left(N_{2}A_{\rho}^{2}N_{3}B_{\mu\nu}^{3}-\frac{1}{2}\eta_{\rho}^{2}N_{3}B_{\mu\nu}^{3}-2\eta_{\mu\nu}^{3}N_{2}A_{\rho}^{2}\right)\right]\\
 & -iN_{2}A_{\rho}^{2}\left[\frac{1}{4\pi}\varepsilon^{\rho\lambda\mu\nu}\partial_{\lambda}B_{\mu\nu}^{2}+\xi_{\rho}^{2}-\Lambda\varepsilon^{\mu\nu\lambda\rho}\left(N_{3}B_{\mu\nu}^{3}N_{1}A_{\lambda}^{1}-\frac{1}{2}\eta_{\lambda}^{1}N_{3}B_{\mu\nu}^{3}-2\eta_{\mu\nu}^{3}N_{1}A_{\lambda}^{1}\right)\right]\\
 & +\frac{1}{8\rho_{3}}\left[\frac{1}{2\pi}\varepsilon^{\mu\nu\lambda\rho}\partial_{\lambda}A_{\rho}^{3}+\xi_{\mu\nu}^{3}-\Lambda\varepsilon^{\mu\nu\lambda\rho}\cdot2N_{1}N_{2}A_{\lambda}^{1}A_{\rho}^{2}\right]^{2}\\
 & +\frac{1}{2\rho_{1}}\left[\frac{1}{4\pi}\varepsilon^{\lambda\rho\mu\nu}\partial_{\rho}B_{\mu\nu}^{1}+\xi_{\lambda}^{1}-\Lambda\varepsilon^{\mu\nu\lambda\rho}\left(N_{2}A_{\rho}^{2}N_{3}B_{\mu\nu}^{3}-\frac{1}{2}\eta_{\rho}^{2}N_{3}B_{\mu\nu}^{3}-2\eta_{\mu\nu}^{3}N_{2}A_{\rho}^{2}\right)\right]^{2}\\
 & +\frac{1}{2\rho_{2}}\left[\frac{1}{4\pi}\varepsilon^{\rho\lambda\mu\nu}\partial_{\lambda}B_{\mu\nu}^{2}+\xi_{\rho}^{2}-\Lambda\varepsilon^{\mu\nu\lambda\rho}\left(N_{3}B_{\mu\nu}^{3}N_{1}A_{\lambda}^{1}-\frac{1}{2}\eta_{\lambda}^{1}N_{3}B_{\mu\nu}^{3}-2\eta_{\mu\nu}^{3}N_{1}A_{\lambda}^{1}\right)\right]^{2}\\
 & +i\eta_{\lambda}^{1}\xi_{\lambda}^{1}+i\eta_{\rho}^{2}\xi_{\rho}^{2}+i\eta_{\mu}^{3}\xi_{\mu\nu}^{3}-iN_{1}N_{2}N_{3}\Lambda\varepsilon^{\mu\nu\lambda\rho}A_{\lambda}^{1}A_{\rho}^{2}B_{\mu\nu}^{3}+\mathcal{L}_{\text{Maxwell}}+\text{boundary terms}
\end{alignedat}
\end{equation}
Let us write $\mathcal{L}=\mathcal{L}_{A^{1}}+\mathcal{L}_{A^{2}}+\mathcal{L}_{B^{3}}-iN_{1}N_{2}N_{3}\Lambda\varepsilon^{\mu\nu\lambda\rho}A_{\lambda}^{1}A_{\rho}^{2}B_{\mu\nu}^{3}+\mathcal{L}_{\text{Maxwell}}$,
where
\begin{equation}
\begin{alignedat}{1}\mathcal{L}_{B^{3}}= & -\frac{iN_{3}}{4\pi}\varepsilon^{\mu\nu\lambda\rho}B_{\mu\nu}^{3}\partial_{\lambda}A_{\rho}^{3}+i\frac{1}{2}\Lambda\varepsilon^{\mu\nu\lambda\rho}N_{3}B_{\mu\nu}^{3}\cdot2N_{1}N_{2}A_{\lambda}^{1}A_{\rho}^{2}\\
 & +\frac{1}{8\rho_{3}}\left[\frac{1}{2\pi}\varepsilon^{\mu\nu\lambda\rho}\partial_{\lambda}A_{\rho}^{3}-\Lambda\varepsilon^{\mu\nu\lambda\rho}\cdot2N_{1}N_{2}A_{\lambda}^{1}A_{\rho}^{2}\right]^{2}+\frac{1}{8\rho_{3}}\left(\xi_{\mu\nu}^{3}\right)^{2}\\
 & +\frac{1}{4\rho_{3}}\xi_{\mu\nu}^{3}\left[i2\rho_{3}\left(2\eta_{\mu\nu}^{3}-N_{3}B_{\mu\nu}^{3}\right)+\frac{1}{2\pi}\varepsilon^{\mu\nu\lambda\rho}\partial_{\lambda}A_{\rho}^{3}-\Lambda\varepsilon^{\mu\nu\lambda\rho}\cdot2N_{1}N_{2}A_{\lambda}^{1}A_{\rho}^{2}\right]\\
\mathcal{L}_{A^{1}}= & -\frac{iN_{1}}{4\pi}\varepsilon^{\lambda\rho\mu\nu}A_{\lambda}^{1}\partial_{\rho}B_{\mu\nu}^{1}+i\Lambda\varepsilon^{\mu\nu\lambda\rho}N_{1}A_{\lambda}^{1}\left(N_{2}A_{\rho}^{2}N_{3}B_{\mu\nu}^{3}-\frac{1}{2}\eta_{\rho}^{2}N_{3}B_{\mu\nu}^{3}-2\eta_{\mu\nu}^{3}N_{2}A_{\rho}^{2}\right)\\
 & +\frac{1}{2\rho_{1}}\left[\frac{1}{4\pi}\varepsilon^{\lambda\rho\mu\nu}\partial_{\rho}B_{\mu\nu}^{1}-\Lambda\varepsilon^{\mu\nu\lambda\rho}\left(N_{2}A_{\rho}^{2}N_{3}B_{\mu\nu}^{3}-\frac{1}{2}\eta_{\rho}^{2}N_{3}B_{\mu\nu}^{3}-2\eta_{\mu\nu}^{3}N_{2}A_{\rho}^{2}\right)\right]^{2}+\frac{1}{2\rho_{1}}\left(\xi_{\lambda}^{1}\right)^{2}\\
 & +\frac{1}{\rho_{1}}\xi_{\lambda}^{1}\left[i\rho_{1}\left(\eta_{\lambda}^{1}-N_{1}A_{\lambda}^{1}\right)+\frac{1}{4\pi}\varepsilon^{\lambda\rho\mu\nu}\partial_{\rho}B_{\mu\nu}^{1}-\Lambda\varepsilon^{\mu\nu\lambda\rho}\left(N_{2}A_{\rho}^{2}N_{3}B_{\mu\nu}^{3}-\frac{1}{2}\eta_{\rho}^{2}N_{3}B_{\mu\nu}^{3}-2\eta_{\mu\nu}^{3}N_{2}A_{\rho}^{2}\right)\right]\\
\mathcal{L}_{A^{2}}= & -\frac{iN_{2}}{4\pi}\varepsilon^{\mu\nu\lambda\rho}A_{\rho}^{2}\partial_{\lambda}B_{\mu\nu}^{2}+i\Lambda\varepsilon^{\mu\nu\lambda\rho}N_{2}A_{\rho}^{2}\left(N_{3}B_{\mu\nu}^{3}N_{1}A_{\lambda}^{1}-\frac{1}{2}\eta_{\lambda}^{1}N_{3}B_{\mu\nu}^{3}-2\eta_{\mu\nu}^{3}N_{1}A_{\lambda}^{1}\right)\\
 & +\frac{1}{2\rho_{2}}\left[\frac{1}{4\pi}\varepsilon^{\rho\lambda\mu\nu}\partial_{\lambda}B_{\mu\nu}^{2}-\Lambda\varepsilon^{\mu\nu\lambda\rho}\left(N_{3}B_{\mu\nu}^{3}N_{1}A_{\lambda}^{1}-\frac{1}{2}\eta_{\lambda}^{1}N_{3}B_{\mu\nu}^{3}-2\eta_{\mu\nu}^{3}N_{1}A_{\lambda}^{1}\right)\right]+\frac{1}{2\rho_{2}}\left(\xi_{\rho}^{2}\right)^{2}\\
 & +\frac{1}{\rho_{2}}\xi_{\rho}^{2}\left[i\rho_{2}\left(\eta_{\rho}^{2}-N_{2}A_{\rho}^{2}\right)+\frac{1}{4\pi}\varepsilon^{\rho\lambda\mu\nu}\partial_{\lambda}B_{\mu\nu}^{2}-\Lambda\varepsilon^{\mu\nu\lambda\rho}\left(N_{3}B_{\mu\nu}^{3}N_{1}A_{\lambda}^{1}-\frac{1}{2}\eta_{\lambda}^{1}N_{3}B_{\mu\nu}^{3}-2\eta_{\mu\nu}^{3}N_{1}A_{\lambda}^{1}\right)\right]
\end{alignedat}
\end{equation}
Integrate out $\xi_{\mu\nu}^{3}$:
\begin{equation}
\begin{alignedat}{1}\mathcal{L}_{B_{3}}= & -\frac{iN_{3}}{4\pi}\varepsilon^{\mu\nu\lambda\rho}B_{\mu\nu}^{3}\partial_{\lambda}A_{\rho}^{3}+i\frac{1}{2}\Lambda\varepsilon^{\mu\nu\lambda\rho}N_{3}B_{\mu\nu}^{3}\cdot2N_{1}N_{2}A_{\lambda}^{1}A_{\rho}^{2}\\
 & +\frac{1}{8\rho_{3}}\left[\frac{1}{2\pi}\varepsilon^{\mu\nu\lambda\rho}\partial_{\lambda}A_{\rho}^{3}-\Lambda\varepsilon^{\mu\nu\lambda\rho}\cdot2N_{1}N_{2}A_{\lambda}^{1}A_{\rho}^{2}\right]^{2}\\
 & -\frac{1}{8\rho_{3}}\left[i2\rho_{3}\left(2\eta_{\mu\nu}^{3}-N_{3}B_{\mu\nu}^{3}\right)+\frac{1}{2\pi}\varepsilon^{\mu\nu\lambda\rho}\partial_{\lambda}N_{3}A_{\rho}^{3}-\Lambda\varepsilon^{\mu\nu\lambda\rho}\cdot2N_{1}N_{2}A_{\lambda}^{1}A_{\rho}^{2}\right]^{2}\\
= & -\frac{iN_{3}}{4\pi}\varepsilon^{\mu\nu\lambda\rho}B_{\mu\nu}^{3}\partial_{\lambda}A_{\rho}^{3}+iN_{1}N_{2}N_{3}\Lambda\varepsilon^{\mu\nu\lambda\rho}B_{\mu\nu}^{3}A_{\lambda}^{1}A_{\rho}^{2}\\
 & +\frac{\rho_{3}}{2}\left(2\eta_{\mu\nu}^{3}-N_{3}B_{\mu\nu}^{3}\right)^{2}-i\frac{1}{2}\left(2\eta_{\mu\mu}^{3}-N_{3}B_{\mu\nu}^{3}\right)\left(\frac{1}{2\pi}\varepsilon^{\mu\nu\lambda\rho}\partial_{\lambda}A_{\rho}^{3}-\Lambda\varepsilon^{\mu\nu\lambda\rho}\cdot2N_{1}N_{2}A_{\lambda}^{1}A_{\rho}^{2}\right)\\
= & \frac{\rho_{3}}{2}\left(2\eta_{\mu\nu}^{3}-N_{3}B_{\mu\nu}^{3}\right)^{2}-\frac{i}{2\pi}\varepsilon^{\mu\nu\lambda\rho}\eta_{\mu\nu}^{3}\partial_{\lambda}A_{\rho}^{3}+i\Lambda\varepsilon^{\mu\nu\lambda\rho}\eta_{\mu\nu}^{3}\cdot2N_{1}N_{2}A_{\lambda}^{1}A_{\rho}^{2}
\end{alignedat}
\end{equation}
Integrate out $\xi_{\lambda}^{1}$:
\begin{equation}
\begin{alignedat}{1}\mathcal{L}_{A^{1}}= & -\frac{iN_{1}}{4\pi}\varepsilon^{\lambda\rho\mu\nu}A_{\lambda}^{1}\partial_{\rho}B_{\mu\nu}^{1}+i\Lambda\varepsilon^{\mu\nu\lambda\rho}N_{1}A_{\lambda}^{1}\left(N_{2}A_{\rho}^{2}N_{3}B_{\mu\nu}^{3}-\frac{1}{2}\eta_{\rho}^{2}N_{3}B_{\mu\nu}^{3}-2\eta_{\mu\nu}^{3}N_{2}A_{\rho}^{2}\right)\\
 & +\frac{1}{2\rho_{1}}\left[\frac{1}{4\pi}\varepsilon^{\lambda\rho\mu\nu}\partial_{\rho}B_{\mu\nu}^{1}-\Lambda\varepsilon^{\mu\nu\lambda\rho}\left(N_{2}A_{\rho}^{2}N_{3}B_{\mu\nu}^{3}-\frac{1}{2}\eta_{\rho}^{2}N_{3}B_{\mu\nu}^{3}-2\eta_{\mu\nu}^{3}N_{2}A_{\rho}^{2}\right)\right]^{2}\\
 & -\frac{1}{2\rho_{1}}\left[i\rho\left(\eta_{\lambda}^{1}-N_{1}A_{\lambda}^{1}\right)+\frac{1}{4\pi}\varepsilon^{\lambda\rho\mu\nu}\partial_{\rho}B_{\mu\nu}^{1}-\Lambda\varepsilon^{\mu\nu\lambda\rho}\left(N_{2}A_{\rho}^{2}N_{3}B_{\mu\nu}^{3}-\frac{1}{2}\eta_{\rho}^{2}N_{3}B_{\mu\nu}^{3}-2\eta_{\mu\nu}^{3}N_{2}A_{\rho}^{2}\right)\right]^{2}\\
= & -\frac{iN_{1}}{4\pi}\varepsilon^{\lambda\rho\mu\nu}A_{\lambda}^{1}\partial_{\rho}B_{\mu\nu}^{1}+i\Lambda\varepsilon^{\mu\nu\lambda\rho}N_{1}A_{\lambda}^{1}\left(N_{2}A_{\rho}^{2}N_{3}B_{\mu\nu}^{3}-\frac{1}{2}\eta_{\rho}^{2}N_{3}B_{\mu\nu}^{3}-2\eta_{\mu\nu}^{3}N_{2}A_{\rho}^{2}\right)\\
 & +\frac{\rho_{1}}{2}\left(\eta_{\lambda}^{1}-N_{1}A_{\lambda}^{1}\right)^{2}-i\left(\eta_{\lambda}^{1}-N_{1}A_{\lambda}^{1}\right)\left[\frac{1}{4\pi}\varepsilon^{\lambda\rho\mu\nu}\partial_{\rho}B_{\mu\nu}^{1}-\Lambda\varepsilon^{\mu\nu\lambda\rho}\left(N_{2}A_{\rho}^{2}N_{3}B_{\mu\nu}^{3}-\frac{1}{2}\eta_{\rho}^{2}N_{3}B_{\mu\nu}^{3}-2\eta_{\mu\nu}^{3}N_{2}A_{\rho}^{2}\right)\right]\\
= & \frac{\rho_{1}}{2}\left(\eta_{\lambda}^{1}-A_{\lambda}^{1}\right)^{2}-\frac{i}{4\pi}\varepsilon^{\lambda\rho\mu\nu}\eta_{\lambda}^{1}\partial_{\rho}B_{\mu\nu}^{1}+i\Lambda\varepsilon^{\mu\nu\lambda\rho}\eta_{\lambda}^{1}\left(N_{2}A_{\rho}^{2}N_{3}B_{\mu\nu}^{3}-\frac{1}{2}\eta_{\rho}^{2}N_{3}B_{\mu\nu}^{3}-2\eta_{\mu\nu}^{3}N_{2}A_{\rho}^{2}\right)
\end{alignedat}
\end{equation}
Integrate out $\xi_{\rho}^{2}$:
\begin{equation}
\begin{alignedat}{1}\mathcal{L}_{A^{2}}= & -\frac{iN_{2}}{4\pi}\varepsilon^{\mu\nu\lambda\rho}A_{\rho}^{2}\partial_{\lambda}B_{\mu\nu}^{2}+i\Lambda\varepsilon^{\mu\nu\lambda\rho}N_{2}A_{\rho}^{2}\left(N_{3}B_{\mu\nu}^{3}N_{1}A_{\lambda}^{1}-\frac{1}{2}\eta_{\lambda}^{1}N_{3}B_{\mu\nu}^{3}-2\eta_{\mu\nu}^{3}N_{1}A_{\lambda}^{1}\right)\\
 & +\frac{1}{2\rho_{2}}\left[\frac{1}{4\pi}\varepsilon^{\rho\lambda\mu\nu}\partial_{\lambda}B_{\mu\nu}^{2}-\Lambda\varepsilon^{\mu\nu\lambda\rho}\left(N_{3}B_{\mu\nu}^{3}N_{1}A_{\lambda}^{1}-\frac{1}{2}\eta_{\lambda}^{1}N_{3}B_{\mu\nu}^{3}-2\eta_{\mu\nu}^{3}N_{1}A_{\lambda}^{1}\right)\right]\\
 & -\frac{1}{2\rho_{2}}\left[i\rho_{2}\left(\eta_{\rho}^{2}-N_{2}A_{\rho}^{2}\right)+\frac{1}{4\pi}\varepsilon^{\rho\lambda\mu\nu}\partial_{\lambda}B_{\mu\nu}^{2}-\Lambda\varepsilon^{\mu\nu\lambda\rho}\left(N_{3}B_{\mu\nu}^{3}N_{1}A_{\lambda}^{1}-\frac{1}{2}\eta_{\lambda}^{1}N_{3}B_{\mu\nu}^{3}-2\eta_{\mu\nu}^{3}N_{1}A_{\lambda}^{1}\right)\right]^{2}\\
= & -\frac{iN_{2}}{4\pi}\varepsilon^{\mu\nu\lambda\rho}A_{\rho}^{2}\partial_{\lambda}B_{\mu\nu}^{2}+i\Lambda\varepsilon^{\mu\nu\lambda\rho}N_{2}A_{\rho}^{2}\left(N_{3}B_{\mu\nu}^{3}N_{1}A_{\lambda}^{1}-\frac{1}{2}\eta_{\lambda}^{1}N_{3}B_{\mu\nu}^{3}-2\eta_{\mu\nu}^{3}N_{1}A_{\lambda}^{1}\right)\\
 & +\frac{\rho_{2}}{2}\left(\eta_{\rho}^{2}-N_{2}A_{\rho}^{2}\right)^{2}-i\left(\eta_{\rho}^{2}-N_{2}A_{\rho}^{2}\right)\left[\frac{1}{4\pi}\varepsilon^{\rho\lambda\mu\nu}\partial_{\lambda}B_{\mu\nu}^{2}-\Lambda\varepsilon^{\mu\nu\lambda\rho}\left(N_{3}B_{\mu\nu}^{3}N_{1}A_{\lambda}^{1}-\frac{1}{2}\eta_{\lambda}^{1}N_{3}B_{\mu\nu}^{3}-2\eta_{\mu\nu}^{3}N_{1}A_{\lambda}^{1}\right)\right]\\
= & \frac{\rho_{2}}{2}\left(\eta_{\rho}^{2}-N_{2}A_{\rho}^{2}\right)^{2}-\frac{i}{4\pi}\varepsilon^{\rho\lambda\mu\nu}\eta_{\rho}^{2}\partial_{\lambda}B_{\mu\nu}^{2}+i\Lambda\varepsilon^{\mu\nu\lambda\rho}\eta_{\rho}^{2}\left(N_{3}B_{\mu\nu}^{3}N_{1}A_{\lambda}^{1}-\frac{1}{2}\eta_{\lambda}^{1}N_{3}B_{\mu\nu}^{3}-2\eta_{\mu\nu}^{3}N_{1}A_{\lambda}^{1}\right)
\end{alignedat}
\end{equation}
We end up with
\begin{equation}
\begin{alignedat}{1}\mathcal{L}= & \frac{\rho_{3}}{2}\left(2\eta_{\mu\nu}^{3}-N_{3}B_{\mu\nu}^{3}\right)^{2}-\frac{i}{2\pi}\varepsilon^{\mu\nu\lambda\rho}\eta_{\mu\nu}^{3}\partial_{\lambda}A_{\rho}^{3}+i\Lambda\varepsilon^{\mu\nu\lambda\rho}\eta_{\mu\nu}^{3}\cdot2N_{1}N_{2}A_{\lambda}^{1}A_{\rho}^{2}\\
 & +\frac{\rho_{1}}{2}\left(\eta_{\lambda}^{1}-N_{1}A_{\lambda}^{1}\right)^{2}-\frac{i}{4\pi}\varepsilon^{\lambda\rho\mu\nu}\eta_{\lambda}^{1}\partial_{\rho}B_{\mu\nu}^{1}+i\Lambda\varepsilon^{\mu\nu\lambda\rho}\eta_{\lambda}^{1}\left(N_{2}A_{\rho}^{2}N_{3}B_{\mu\nu}^{3}-\frac{1}{2}\eta_{\rho}^{2}N_{3}B_{\mu\nu}^{3}-2\eta_{\mu\nu}^{3}N_{2}A_{\rho}^{2}\right)\\
 & +\frac{\rho_{2}}{2}\left(\eta_{\rho}^{2}-N_{2}A_{\rho}^{2}\right)^{2}-\frac{i}{4\pi}\varepsilon^{\rho\lambda\mu\nu}\eta_{\rho}^{2}\partial_{\lambda}B_{\mu\nu}^{2}+i\Lambda\varepsilon^{\mu\nu\lambda\rho}\eta_{\rho}^{2}\left(N_{3}B_{\mu\nu}^{3}N_{1}A_{\lambda}^{1}-\frac{1}{2}\eta_{\lambda}^{1}N_{3}B_{\mu\nu}^{3}-2\eta_{\mu\nu}^{3}N_{1}A_{\lambda}^{1}\right)\\
 & -iN_{1}N_{2}N_{3}\Lambda\varepsilon^{\mu\nu\lambda\rho}A_{\lambda}^{1}A_{\rho}^{2}B_{\mu\nu}^{3}+\mathcal{L}_{\text{Maxwell}}\\
= & -\frac{i}{2\pi}\varepsilon^{\mu\nu\lambda\rho}\eta_{\mu\nu}^{3}\partial_{\lambda}A_{\rho}^{3}-\frac{i}{4\pi}\varepsilon^{\lambda\rho\mu\nu}\eta_{\lambda}^{1}\partial_{\rho}B_{\mu\nu}^{1}-\frac{i}{4\pi}\varepsilon^{\rho\lambda\mu\nu}\eta_{\rho}^{2}\partial_{\lambda}B_{\mu\nu}^{2}\\
 & +\frac{\rho_{3}}{2}\left(2\eta_{\mu\nu}^{3}-N_{3}B_{\mu\nu}^{3}\right)^{2}+\frac{\rho_{1}}{2}\left(\eta_{\lambda}^{1}-N_{1}A_{\lambda}^{1}\right)^{2}+\frac{\rho_{2}}{2}\left(\eta_{\rho}^{2}-N_{2}A_{\rho}^{2}\right)^{2}\\
 & +i\Lambda\varepsilon^{\mu\nu\lambda\rho}\left[2\left(\eta_{\lambda}^{1}-N_{1}A_{\lambda}^{1}\right)\left(\eta_{\rho}^{2}-N_{2}A_{\rho}^{2}\right)\left(\eta_{\mu\nu}^{3}-\frac{1}{2}N_{3}B_{\mu\nu}^{3}\right)-2N_{1}N_{2}N_{3}\eta_{\lambda}^{1}\eta_{\rho}^{2}\eta_{\mu\nu}^{3}\right]+\mathcal{L}_{\text{Maxwell}}
\end{alignedat}
\end{equation}
Since we consider the vortexline condensation for $\rho_{3}$ and
particle condensation for $\rho_{1}$ and $\rho_{2}$, i.e., $\rho_{3}\rightarrow\infty$,
$\rho_{1}\rightarrow\infty$ and $\rho_{2}\rightarrow\infty$ are
taken. These limit conditions enforce that $\eta_{\lambda}^{1}=N_{1}A_{\lambda}^{1}$,
$\eta_{\rho}^{2}=N_{2}A_{\rho}^{2}$ and $\eta_{\mu\nu}^{3}=\frac{1}{2}N_{3}B_{\mu\nu}^{3}$.
So we obtain
\begin{equation}
\begin{alignedat}{1}\mathcal{L}= & -\frac{i}{2\pi}\varepsilon^{\mu\nu\lambda\rho}\frac{1}{2}N_{3}B_{\mu\nu}^{3}\partial_{\lambda}A_{\rho}^{3}-\frac{i}{4\pi}\varepsilon^{\lambda\rho\mu\nu}N_{1}A_{\lambda}^{1}\partial_{\rho}B_{\mu\nu}^{1}-\frac{i}{4\pi}\varepsilon^{\rho\lambda\mu\nu}N_{2}A_{\rho}^{2}\partial_{\lambda}B_{\mu\nu}^{2}-iN_{1}N_{2}N_{3}\Lambda\varepsilon^{\mu\nu\lambda\rho}A_{\lambda}^{1}A_{\rho}^{2}B_{\mu\nu}^{3}+\mathcal{L}_{\text{Maxwell}}\\
= & -\frac{iN_{3}}{4\pi}\varepsilon^{\mu\nu\lambda\rho}B_{\mu\nu}^{3}\partial_{\lambda}A_{\rho}^{3}+\frac{iN_{1}}{4\pi}\varepsilon^{\lambda\rho\mu\nu}B_{\mu\nu}^{1}\partial_{\rho}A_{\lambda}^{1}+\frac{iN_{2}}{4\pi}\varepsilon^{\rho\lambda\mu\nu}B_{\mu\nu}^{2}\partial_{\lambda}A_{\rho}^{2}-iN_{1}N_{2}N_{3}\Lambda\varepsilon^{\mu\nu\lambda\rho}A_{\lambda}^{1}A_{\rho}^{2}B_{\mu\nu}^{3}+\mathcal{L}_{\text{Maxwell}}\\
= & -\frac{iN_{3}}{4\pi}\varepsilon^{\mu\nu\lambda\rho}B_{\mu\nu}^{3}\partial_{\lambda}A_{\rho}^{3}-\frac{iN_{1}}{4\pi}\varepsilon^{\mu\nu\lambda\rho}B_{\mu\nu}^{1}\partial_{\lambda}A_{\rho}^{1}-\frac{iN_{2}}{4\pi}\varepsilon^{\mu\nu\lambda\rho}B_{\mu\nu}^{2}\partial_{\lambda}A_{\rho}^{2}-iN_{1}N_{2}N_{3}\Lambda\varepsilon^{\mu\nu\lambda\rho}A_{\mu}^{1}A_{\nu}^{2}B_{\lambda\rho}^{3}+\mathcal{L}_{\text{Maxwell}}.
\end{alignedat}
\end{equation}
The Maxwell kinetic term $\mathcal{L}_{\text{Maxwell}}$ can be neglected since its scaling dimension is more irrelevant than the other topological terms. Finally, we can drop the overall minus sign by relabeling indices and then obtain
\begin{equation}
\mathcal{L}=\frac{iN_{1}}{4\pi}\varepsilon^{\mu\nu\lambda\rho}B_{\mu\nu}^{1}\partial_{\lambda}A_{\rho}^{1}+\frac{iN_{2}}{4\pi}\varepsilon^{\mu\nu\lambda\rho}B_{\mu\nu}^{2}\partial_{\lambda}A_{\rho}^{2}+\frac{iN_{3}}{4\pi}\varepsilon^{\mu\nu\lambda\rho}B_{\mu\nu}^{3}\partial_{\lambda}A_{\rho}^{3}+iN_{1}N_{2}N_{3}\Lambda\varepsilon^{\mu\nu\lambda\rho}A_{\mu}^{1}A_{\nu}^{2}B_{\lambda\rho}^{3}.
\end{equation}

\section{All legitimate TQFT actions when $G=\prod_{i=1}^{4}\mathbb{Z}_{N_{i}}$}\label{sec_all_actions_zn1zn2zn3zn4}

\begin{table*}
\caption{\textbf{General expressions of compatible braiding phases, TQFT actions and
gauge transformations when $G=\prod_{i=1}^{4}\mathbb{Z}_{N_{i}}$}.
The definitions of coefficients and the Levi-Civita symbol are the
same as those in Table~\ref{table_zn1zn2zn3zn4}. By properly
assigning $\left(i,j,k,l\right)$, one can obtain all legitimate TQFT
actions for $G=\prod_{i=1}^{4}\mathbb{Z}_{N_{i}}$.\\
\textbf{The \nth{1} row}: since no $AAB$ terms in the TQFT action, no assignment for $\left(i,j,k,l\right)$
is needed. \\
\textbf{The \nth{2} row}: for actions with $\left(A^{i}A^{j}B^{l}+A^{i}A^{k}B^{l}+A^{j}A^{k}B^{l}\right)$
terms, the assignments for $\left(i,j,k,l\right)$
are $\left(1,2,3,4\right)$, $\left(2,3,4,1\right)$, $\left(3,4,1,2\right)$
and $\left(4,1,2,3\right)$.
For example, by taking $\left(i,j,k,l\right)=\left(4,1,2,3\right)$, we re-obtain
the \nth{2} TQFT action in Table~\ref{table_zn1zn2zn3zn4}. \\
\textbf{The \nth{3} row}: for actions with
$\left(A^{i}A^{j}B^{k}+A^{i}A^{j}B^{l}\right)$ terms, there
are 6 assignments for $\left(i,j,k,l\right)$: $\left(3,4,1,2\right)$,
$\left(2,4,1,3\right)$, $\left(2,3,1,4\right)$, $\left(4,1,2,3\right)$,
$\left(3,1,2,4\right)$ and $\left(1,2,3,4\right)$. By taking $\left(i,j,k,l\right)=\left(1,2,3,4\right)$,
we re-obtain the \nth{3} TQFT action in Table~\ref{table_zn1zn2zn3zn4}. \label{table_ALL_zn1zn2zn3zn4}}

\begin{tabular*}{\textwidth}{@{\extracolsep{\fill}}ccc}
\hline

\hline

\hline
Compatible braiding phases & TQFT actions & Gauge transformations\tabularnewline
\hline
$\begin{alignedat}{1}\Theta_{r}^{\text{H}}= & \frac{2\pi}{N_{r}}\\
\Theta_{s,s|r}^{\text{3L}}= & -2\cdot\Theta_{r,s|s}^{\text{3L}}=\frac{4\pi q_{rss}}{N_{r}N_{s}}\\
\Theta_{r,r|s}^{\text{3L}}= & -2\cdot\Theta_{s,r|r}^{\text{3L}}=\frac{4\pi q_{srr}}{N_{s}N_{r}}\\
\Theta_{s,t|r}^{\text{3L}}= & -\Theta_{r,t|s}^{\text{3L}}=\frac{2\pi q_{rst}}{N_{r}N_{s}}\\
\Theta_{t,r|s}^{\text{3L}}= & -\Theta_{s,r|t}^{\text{3L}}=\frac{2\pi q_{str}}{N_{s}N_{t}}\\
\Theta_{1,2,3,4}^{\text{4L}}= & \frac{2\pi q_{1234}}{N_{1}N_{2}N_{3}N_{4}}
\end{alignedat}
$ & $\begin{alignedat}{1} & \int\sum_{r=1}^{4}\frac{N_{r}}{2\pi}B^{r}dA^{r}\\
+ & \sum_{r<s}\left[\frac{q_{rss}}{\left(2\pi\right)^{2}}A^{r}A^{s}dA^{s}+\frac{q_{srr}}{\left(2\pi\right)^{2}}A^{s}A^{r}dA^{r}\right]\\
+ & \sum_{r<s<t}\left[\frac{q_{rst}}{\left(2\pi\right)^{2}}A^{r}A^{s}dA^{t}+\frac{q_{str}}{\left(2\pi\right)^{2}}A^{s}A^{t}dA^{r}\right]\\
+ & \frac{q_{1234}}{\left(2\pi\right)^{3}}A^{1}A^{2}A^{3}A^{4}
\end{alignedat}
$ & $\begin{alignedat}{1}A^{r}\rightarrow & A^{r}+d\chi^{r}\\
B^{r}\rightarrow & B^{r}+dV^{r}\\
 & +\sum_{s}\left(\frac{q_{rss}}{2\pi N_{r}}d\chi^{s}A^{s}-\frac{q_{srr}}{2\pi N_{r}}d\chi^{s}A^{r}\right)\\
 & +\sum_{a<b<c}\frac{q_{abc}}{2\pi N_{r}}\left(\delta_{r,a}d\chi^{b}A^{c}-\delta_{r,b}d\chi^{a}A^{c}\right)\\
 & +\sum_{a<b<c}\frac{q_{bac}}{2\pi N_{r}}\left(\delta_{r,b}d\chi^{c}A^{b}-\delta_{r,c}d\chi^{b}A^{a}\right)\\
 & -\frac{1}{2}\sum_{s,t,u}\frac{q_{1234}}{\left(2\pi\right)^{2}N_{r}}\epsilon^{rstu}A^{s}A^{t}\chi^{u}\\
 & +\frac{1}{2}\sum_{s,t,u}\frac{q_{1234}}{\left(2\pi\right)^{2}N_{r}}\epsilon^{rstu}A^{s}\chi^{t}d\chi^{u}\\
 & +\frac{1}{6}\sum_{s,t,u}\frac{q_{1234}}{\left(2\pi\right)^{2}N_{r}}\epsilon^{rstu}\chi^{s}d\chi^{t}d\chi^{u}
\end{alignedat}
$\tabularnewline
\hline
$\begin{alignedat}{1}\Theta_{r}^{\text{H}}= & \frac{2\pi}{N_{r}}\\
\Theta_{s,s|r}^{\text{3L}}= & -2\cdot\Theta_{r,s|s}^{\text{3L}}=\frac{4\pi q_{rss}}{N_{r}N_{s}}\\
\mathsf{\Theta}_{j,k|i}^{\text{3L}}= & -\mathsf{\Theta}_{i,k|j}^{\text{3L}}=\frac{2\pi q_{ijk}}{N_{i}N_{j}}\\
\mathsf{\Theta}_{k,i|j}^{\text{3L}}= & -\mathsf{\Theta}_{j,i|k}^{\text{3L}}=\frac{2\pi q_{jki}}{N_{j}N_{k}}\\
\Theta_{i,j|l}^{\text{BR}}= & \frac{2\pi p_{ij,l}}{N_{i}N_{j}N_{l}}\\
\Theta_{i,k|l}^{\text{BR}}= & \frac{2\pi p_{ik,l}}{N_{i}N_{k}N_{l}}\\
\Theta_{j,k|l}^{\text{BR}}= & \frac{2\pi p_{jk,l}}{N_{j}N_{k}N_{l}}
\end{alignedat}
$ & $\begin{alignedat}{1} & \int\sum_{r=1}^{4}\frac{N_{r}}{2\pi}B^{r}dA^{r}\\
\\
+ & \sum_{r\neq l,s\neq l}\frac{q_{rss}}{\left(2\pi\right)^{2}}A^{r}A^{s}dA^{s}\\
+ & \frac{q_{ijk}}{\left(2\pi\right)^{2}}A^{i}A^{j}dA^{k}+\frac{q_{jki}}{\left(2\pi\right)^{2}}A^{j}A^{k}dA^{i}\\
+ & \frac{p_{ij,l}}{\left(2\pi\right)^{2}}A^{i}A^{j}B^{l}\\
+ & \frac{p_{ik,l}}{\left(2\pi\right)^{2}}A^{i}A^{k}B^{l}\\
+ & \frac{p_{jk,l}}{\left(2\pi\right)^{2}}A^{j}A^{k}B^{l}
\end{alignedat}
$ & $\begin{alignedat}{1}A^{r}\rightarrow & A^{r}+d\chi^{r}\\
 & -\sum_{a,b}\frac{p_{ab,l}}{2\pi N_{4}}\delta_{r,l}\left(\chi^{a}A^{b}+\frac{1}{2}\chi^{a}d\chi^{b}\right)\\
B^{r}\rightarrow & B^{r}+dV^{r}\\
 & +\sum_{r\neq l,s\neq l}\left[\frac{q_{rss}}{2\pi N_{r}}d\chi^{s}A^{s}-\frac{q_{srr}}{2\pi N_{r}}d\chi^{s}A^{r}\right]\\
 & +\sum_{r\neq l,s\neq l}\left(\frac{q_{ijk}}{2\pi N_{r}}\epsilon^{rsk}d\chi^{s}A^{k}+\frac{q_{jki}}{2\pi N_{r}}\epsilon^{rsi}d\chi^{s}A^{i}\right)\\
 & -\sum_{r\neq l,j\neq l}\frac{p_{rs,l}}{2\pi N_{r}}\left(\chi^{s}B^{l}-A^{s}V^{l}+\chi^{s}dV^{l}\right)
\end{alignedat}
$\tabularnewline
\hline
$\begin{alignedat}{1}\Theta_{r}^{\text{H}}= & \frac{2\pi}{N_{r}}\\
\Theta_{j,j|i}^{\text{3L}}= & -2\cdot\Theta_{i,j|j}^{\text{3L}}=\frac{4\pi q_{ijj}}{N_{i}N_{j}}\\
\Theta_{i,i|j}^{\text{3L}}= & -2\cdot\Theta_{j,i|i}^{\text{3L}}=\frac{4\pi q_{jii}}{N_{j}N_{i}}\\
\Theta_{i,j|k}^{\text{BR}}= & \frac{2\pi p_{ij,k}}{N_{i}N_{j}N_{k}}\\
\Theta_{i,j|l}^{\text{BR}}= & \frac{2\pi p_{ij,l}}{N_{i}N_{j}N_{l}}
\end{alignedat}
$ & $\begin{alignedat}{1} & \int\sum_{r=1}^{4}\frac{N_{r}}{2\pi}B^{r}dA^{r}\\
+ & \frac{q_{ijj}}{\left(2\pi\right)^{2}}A^{i}A^{j}dA^{j}+\frac{q_{jii}}{\left(2\pi\right)^{2}}A^{j}A^{i}dA^{i}\\
+ & \frac{p_{ij,k}}{\left(2\pi\right)^{2}}A^{i}A^{j}B^{k}+\frac{p_{ij,l}}{\left(2\pi\right)^{2}}A^{i}A^{j}B^{l}
\end{alignedat}
$ & $\begin{alignedat}{1}A^{r}\rightarrow & A^{r}+d\chi^{r}\\
 & -\left(\frac{p_{ij,r}}{2\pi N_{r}}\delta_{r,k}+\frac{p_{ij,r}}{2\pi N_{r}}\delta_{r,l}\right)\left(\chi^{i}A^{j}+\frac{1}{2}\chi^{i}d\chi^{j}\right)\\
 & +\left(\frac{p_{ij,r}}{2\pi N_{r}}\delta_{r,k}+\frac{p_{ij,r}}{2\pi N_{r}}\delta_{r,l}\right)\left(\chi^{j}A^{i}+\frac{1}{2}\chi^{j}d\chi^{i}\right)\\
B^{r}\rightarrow & B^{r}+dV^{r}\\
 & +\sum_{s\neq k,l}\left(\frac{q_{rss}}{2\pi N_{r}}d\chi^{s}A^{s}-\frac{q_{jrr}}{2\pi N_{r}}d\chi^{s}A^{r}\right)\\
 & -\sum_{s}\frac{p_{ij,s}}{2\pi N_{r}}\delta_{r,i}\left(\chi^{j}B^{s}-A^{j}V^{s}+\chi^{j}dV^{s}\right)\\
 & +\sum_{s}\frac{p_{ij,s}}{2\pi N_{r}}\delta_{r,j}\left(\chi^{i}B^{s}-A^{i}V^{s}+\chi^{i}dV^{s}\right)
\end{alignedat}
$\tabularnewline
\hline

\hline

\hline
\end{tabular*}
\end{table*}
The general expression of legitimate TQFT actions and corresponding gauge transformations are listed in Table~\ref{table_ALL_zn1zn2zn3zn4}. By properly reassigning the indices $\left\{i,j,k,l\right\}$ (see the captions in Table~\ref{table_ALL_zn1zn2zn3zn4}), we can obtain all possible legitimate TQFT actions.  As discussed in Sec.~\ref{subsec_zn1zn2zn3zn4}, when $G=\prod_{i=1}^{4}\mathbb{Z}_{N_{i}}$, the legitimate TQFT actions can be classified by the properties of BR braidings they describe:
\begin{enumerate}
\item $S$ with no $AAB$ terms. The set of compatible
braiding phases is
\begin{equation}
\left\{ \Theta_{r}^{\text{H}};\underline{\Theta_{s,s|r}^{\text{3L}},\Theta_{r,s|s}^{\text{3L}}},\underline{\Theta_{r,r|s}^{\text{3L}},\Theta_{s,r|r}^{\text{3L}}},\underline{\Theta_{s,t|r}^{\text{3L}},\Theta_{r,t|s}^{\text{3L}}},\underline{\mathsf{\Theta}_{t,r|s}^{\text{3L}},\mathsf{\Theta}_{s,r|t}^{\text{3L}}};\mathsf{\Theta}_{1,2,3,4}^{\text{4L}}\right\} _{\prod_{r=1}^{4}\mathbb{Z}_{N_{r}}}
\end{equation}
where $r<s<t$ and $\left\{ r,s,t\right\} \subset\left\{ 1,2,3,4\right\} $. The underlines denote the linear dependence between braiding phases:
$\Theta_{s,s|r}^{\text{3L}}=-2\cdot\Theta_{r,s|s}^{\text{3L}}$,
$\Theta_{s,t|r}^{\text{3L}}=-\Theta_{r,t|s}^{\text{3L}}$, etc.
\item $S$ with $AAB$ terms which involve one flavor of $B^{i}$. The set
of compatible braiding phases is
\begin{equation}
\left\{ \Theta_{r}^{\text{H}};\underline{\Theta_{s,s|r}^{\text{3L}},\Theta_{r,s|s}^{\text{3L}}},\underline{\Theta_{j,k|i}^{\text{3L}},\Theta_{i,k|j}^{\text{3L}}},\underline{\Theta_{k,i|j}^{\text{3L}},\Theta_{j,i|k}^{\text{3L}}};\mathsf{\Theta}_{i,j|l}^{\text{BR}},\mathsf{\Theta}_{i,k|l}^{\text{BR}},\mathsf{\Theta}_{j,k|l}^{\text{BR}}\right\} _{\prod_{r=1}^{4}\mathbb{Z}_{r}}
\end{equation}
where $r\neq s$, $r\neq l$, $s\neq l$, and the assignments for $\left(i,j,k,l\right)$
are $\left(1,2,3,4\right)$, $\left(2,3,4,1\right)$, $\left(3,4,1,2\right)$
and $\left(4,1,2,3\right)$.
\item $S$ with $AAB$ terms which involve two flavors of $B^{i}$. The
set of compatible braiding phases is
\begin{equation}
\left\{ \Theta_{r}^{\text{H}};\underline{\Theta_{j,j|i}^{\text{3L}},\Theta_{i,j|j}^{\text{3L}}},\underline{\Theta_{i,i|j}^{\text{3L}},\Theta_{j,i|i}^{\text{3L}}};\mathsf{\Theta}_{i,j|k}^{\text{BR}},\mathsf{\Theta}_{i,j|l}^{\text{BR}}\right\} _{\prod_{r=1}^{4}\mathbb{Z}_{r}}
\end{equation}
where the assignments for $\left(i,j,k,l\right)$ are: $\left(3,4,1,2\right)$,
$\left(2,4,1,3\right)$, $\left(2,3,1,4\right)$, $\left(4,1,2,3\right)$,
$\left(3,1,2,4\right)$ and $\left(1,2,3,4\right)$.
\end{enumerate}

\section{Some technical details}

\subsection{Derivation for the incompatibility of $A^{1}A^{2}B^{3}$ and $A^{2}A^{3}B^{1}$\label{subsec_Derivation_A1A2B3+A2A3B1}}

We try to find proper gauge transformations for this action:
\begin{equation}
S=\int\sum_{i=1}^{3}\frac{N_{i}}{2\pi}B^{i}dA^{i}+\frac{p_{12,3}}{\left(2\pi\right)^{2}}A^{1}A^{2}B^{3}+\frac{p_{23,1}}{\left(2\pi\right)^{2}}A^{2}A^{3}B^{1}.
\label{eq_action_Appendix_A1A2B3+A2A3B1}
\end{equation}

We list all 4 possible test gauge transformations respecting the $\mathbb{\mathbb{\mathbb{Z}}}_{N_{1}}$
and $\mathbb{\mathbb{\mathbb{Z}}}_{N_{3}}$ cyclic group structure and corresponding $\Delta S$'s in Table~\ref{table_test_GT-A1A2B3+A2A3B1_A1_normal} and Table~\ref{table_test_GT-A1A2B3+A2A3B1_B1_normal}.
We point out that there always exist ``stubborn terms'' which cannot
be eliminated by subtraction not be absorbed into a total derivative
term, making $\Delta S$ nonvanishing. If we impose that $\Delta S=0\mod2\pi$, the $\mathbb{\mathbb{\mathbb{Z}}}_{N_{1}}$ and
$\mathbb{\mathbb{\mathbb{Z}}}_{N_{3}}$ cyclic group structure are no longer
respected. Such dilemma indicates that (\ref{eq_action_Appendix_A1A2B3+A2A3B1})
is not a legitimate TQFT action, thus $A^{1}A^{2}B^{3}$ and $A^{2}A^{3}B^{1}$
are incompatible.

For the \nth{2} column of Table~\ref{table_test_GT-A1A2B3+A2A3B1_B1_normal},
it may not so straightforward to see that $\Delta S$ is nonvanishing.
Below we make some illustration. The gauge transformations are
\begin{equation}
\begin{alignedat}{1}A^{1}\rightarrow & A^{1}+d\chi^{1}+X^{1},\\
A^{2}\rightarrow & A^{2}+d\chi^{2},\\
A^{3}\rightarrow & A^{3}+d\chi^{3}+X^{3},\\
B^{1}\rightarrow & B^{1}+dV^{1},\\
B^{2}\rightarrow & B^{2}+dV^{2}+Y^{2},\\
B^{3}\rightarrow & B^{3}+dV^{3}.
\end{alignedat}
\label{eq_test_GT_A1A2B3+A2A2B1_B1_B3_normal}
\end{equation}
The variation of action (boundary terms neglected) is
\begin{equation}
\begin{alignedat}{1}\Delta S= & \int\frac{N_{1}}{2\pi}B^{1}dX^{1}+\frac{N_{2}}{2\pi}Y^{2}dA^{2}+\frac{N_{3}}{2\pi}B^{3}dX^{3}\\
 & +\frac{p_{12,3}}{\left(2\pi\right)^{2}}\left(A^{1}d\chi^{2}B^{3}+d\chi^{1}A^{2}B^{3}+d\chi^{1}d\chi^{2}B^{3}+X^{1}A^{2}B^{3}+X^{1}d\chi^{2}B^{3}\right)\\
 & +\frac{p_{12,3}}{\left(2\pi\right)^{2}}\left(A^{1}A^{2}dV^{3}+A^{1}d\chi^{2}dV^{3}+d\chi^{1}A^{2}dV^{3}+X^{1}A^{2}dV^{3}+X^{1}d\chi^{2}dV^{3}\right)\\
 & +\frac{p_{23,1}}{\left(2\pi\right)^{2}}\left(d\chi^{2}A^{3}B^{1}+A^{2}d\chi^{3}B^{1}+d\chi^{2}d\chi^{3}B^{1}+A^{2}X^{3}B^{1}+d\chi^{2}X^{3}B^{1}\right)\\
 & +\frac{p_{23,1}}{\left(2\pi\right)^{2}}\left(A^{2}A^{3}dV^{1}+d\chi^{2}A^{3}dV^{1}+A^{2}d\chi^{3}dV^{1}+A^{2}X^{3}dV^{1}+d\chi^{2}X^{3}dV^{1}\right).
\end{alignedat}
\end{equation}
If we want to eliminate the $A^{1}A^{2}dV^{3}$ term by subtraction, we can only require that
\begin{equation}
\begin{alignedat}{1}\frac{N_{3}}{2\pi}dV^{3}dX^{3}+\frac{p_{12,3}}{\left(2\pi\right)^{2}}A^{1}A^{2}dV^{3}= & \frac{p_{12,3}}{\left(2\pi\right)^{2}}\left[dV^{3}\left(\frac{N_{3}}{2\pi}\cdot\frac{\left(2\pi\right)^{2}}{p_{12,3}}dX^{3}+A^{1}A^{2}\right)\right]\\
= & \frac{p_{12,3}}{\left(2\pi\right)^{2}}\left[dV^{3}\left(\frac{N_{3}}{2\pi}\cdot\frac{\left(2\pi\right)^{2}}{p_{12,3}}\cdot\left(-A^{1}A^{2}+\cdots\right)+A^{1}A^{2}\right)\right]\\
= & \frac{N_{3}}{2\pi}dV^{3}\left(\cdots\right),
\end{alignedat}
\end{equation}
which requires
\begin{equation}
dX^{3}=-A^{1}A^{2}+\cdots.
\label{eq_A1A2B3+A2A3B1_dX3}
\end{equation}
However (\ref{eq_A1A2B3+A2A3B1_dX3}) is impossible since $A^{1}A^{2}$ is not exact in general.
If we want to absorb $A^{1}A^{2}dV^{3}$ into a total derivative term,
we would need a $\left(dA^{1}A^{2}V^{3}-A^{1}dA^{2}V^{3}\right)$ term since $d\left(A^{1}A^{2}V^{3}\right)=dA^{1}A^{2}V^{3}-A^{1}dA^{2}V^{3}+A^{1}A^{2}dV^{3}$.
We may have a $-A^{1}dA^{2}V^{3}$ term contributed by $\frac{N_{2}}{2\pi}Y^{2}dA^{2}$,
but we do not have a $dA^{1}A^{2}V^{3}$ term since no term containing $dA^{1}$ in $\Delta S$. Therefore, action $S=\int\sum_{i=1}^{3}\frac{N_{i}}{2\pi}B^{i}dA^{i}+\frac{p_{12,3}}{\left(2\pi\right)^{2}}A^{1}A^{2}B^{3}+\frac{p_{23,1}}{\left(2\pi\right)^{2}}A^{2}A^{3}B^{1}$
is not gauge invariant under (\ref{eq_test_GT_A1A2B3+A2A2B1_B1_B3_normal}).

\begin{table*}
\caption{Test gauge transformations with $A^{1}\rightarrow A^{1}+d\chi^{1}$
for $S=\int\sum_{i=1}^{3}\frac{N_{i}}{2\pi}B^{i}dA^{i}+\frac{p_{12,3}}{\left(2\pi\right)^{2}}A^{1}A^{2}B^{3}+\frac{p_{23,1}}{\left(2\pi\right)^{2}}A^{2}A^{3}B^{1}$.
The ``stubborn terms'' are terms which cannot be eliminated by subtraction nor be absorbed into a total derivative term,
making $\Delta S\neq 0\!\!\mod 2\pi$.
\label{table_test_GT-A1A2B3+A2A3B1_A1_normal}}
\begin{tabular*}{\textwidth}{@{\extracolsep{\fill}}ccc}
\hline

\hline

\hline
Test gauge transformations & $\Delta S$ & Stubborn terms\tabularnewline
\hline
$\begin{alignedat}{1}A^{1}\rightarrow & A^{1}+d\chi^{1}\\
A^{2}\rightarrow & A^{2}+d\chi^{2}\\
A^{3}\rightarrow & A^{3}+d\chi^{3}\\
B^{1}\rightarrow & B^{1}+dV^{1}+Y^{1}\\
B^{2}\rightarrow & B^{2}+dV^{2}+Y^{2}\\
B^{3}\rightarrow & B^{3}+dV^{3}+Y^{3}
\end{alignedat}
$ & $\begin{alignedat}{1} & \int\sum_{i=1}^{3}\frac{N_{i}}{2\pi}Y^{i}dA^{i}+\frac{p_{12,3}}{\left(2\pi\right)^{2}}\left(d\chi^{1}A^{2}B^{3}+A^{1}d\chi^{2}B^{3}+d\chi^{1}d\chi^{2}B^{3}\right)\\
+ & \frac{p_{12,3}}{\left(2\pi\right)^{2}}\left(A^{1}A^{2}dV^{3}+d\chi^{1}A^{2}dV^{3}+A^{1}d\chi^{2}dV^{3}+d\chi^{1}d\chi^{2}dV^{3}\right)\\
+ & \frac{p_{12,3}}{\left(2\pi\right)^{2}}\left(A^{1}A^{2}Y^{3}+d\chi^{1}A^{2}Y^{3}+A^{1}d\chi^{2}Y^{3}+d\chi^{1}d\chi^{2}Y^{3}\right)\\
+ & \frac{p_{23,1}}{\left(2\pi\right)^{2}}\left(d\chi^{2}A^{3}B^{1}+A^{2}d\chi^{3}B^{1}+d\chi^{2}d\chi^{3}B^{1}\right)\\
+ & \frac{p_{23,1}}{\left(2\pi\right)^{2}}\left(A^{2}A^{3}dV^{1}+d\chi^{2}A^{3}dV^{1}+A^{2}d\chi^{3}dV^{1}+d\chi^{2}d\chi^{3}dV^{1}\right)\\
+ & \frac{p_{23,1}}{\left(2\pi\right)^{2}}\left(A^{2}A^{3}Y^{1}+d\chi^{2}A^{3}Y^{1}+A^{2}d\chi^{3}Y^{1}+d\chi^{2}d\chi^{3}Y^{1}\right)
\end{alignedat}
$ & $\begin{alignedat}{1}d\chi^{1}d\chi^{2}B^{3},\\
d\chi^{2}d\chi^{3}B^{1}
\end{alignedat}
$\tabularnewline
\hline
$\begin{alignedat}{1}A^{1}\rightarrow & A^{1}+d\chi^{1}\\
A^{2}\rightarrow & A^{2}+d\chi^{2}\\
A^{3}\rightarrow & A^{3}+d\chi^{3}+X^{3}\\
B^{1}\rightarrow & B^{1}+dV^{1}+Y^{1}\\
B^{2}\rightarrow & B^{2}+dV^{2}+Y^{2}\\
B^{3}\rightarrow & B^{3}+dV^{3}
\end{alignedat}
$ & $\begin{alignedat}{1} & \int\sum_{i=1}^{2}\frac{N_{i}}{2\pi}Y^{i}dA^{i}+\frac{N_{3}}{2\pi}\left(dV^{3}dA^{3}+B^{3}dX^{3}+dV^{3}dX^{3}\right)\\
+ & \frac{p_{12,3}}{\left(2\pi\right)^{2}}\left(d\chi^{1}A^{2}B^{3}+A^{1}d\chi^{1}B^{3}+d\chi^{1}d\chi^{2}B^{3}\right)\\
+ & \frac{p_{12,3}}{\left(2\pi\right)^{2}}\left(A^{1}A^{2}dV^{3}+d\chi^{1}A^{2}dV^{3}+A^{1}d\chi^{1}dV^{3}+d\chi^{1}d\chi^{2}dV^{3}\right)\\
+ & \frac{p_{23,1}}{\left(2\pi\right)^{2}}\left(d\chi^{2}A^{3}B^{1}+A^{2}d\chi^{3}B^{1}+d\chi^{2}d\chi^{3}B^{1}\right)\\
+ & \frac{p_{23,1}}{\left(2\pi\right)^{2}}\left(A^{2}A^{3}dV^{1}+d\chi^{2}A^{3}dV^{1}+A^{2}d\chi^{3}dV^{1}+d\chi^{2}d\chi^{3}dV^{1}\right)\\
+ & \frac{p_{23,1}}{\left(2\pi\right)^{2}}\left(A^{2}A^{3}Y^{1}+d\chi^{2}A^{3}Y^{1}+A^{2}d\chi^{3}Y^{1}+d\chi^{2}d\chi^{3}Y^{1}\right)
\end{alignedat}
$ & $d\chi^{2}d\chi^{3}B^{1}$\tabularnewline
\hline

\hline

\hline
\end{tabular*}
\end{table*}

\begin{table*}
\caption{Test gauge transformations with $B^{1}\rightarrow B^{1}+dV^{1}$ for
$S=\int\sum_{i=1}^{3}\frac{N_{i}}{2\pi}B^{i}dA^{i}+\frac{p_{12,3}}{\left(2\pi\right)^{2}}A^{1}A^{2}B^{3}+\frac{p_{23,1}}{\left(2\pi\right)^{2}}A^{2}A^{3}B^{1}$.
The ``stubborn terms'' are terms which cannot be eliminated by subtraction nor be absorbed into a total derivative term,
making $\Delta S\neq 0\!\!\mod 2\pi$. \label{table_test_GT-A1A2B3+A2A3B1_B1_normal}}
\begin{tabular*}{\textwidth}{@{\extracolsep{\fill}}ccc}
\hline

\hline

\hline

Test gauge transformations & $\Delta S$ & Stubborn terms\tabularnewline
\hline
$\begin{alignedat}{1}A^{1}\rightarrow & A^{1}+d\chi^{1}+X^{1}\\
A^{2}\rightarrow & A^{2}+d\chi^{2}\\
A^{3}\rightarrow & A^{3}+d\chi^{3}\\
B^{1}\rightarrow & B^{1}+dV^{1}\\
B^{2}\rightarrow & B^{2}+dV^{2}+Y^{2}\\
B^{3}\rightarrow & B^{3}+dV^{3}+Y^{3}
\end{alignedat}
$ & $\begin{alignedat}{1} & \int\frac{N_{1}}{2\pi}dV^{1}dA^{1}+\frac{N_{1}}{2\pi}B^{1}dX^{1}+\frac{N_{1}}{2\pi}dV^{1}dX^{1}+\sum_{i=2}^{3}\frac{N_{i}}{2\pi}Y^{i}dA^{i}\\
+ & \frac{p_{12,3}}{\left(2\pi\right)^{2}}\left(A^{1}d\chi^{2}B^{3}+d\chi^{1}A^{2}B^{3}+d\chi^{1}d\chi^{2}B^{3}+X^{1}A^{2}B^{3}+X^{1}d\chi^{2}B^{3}\right)\\
+ & \frac{p_{12,3}}{\left(2\pi\right)^{2}}\left(A^{1}A^{2}dV^{3}+A^{1}d\chi^{2}dV^{3}+d\chi^{1}A^{2}dV^{3}+d\chi^{1}d\chi^{2}dV^{3}+X^{1}A^{2}dV^{3}+X^{1}d\chi^{2}dV^{3}\right)\\
+ & \frac{p_{12,3}}{\left(2\pi\right)^{2}}\left(A^{1}A^{2}Y^{3}+A^{1}d\chi^{2}Y^{3}+d\chi^{1}A^{2}Y^{3}+d\chi^{1}d\chi^{2}Y^{3}+X^{1}A^{2}Y^{3}+X^{1}d\chi^{2}Y^{3}\right)\\
+ & \frac{p_{23,1}}{\left(2\pi\right)^{2}}\left(A^{2}A^{3}+d\chi^{2}A^{3}B^{1}+A^{2}d\chi^{3}B^{1}+d\chi^{2}d\chi^{3}B^{1}\right)\\
+ & \frac{p_{23,1}}{\left(2\pi\right)^{2}}\left(A^{2}A^{3}dV^{1}+d\chi^{2}A^{3}dV^{1}+A^{2}d\chi^{3}dV^{1}+d\chi^{2}d\chi^{3}dV^{1}\right)\\
+ & \frac{p_{23,1}}{\left(2\pi\right)^{2}}\left(A^{2}A^{3}Y^{1}+d\chi^{2}A^{3}Y^{1}+A^{2}d\chi^{3}Y^{1}+d\chi^{2}d\chi^{3}Y^{1}\right)
\end{alignedat}
$ & $d\chi^{1}d\chi^{2}B^{3}$\tabularnewline
\hline
$\begin{alignedat}{1}A^{1}\rightarrow & A^{1}+d\chi^{1}+X^{1}\\
A^{2}\rightarrow & A^{2}+d\chi^{2}\\
A^{3}\rightarrow & A^{3}+d\chi^{3}+X^{3}\\
B^{1}\rightarrow & B^{1}+dV^{1}\\
B^{2}\rightarrow & B^{2}+dV^{2}+Y^{2}\\
B^{3}\rightarrow & B^{3}+dV^{3}
\end{alignedat}
$ & $\begin{alignedat}{1} & \int\frac{N_{1}}{2\pi}B^{1}dX^{1}+\frac{N_{1}}{2\pi}dV^{1}dX^{1}+\frac{N_{2}}{2\pi}Y^{2}dA^{2}+\frac{N_{3}}{2\pi}B^{3}dX^{3}+\frac{N_{3}}{2\pi}dV^{3}dX^{3}\\
+ & \frac{p_{12,3}}{\left(2\pi\right)^{2}}\left(A^{1}d\chi^{2}B^{3}+d\chi^{1}A^{2}B^{3}+d\chi^{1}d\chi^{2}B^{3}+X^{1}A^{2}B^{3}+X^{1}d\chi^{2}B^{3}\right)\\
+ & \frac{p_{12,3}}{\left(2\pi\right)^{2}}\left(A^{1}A^{2}dV^{3}+A^{1}d\chi^{2}dV^{3}+d\chi^{1}A^{2}dV^{3}+d\chi^{1}d\chi^{2}dV^{3}+X^{1}A^{2}dV^{3}+X^{1}d\chi^{2}dV^{3}\right)\\
+ & \frac{p_{23,1}}{\left(2\pi\right)^{2}}\left(d\chi^{2}A^{3}B^{1}+A^{2}d\chi^{3}B^{1}+d\chi^{2}d\chi^{3}B^{1}+A^{2}X^{3}B^{1}+d\chi^{2}X^{3}B^{1}\right)\\
+ & \frac{p_{23,1}}{\left(2\pi\right)^{2}}\left(A^{2}A^{3}dV^{1}+d\chi^{2}A^{3}dV^{1}+A^{2}d\chi^{3}dV^{1}+d\chi^{2}d\chi^{3}dV^{1}+A^{2}X^{3}dV^{1}+d\chi^{2}X^{3}dV^{1}\right)
\end{alignedat}
$ & $\begin{alignedat}{1}A^{1}A^{2}dV^{3},\\
A^{2}A^{3}dV^{1}
\end{alignedat}
$\tabularnewline
\hline

\hline

\hline

\end{tabular*}
\end{table*}

\subsection{Derivation of the incompatibility between $A^{1}A^{3}dA^{3}$ and
$A^{1}A^{2}B^{3}$}\label{subsec_derivation_incompatibility_A1A3dA3+A1A2B3}
The TQFT action is assumed to be
\begin{equation}
S=\int\sum_{i=1}^{3}\frac{N_{i}}{2\pi}B^{i}dA^{i}+\frac{q_{133}}{\left(2\pi\right)^{2}}A^{1}A^{3}dA^{3}+\frac{p_{12,3}}{\left(2\pi\right)^{2}}A^{1}A^{2}B^{3}.\label{eq_appendix_action_A1A3dA3+A1A2B3}
\end{equation}

First, we assume that the gauge transformations are
\begin{equation}
\begin{alignedat}{1}A^{i}\rightarrow & A^{i}+d\chi^{i},\\
B^{i}\rightarrow & B^{i}+dV^{i}+Y^{i},
\end{alignedat}
\label{eq_appendix_GT-A1A3dA3+A1A2B3-A3-normal}
\end{equation}
 i.e., the $\mathbb{Z}_{N_{3}}$ cyclic group structure is encoded in
\begin{equation}
\oint A^{3}\in\frac{2\pi}{N_{3}}\mathbb{Z}_{N_{3}}.
\end{equation}
 Under the gauge transformations, the variation of action is (boundary
terms neglected) is
\begin{equation}
\begin{alignedat}{1}\Delta S= & \int\frac{N_{1}}{2\pi}Y^{1}dA^{1}+\frac{N_{2}}{2\pi}Y^{2}dA^{2}+\frac{N_{3}}{2\pi}Y^{3}dA^{3}\\
+ & \frac{q_{133}}{\left(2\pi\right)^{2}}\left(d\chi^{1}A^{3}dA^{3}+A^{1}d\chi^{3}dA^{3}\right)\\
+ & \frac{p_{12,3}}{\left(2\pi\right)^{2}}\left[\left(A^{1}d\chi^{2}B^{3}+d\chi^{1}A^{2}B^{3}+d\chi^{1}d\chi^{2}B^{3}\right)\right.\\
+ & \left(A^{1}A^{2}dV^{3}+A^{1}d\chi^{2}dV^{3}+d\chi^{1}A^{2}dV^{3}\right)\\
+ & \left.\left(A^{1}A^{2}Y^{3}+A^{1}d\chi^{2}Y^{3}+d\chi^{1}A^{2}Y^{3}+d\chi^{1}d\chi^{2}Y^{3}\right)\right].
\end{alignedat}
\end{equation}
$\Delta S$ is expected to be boundary terms. However, we are going
to prove that this is impossible: the $d\chi^{1}d\chi^{2}B^{3}$ term
cannot be eliminated by subtraction nor be absorbed into a total derivative term. If $d\chi^{1}d\chi^{2}B^{3}$ can be eliminated by subtraction,
it is required that
\begin{equation}
\begin{alignedat}{1} & d\chi^{1}d\chi^{2}B^{3}+d\chi^{1}d\chi^{2}Y^{3}\\
= & d\chi^{1}d\chi^{2}\left(B^{3}-B^{3}+\cdots\right),
\end{alignedat}
\end{equation}
i.e.,
\begin{equation}
Y^{3}=-B^{3}+\cdots,
\end{equation}
but this is not allowed otherwise the gauge transformation of $B^{3}$
is ill-defined. If $d\chi^{1}d\chi^{2}B^{3}$ can be absorb into a
total derivative term, $\Delta S$ should (but in fact does not) contain a $-\chi^{1}d\chi^{2}dB^{3}$
term since
\begin{equation}
d\left(\chi^{1}d\chi^{2}B^{3}\right)=d\chi^{1}d\chi^{2}B^{3}-\chi^{1}d\chi^{2}dB^{3}.
\end{equation}
Therefore, the action (\ref{eq_appendix_action_A1A3dA3+A1A2B3}) is
impossible to be gauge invariant under gauge transformation~(\ref{eq_appendix_GT-A1A3dA3+A1A2B3-A3-normal}),
hence it is not a legitimate TQFT action.

Next, we assume that the gauge transformations are
\begin{equation}
\begin{alignedat}{1}A^{1,2}\rightarrow & A^{1,2}+d\chi^{1,2},\\
A^{3}\rightarrow & A^{3}+d\chi^{3}+X^{3},\\
B^{1,2}\rightarrow & B^{1,2}+dV^{1,2}+Y^{1,2},\\
B^{3}\rightarrow & B^{3}+dV^{3}.
\end{alignedat}
\label{eq_GT-A1A3dA3+A1A2B3-B3-normal-1-appendix}
\end{equation}
The $\mathbb{Z}_{N_{3}}$ cyclic group structure is encoded in
\begin{equation}
\oint B^{3}\in\frac{2\pi}{N_{3}}\mathbb{Z}_{N_{3}}.
\end{equation}
Under the gauge transformation, the variation of action is (boundary
terms neglected) is
\begin{equation}
\begin{alignedat}{1}\Delta S= & \int\frac{N_{1}}{2\pi}Y^{1}dA^{1}+\frac{N_{2}}{2\pi}Y^{2}dA^{2}+\frac{N_{3}}{2\pi}B^{3}dX^{3}\\
 & +\frac{q_{133}}{\left(2\pi\right)^{2}}\left(d\chi^{1}A^{3}dA^{3}+A^{1}d\chi^{3}dA^{3}\right)\\
 & +\frac{q_{133}}{\left(2\pi\right)^{2}}\left(A^{1}X^{3}dA^{3}+d\chi^{1}X^{3}dA^{3}\right)\\
 & +\frac{q_{133}}{\left(2\pi\right)^{2}}\left(A^{1}A^{3}dX^{3}+d\chi^{1}A^{3}dX^{3}+A^{1}d\chi^{3}dX^{3}\right)\\
 & +\frac{q_{133}}{\left(2\pi\right)^{2}}\left(A^{1}X^{3}dX^{3}+d\chi^{1}X^{3}dX^{3}\right)\\
 & +\frac{p_{12,3}}{\left(2\pi\right)^{2}}\left(d\chi^{1}A^{2}B^{3}+A^{1}d\chi^{2}B^{3}+d\chi^{1}d\chi^{2}B^{3}\right)\\
 & +\frac{p_{12,3}}{\left(2\pi\right)^{2}}\left(A^{1}A^{2}dV^{3}+d\chi^{1}A^{2}dV^{3}+A^{1}d\chi^{2}dV^{3}\right).
\end{alignedat}
\end{equation}
Similarly, $\Delta S$ is impossible to be boundary terms since the
term $d\chi^{1}A^{3}dA^{3}$ cannot be eliminated by subtraction nor
be absorbed into a total derivative term. The only way to eliminate
$d\chi^{1}A^{3}dA^{3}$ by subtraction is to require that
\begin{equation}
d\chi^{1}A^{3}dA^{3}+d\chi^{1}X^{3}dA^{3}=0,
\end{equation}
which means that
\begin{equation}
X^{3}=-A^{3}+\cdots.
\end{equation}
If so, the gauge transformation of $A^{3}$ is
\begin{equation}
A^{3}\rightarrow d\chi^{3}+\cdots
\end{equation}
which is not well-defined. If $d\chi^{1}A^{3}dA^{3}$
can be absorbed into a total derivative term, $\Delta S$ should (but in fact does not) contain
a $\chi^{1}dA^{3}dA^{3}$ term since
\begin{equation}
d\left(\chi^{1}A^{3}dA^{3}\right)=d\chi^{1}A^{3}dA^{3}+\chi^{1}dA^{3}dA^{3},
\end{equation}
Therefore, (\ref{eq_GT-A1A3dA3+A1A2B3-B3-normal-1-appendix}) cannot be the
proper gauge transformations for (\ref{eq_action-A1A3dA3+A1A2B3}).

So far, we have seen that if the $\mathbb{\mathbb{\mathbb{Z}}}_{N_{3}}$ cyclic group structure is respected, we cannot find a proper set of gauge transformations
under which the action~(\ref{eq_action-A1A3dA3+A1A2B3}) is invariant up to boundary
terms. If the action is imposed to be gauge invariant up to boundary terms, the
$\mathbb{Z}_{N_{3}}$ cyclic group structure would be violated. This dilemma reveals
that the action~(\ref{eq_action-A1A3dA3+A1A2B3}) is not a legitimate TQFT theory.
In other words, $A^{1}A^{3}dA^{3}$ is incompatible with $A^{1}A^{2}B^{3}$, i.e., $\mathsf{\Theta}_{3,3|1}^{\text{3L}}$
is incompatible with $\mathsf{\Theta}_{1,2|3}^{\text{BR}}$.

\subsection{Derivation of gauge transformations for $S=\int\sum_{i=1}^{3}\frac{N_{i}}{2\pi}B^{i}dA^{i}+\frac{q_{123}}{\left(2\pi\right)^{2}}A^{1}A^{2}dA^{3}+\frac{p_{12,3}}{\left(2\pi\right)^{2}}A^{1}A^{2}B^{3}$ \label{subsec_Derivation-GT-A1A2dA3+A1A2B3}}
The action is
\begin{equation}
S=\int\sum_{i=1}^{3}\frac{N_{i}}{2\pi}B^{i}dA^{i}+\frac{q_{123}}{\left(2\pi\right)^{2}}A^{1}A^{2}dA^{3}+\frac{p_{12,3}}{\left(2\pi\right)^{2}}A^{1}A^{2}B^{3}
\end{equation}

First, we assume $A^{3}\rightarrow A^{3}+d\chi^{3}$. The test gauge
transformations are
\begin{equation}
\begin{alignedat}{1}A^{i}\rightarrow & A^{i}+d\chi^{i},\\
B^{i}\rightarrow & B^{i}+dV^{i}+Y^{i}.
\end{alignedat}
\label{eq_test-GT-A1A2dA3+A1A2B3-A3-normal}
\end{equation}
Under (\ref{eq_test-GT-A1A2dA3+A1A2B3-A3-normal}), the variation
of action is (boundary terms neglected)
\begin{equation}
\begin{alignedat}{1}\Delta S= & \int\sum_{i=1}^{3}\frac{N_{i}}{2\pi}Y^{i}dA^{i}\\
 & +\frac{q_{123}}{\left(2\pi\right)^{2}}\left(A^{1}d\chi^{2}dA^{3}+d\chi^{1}A^{2}dA^{3}+d\chi^{1}d\chi^{2}dA^{3}\right)\\
 & +\frac{p_{12,3}}{\left(2\pi\right)^{2}}\left[\left(d\chi^{1}A^{2}B^{3}+A^{1}d\chi^{2}B^{3}+d\chi^{1}d\chi^{2}B^{3}\right)\right.\\
 & +\left(A^{1}A^{2}dV^{3}+d\chi^{1}A^{2}dV^{3}+A^{1}d\chi^{2}dV^{3}\right)\\
 & +\left.\left(A^{1}A^{2}Y^{3}+d\chi^{1}A^{2}Y^{3}+A^{1}d\chi^{2}Y^{3}+d\chi^{1}d\chi^{2}Y^{3}\right)\right].
\end{alignedat}
\end{equation}
Notice that the term $d\chi^{1}d\chi^{2}B^{3}$ cannot be eliminated by subtraction
nor absorbed into a total derivative term. Therefore, action (\ref{eq_action-A1A2dA3+A1A2B3})
is not gauge invariant under transformation (\ref{eq_test-GT-A1A2dA3+A1A2B3-A3-normal}).

Second, we assume $B^{3}\rightarrow B^{3}+dV^{3}$. The test gauge transformations are
\begin{equation}
\label{eq_test-GT-A1A2dA3+A1A2B3-B3-normal}
\begin{alignedat}{1}A^{1,2}\rightarrow & A^{1,2}+d\chi^{1,2}\\
A^{3}\rightarrow & A^{3}+d\chi^{3}+X^{3}\\
B^{1,2}\rightarrow & B^{1,2}+dV^{1,2}+Y^{1,2}\\
B^{3}\rightarrow & B^{3}+dV^{3}.
\end{alignedat}
\end{equation}
Under~(\ref{eq_test-GT-A1A2dA3+A1A2B3-B3-normal}), the variation of action is (boundary terms
neglected)
\begin{equation}
\begin{alignedat}{1}\Delta S= & \int\frac{N_{1}}{2\pi}Y^{1}dA^{1}+\frac{N_{2}}{2\pi}Y^{2}dA^{2}+\frac{N_{3}}{2\pi}B^{3}dX^{3}\\
 & +\frac{q_{123}}{\left(2\pi\right)^{2}}\left(\underbrace{A^{1}d\chi^{2}dA^{3}+d\chi^{1}A^{2}dA^{3}+d\chi^{1}d\chi^{2}dA^{3}}_{M}\right)\\
 & +\underbrace{\frac{q_{123}}{\left(2\pi\right)^{2}}\left(A^{1}A^{2}dX^{3}+A^{1}d\chi^{2}dX^{3}+d\chi^{1}A^{2}dX^{3}+d\chi^{1}d\chi^{2}dX^{3}\right)}_{N}\\
 & +\frac{p_{12,3}}{\left(2\pi\right)^{2}}\left(\underbrace{d\chi^{1}A^{2}B^{3}+A^{1}d\chi^{2}B^{3}+d\chi^{1}d\chi^{2}B^{3}}_{O}\right)\\
 & +\frac{p_{12,3}}{\left(2\pi\right)^{2}}\left(\underbrace{A^{1}A^{2}dV^{3}+d\chi^{1}A^{2}dV^{3}+A^{1}d\chi^{2}dV^{3}}_{P}\right).
\end{alignedat}
\end{equation}
We expect that $\Delta S$ is an integral of total derivative terms. For this purpose, we need to properly construct the shift terms, i.e., $X^{3}$, $Y^{1}$ and $Y^{2}$, such that $\Delta S$ is zero up to boundary terms.

For the $M$ term, notice that
\begin{equation}\label{eq_GT_A1A2dA3+A1A2B3_M_term}
\begin{alignedat}{1} & A^{1}d\chi^{2}dA^{3}+d\chi^{1}A^{2}dA^{3}+d\chi^{2}A^{3}dA^{1}-d\chi^{1}A^{3}dA^{2}\\
= & d\chi^{2}d\left(A^{1}A^{3}\right)+d\chi^{1}d\left(A^{3}A^{2}\right),
\end{alignedat}
\end{equation}
thus we can let
\begin{equation}
Y^{1}=\cdots+\frac{q_{123}}{\left(2\pi\right)^{2}}\frac{2\pi}{N_{1}}d\chi^{2}A^{3}+\cdots\label{eq_shift-Y1-M}
\end{equation}
and
\begin{equation}
Y^{2}=\cdots+\frac{q_{123}}{\left(2\pi\right)^{2}}\frac{2\pi}{N_{2}}\left(-d\chi^{1}A^{3}\right)+\cdots\label{eq_shift-Y2-M}
\end{equation}
in order to construct the total derivative terms in Eq.~(\ref{eq_GT_A1A2dA3+A1A2B3_M_term}).

For the $O$ term, it cannot be absorbed into a total derivative term
since there is no terms containing $dB^{3}$ in $\Delta S$. Therefore we have to eliminate the $O$ term by subtraction. To do this, one can assume
that
\begin{equation}
\frac{N_{3}}{2\pi}B^{3}dX^{3}+\frac{p_{12,3}}{\left(2\pi\right)^{2}}\left(d\chi^{1}A^{2}B^{3}+A^{1}d\chi^{2}B^{3}+d\chi^{1}d\chi^{2}B^{3}\right)=0
\end{equation}
which leads to
\begin{equation}
dX^{3}=\frac{p_{12,3}}{\left(2\pi\right)^{2}}\frac{2\pi}{N_{3}}\left(-d\chi^{1}A^{2}-A^{1}d\chi^{2}-d\chi^{1}d\chi^{2}\right).\label{eq_dX3-condition-1}
\end{equation}
However, we cannot find a $X^{3}$ satisfying Eq. (\ref{eq_dX3-condition-1}).
Alternatively, we can add terms containing $dA^{1}$ or $dA^{2}$. We can
let
\begin{equation}
X=\frac{p_{12,3}}{\left(2\pi\right)^{2}}\frac{2\pi}{N_{3}}\left(-\chi^{1}A^{2}+A^{1}\chi^{2}-\chi^{1}d\chi^{2}\right)
\end{equation}
hence
\begin{equation}\label{eq_GT_A1A2dA3+A1A2B3_dX3_term}
dX^{3}=\frac{p_{12,3}}{\left(2\pi\right)^{2}}\frac{2\pi}{N_{3}}\left(-d\chi^{1}A^{2}-\chi^{1}dA^{2}+dA^{1}\chi^{2}-A^{1}d\chi^{2}-d\chi^{1}d\chi^{2}\right).
\end{equation}
Then we notice that
\begin{equation}
\begin{alignedat}{1}0= & \frac{N_{3}}{2\pi}B^{3}dX^{3}+\frac{p_{12,3}}{\left(2\pi\right)^{2}}\left(d\chi^{1}A^{2}B^{3}+A^{1}d\chi^{2}B^{3}+d\chi^{1}d\chi^{2}B^{3}\right)\\
 & +\frac{p_{12,3}}{\left(2\pi\right)^{2}}\left(\chi^{1}B^{3}dA^{2}-\chi^{2}B^{3}dA^{1}\right)
\end{alignedat}
\label{eq_dX3-total derivative}
\end{equation}
Therefore, we can let
\begin{equation}
Y^{1}=\cdots+\frac{p_{12,3}}{\left(2\pi\right)^{2}}\frac{2\pi}{N_{1}}\left(-\chi^{2}B^{3}\right)+\cdots\label{eq_shift-Y1-O}
\end{equation}
and
\begin{equation}
Y^{2}=\cdots+\frac{p_{12,3}}{\left(2\pi\right)^{2}}\frac{2\pi}{N_{2}}\left(\chi^{1}B^{3}\right)+\cdots\label{eq_shift-Y2-O}
\end{equation}
in order to contribute the $\frac{p_{12,3}}{\left(2\pi\right)^{2}}\left(\chi^{1}B^{3}dA^{2}-\chi^{2}B^{3}dA^{1}\right)$ term in Eq.~(\ref{eq_dX3-total derivative}).

The $N$ term, with $dX^{3}$ known in Eq.~(\ref{eq_GT_A1A2dA3+A1A2B3_dX3_term}), is (the underlines and underbraces are used to identify terms between equal signs)
\begin{equation}
\begin{alignedat}{1}N= & \frac{q_{123}}{\left(2\pi\right)^{2}}\left(A^{1}A^{2}+A^{1}d\chi^{2}+d\chi^{1}A^{2}+d\chi^{1}d\chi^{2}\right)dX^{3}\\
= & \frac{q_{123}}{\left(2\pi\right)^{2}}\frac{p_{12,3}}{\left(2\pi\right)^{2}}\frac{2\pi}{N_{3}}\left(\underbrace{A^{1}A^{2}}_{1st}+\underbrace{A^{1}d\chi^{2}}_{2nd}+\underbrace{d\chi^{1}A^{2}}_{3rd}+\underbrace{d\chi^{1}d\chi^{2}}_{4th}\right)\left(-A^{1}d\chi^{2}-d\chi^{1}A^{2}-\chi^{1}dA^{2}+\chi^{2}dA^{1}-d\chi^{1}d\chi^{2}\right)\\
= & \frac{q_{123}}{\left(2\pi\right)^{2}}\frac{p_{12,3}}{\left(2\pi\right)^{2}}\frac{2\pi}{N_{3}}\left[\left(\underbrace{\underline{-A^{1}A^{2}\chi^{1}dA^{2}}+A^{1}A^{2}\chi^{2}dA^{1}-A^{1}A^{2}d\chi^{1}d\chi^{2}}_{1st}\right)+\left(\underbrace{-A^{1}d\chi^{2}d\chi^{1}A^{2}\underline{-A^{1}d\chi^{2}\chi^{1}dA^{2}}+A^{1}d\chi^{2}\chi^{2}dA^{1}}_{2nd}\right)\right.\\
 & \left.+\left(\underbrace{-d\chi^{1}A^{2}A^{1}d\chi^{2}\underline{-d\chi^{1}A^{2}\chi^{1}dA^{2}}+d\chi^{1}A^{2}\chi^{2}dA^{1}}_{3rd}\right)+\left(\underbrace{\underline{-d\chi^{1}d\chi^{2}\chi^{1}dA^{2}}+d\chi^{1}d\chi^{2}\chi^{2}dA^{1}}_{4th}\right)\right]\\
= & \frac{q_{123}}{\left(2\pi\right)^{2}}\frac{p_{12,3}}{\left(2\pi\right)^{2}}\frac{2\pi}{N_{3}}\left[-A^{1}d\chi^{2}d\chi^{1}A^{2}+\left(A^{1}A^{2}\chi^{2}+A^{1}d\chi^{2}\chi^{2}+d\chi^{1}A^{2}\chi^{2}+d\chi^{1}d\chi^{2}\chi^{2}\right)dA^{1}\right.\\
 & \left.+\underline{\left(-A^{1}A^{2}\chi^{1}-A^{1}d\chi^{2}\chi^{1}-d\chi^{1}A^{2}\chi^{1}-d\chi^{1}d\chi^{2}\chi^{1}\right)dA^{2}}\right].
\end{alignedat}
\end{equation}
Notice that
\begin{equation}
-A^{1}d\chi^{2}d\chi^{1}A^{2}=A^{1}A^{2}d\chi^{1}d\chi^{2}
\end{equation}
and
\begin{equation}
\begin{alignedat}{1} & A^{1}A^{2}d\chi^{1}d\chi^{2}+\chi^{1}A^{2}d\chi^{2}dA^{1}-d\left(A^{1}\chi^{1}\right)\chi^{2}dA^{2}\\
= & -A^{1}d\chi^{1}A^{2}d\chi^{2}+dA^{1}\chi^{1}A^{2}d\chi^{2}-d\left(A^{1}\chi^{1}\right)\chi^{2}dA^{2}\\
= & d\left(A^{1}\chi^{1}\right)A^{2}d\chi^{2}-d\left(A^{1}\chi^{1}\right)\chi^{2}dA^{2}\\
= & -d\left(A^{1}\chi^{1}\right)d\left(A^{2}\chi^{2}\right),
\end{alignedat}
\end{equation}
we could  have $\frac{N_{1}}{2\pi}Y^{1}dA^{1}+\frac{N_{2}}{2\pi}Y^{2}dA^{2}+\text{term }N=\text{total derivative terms}$, if
\begin{equation}
Y^{1}=\cdots+\frac{q_{123}}{\left(2\pi\right)^{2}}\frac{p_{12,3}}{\left(2\pi\right)^{2}}\frac{2\pi}{N_{3}}\frac{2\pi}{N_{1}}\left[\chi^{1}A^{2}d\chi^{2}-\left(A^{1}A^{2}\chi^{2}+A^{1}d\chi^{2}\chi^{2}+d\chi^{1}A^{2}\chi^{2}+d\chi^{1}d\chi^{2}\chi^{2}\right)\right]+\cdots\label{eq_shift-Y1-N}
\end{equation}
and
\begin{equation}
Y^{2}=\cdots+\frac{q_{123}}{\left(2\pi\right)^{2}}\frac{p_{12,3}}{\left(2\pi\right)^{2}}\frac{2\pi}{N_{3}}\frac{2\pi}{N_{2}}\left[-d\left(A^{1}\chi^{1}\right)\chi^{2}+\left(A^{1}A^{2}\chi^{1}+A^{1}d\chi^{2}\chi^{1}+d\chi^{1}A^{2}\chi^{1}+d\chi^{1}d\chi^{2}\chi^{1}\right)\right]+\cdots.\label{eq_shift-Y2-N}
\end{equation}

For the $P$ term, notice that
\begin{equation}
\begin{alignedat}{1} & d\left(A^{1}A^{2}V^{3}\right)+d\left(\chi^{1}A^{2}dV^{3}\right)-d\left(A^{1}\chi^{2}dV^{3}\right)\\
= & dA^{1}A^{2}V^{3}-A^{1}dA^{2}V^{3}+\chi^{1}dA^{2}dV^{3}-dA^{1}\chi^{2}dV^{3}+\underbrace{A^{1}A^{2}dV^{3}+d\chi^{1}A^{2}dV^{3}+A^{1}d\chi^{2}dV^{3}}_{P},
\end{alignedat}
\label{eq_P-term-total derivative}
\end{equation}
thus we can let
\begin{equation}
Y^{1}=\cdots+\frac{p_{12,3}}{\left(2\pi\right)^{2}}\frac{2\pi}{N_{1}}\left(A^{2}V^{3}-\chi^{2}dV^{3}\right)+\cdots\label{eq_shift-Y1-P}
\end{equation}
 and
\begin{equation}
Y^{2}=\cdots+\frac{p_{12,3}}{\left(2\pi\right)^{2}}\frac{2\pi}{N_{2}}\left(-A^{1}V^{3}+\chi^{1}dV^{3}\right)+\cdots\label{eq_shift-Y2-P}
\end{equation}
in order to provide the $\left(dA^{1}A^{2}V^{3}-A^{1}dA^{2}V^{3}+\chi^{1}dA^{2}dV^{3}-dA^{1}\chi^{2}dV^{3}\right)$ term in Eq.~(\ref{eq_P-term-total derivative}).

According to Eqs.~(\ref{eq_shift-Y1-M}), (\ref{eq_shift-Y2-M}),
(\ref{eq_shift-Y1-O}), (\ref{eq_shift-Y2-O}), (\ref{eq_shift-Y1-N}),
(\ref{eq_shift-Y2-N}), (\ref{eq_shift-Y1-P}) and (\ref{eq_shift-Y2-P})
, $\Delta S=\int\left(\text{total derivative terms}\right)$ or $S$ is invariant up to total derivative terms under transformation
\begin{equation}
\begin{alignedat}{1}A^{1}\rightarrow & A^{1}+d\chi^{1},\\
A^{2}\rightarrow & A^{2}+d\chi^{2},\\
A^{3}\rightarrow & A^{3}+d\chi^{3}+X^{3},\\
B^{1}\rightarrow & B^{1}+dV^{1}+Y^{1},\\
B^{2}\rightarrow & B^{2}+dV^{2}+Y^{2},\\
B^{3}\rightarrow & B^{3}+dV^{3};
\end{alignedat}
\end{equation}
where

\begin{equation}
\begin{alignedat}{1}X^{3}= & -\frac{p_{12,3}}{\left(2\pi\right)N_{3}}\left(\chi^{1}A^{2}+\frac{1}{2}\chi^{1}d\chi^{2}\right)+\frac{p_{12,3}}{\left(2\pi\right)N_{3}}\left(\chi^{2}A^{1}+\frac{1}{2}\chi^{2}d\chi^{1}\right),\\
Y^{1}= & \frac{q_{123}}{2\pi N_{1}}d\chi^{2}A^{3}-\frac{p_{12,3}}{\left(2\pi\right)N_{1}}\left(\chi^{2}B^{3}-A^{2}V^{3}+\chi^{2}dV^{3}\right)\\
 & +\frac{q_{123}}{\left(2\pi\right)^{2}}\cdot\frac{p_{12,3}}{\left(2\pi\right)N_{3}}\cdot\frac{2\pi}{N_{1}}\cdot\left[\chi^{1}A^{2}d\chi^{2}-\left(A^{1}A^{2}\chi^{2}+A^{1}d\chi^{2}\chi^{2}+d\chi^{1}A^{2}\chi^{2}+d\chi^{1}d\chi^{2}\chi^{2}\right)\right],\\
Y^{2}= & -\frac{q_{123}}{2\pi N_{2}}d\chi^{1}A^{3}+\frac{p_{12,3}}{\left(2\pi\right)N_{2}}\left(\chi^{1}B^{3}-A^{1}V^{3}+\chi^{1}dV^{3}\right)\\
 & +\frac{q_{123}}{\left(2\pi\right)^{2}}\cdot\frac{p_{12,3}}{\left(2\pi\right)N_{3}}\cdot\frac{2\pi}{N_{2}}\cdot\left[-d\left(A^{1}\chi^{1}\right)\chi^{2}+\left(A^{1}A^{2}\chi^{1}+A^{1}d\chi^{2}\chi^{1}+d\chi^{1}A^{2}\chi^{1}+d\chi^{1}d\chi^{2}\chi^{1}\right)\right].
\end{alignedat}
\end{equation}

\subsection{Derivation of incompatibility between $A^{1}A^{2}A^{3}A^{4}$ and
$A^{1}A^{2}B^{4}$\label{subsec_derivation_incompatibility-A1A2A3A4_A1A2B4}}

If we assume that $A^{1}A^{2}A^{3}A^{4}$ are compatible with $A^{1}A^{2}B^{4}$,
the TQFT action should be
\begin{equation}
S=\int\sum_{i=1}^{4}\frac{N_{i}}{2\pi}B^{i}dA^{i}+\frac{q_{1234}}{\left(2\pi\right)^{3}}A^{1}A^{2}A^{3}A^{4}+\frac{p_{12,4}}{\left(2\pi\right)^{2}}A^{1}A^{2}B^{4}.\label{eq_action_appendix_A1A2A3A4+A1A2B3}
\end{equation}
In the action (\ref{eq_action_appendix_A1A2A3A4+A1A2B3}), $B^{1}$,
$B^{2}$ and $B^{3}$ serve as Lagrange multipliers, imposing $dA^{1}=dA^{2}=dA^{3}=0$,
i.e., the gauge transformations of $A^{1}$, $A^{2}$ and $A^{3}$
are $A^{1,2,3}\rightarrow A^{1,2,3}+d\chi^{1,2,3}$. The $\mathbb{Z}_{N_{1}}$,
$\mathbb{Z}_{N_{2}}$ and $\mathbb{Z}_{N_{3}}$ cyclic group structures
are encoded in $\oint A^{I}\in\frac{2\pi}{N_{I}}\mathbb{Z}_{N_{I}}$,
where $I=1,2,3$. The remaining $\mathbb{Z}_{N_{4}}$ cyclic group structure
can be encoded in $\oint A^{4}\in\frac{2\pi}{N_{4}}\mathbb{Z}_{N_{4}}$
or $\oint B^{4}\in\frac{2\pi}{N_{4}}\mathbb{Z}_{N_{4}}$, corresponding
to $A^{4}\rightarrow A^{4}+d\chi^{4}$ or $B^{4}\rightarrow B^{4}+dV^{4}$
respectively. In the following text, we are going to examine both cases.However, we
will find that neither of them would result in gauge transformations
under which the action (\ref{eq_action_appendix_A1A2A3A4+A1A2B3})
is gauge invariant up to boundary terms.

First, we assume that the gauge transformations are
\begin{equation}
\begin{alignedat}{1}A^{1}\rightarrow & A^{1}+d\chi^{1},\\
A^{2}\rightarrow & A^{2}+d\chi^{2},\\
A^{3}\rightarrow & A^{3}+d\chi^{3},\\
A^{4}\rightarrow & A^{4}+d\chi^{4},\\
B^{1}\rightarrow & B^{1}+dV^{1}+Y^{1},\\
B^{2}\rightarrow & B^{2}+dV^{2}+Y^{2},\\
B^{3}\rightarrow & B^{3}+dV^{3}+Y^{3},\\
B^{4}\rightarrow & B^{4}+dV^{4}+Y^{4}.
\end{alignedat}
\label{eq_GT_A1A2A3A4+A1A2B3-A3-normal}
\end{equation}
Under (\ref{eq_GT_A1A2A3A4+A1A2B3-A3-normal}), the variation of
action is (boundary terms neglected)
\begin{equation}
\begin{alignedat}{1}\Delta S= & \int\sum_{i=1}^{4}\frac{N_{i}}{2\pi}Y^{i}dA^{i}\\
 & +\frac{q_{1234}}{\left(2\pi\right)^{3}}\left(d\chi^{1}A^{2}A^{3}A^{4}+A^{1}d\chi^{2}A^{3}A^{4}+A^{1}A^{2}d\chi^{3}A^{4}+A^{1}A^{2}A^{3}d\chi^{4}\right.\\
 & +d\chi^{1}d\chi^{2}A^{3}A^{4}+d\chi^{1}A^{2}d\chi^{3}A^{4}+A^{1}d\chi^{2}d\chi^{3}A^{4}+d\chi^{1}A^{2}A^{3}d\chi^{4}+A^{1}d\chi^{2}A^{3}d\chi^{4}+A^{1}A^{2}d\chi^{3}d\chi^{4}\\
 & +A^{1}d\chi^{2}d\chi^{3}d\chi^{4}+d\chi^{1}A^{2}d\chi^{3}d\chi^{4}+d\chi^{1}d\chi^{2}A^{3}d\chi^{4}+d\chi^{1}d\chi^{2}d\chi^{3}A^{4}\\
 & +\left.d\chi^{1}d\chi^{2}d\chi^{3}d\chi^{4}\right)\\
 & +\frac{p_{12,4}}{\left(2\pi\right)^{2}}\left(A^{1}d\chi^{2}B^{4}+d\chi^{1}A^{2}B^{4}+d\chi^{1}d\chi^{2}B^{4}\right)\\
 & +\frac{p_{12,4}}{\left(2\pi\right)^{2}}\left(A^{1}A^{2}dV^{4}+A^{1}d\chi^{2}dV^{4}+d\chi^{1}A^{2}dV^{4}+d\chi^{1}d\chi^{2}dV^{4}\right)\\
 & +\frac{p_{12,4}}{\left(2\pi\right)^{2}}\left(A^{1}A^{2}Y^{4}+A^{1}d\chi^{2}Y^{4}+d\chi^{1}A^{2}Y^{4}+d\chi^{1}d\chi^{2}Y^{4}\right).
\end{alignedat}
\end{equation}
We recognize that $d\chi^{1}d\chi^{2}B^{4}$ is a stubborn term, i.e.,
it cannot be eliminated by substraction or be absorbed into a total
derivative term. Due to the existence of stubborn term, $\Delta S$
cannot be written as an integral of total derivative terms.
Therefore, the action~(\ref{eq_action_appendix_A1A2A3A4+A1A2B3})
is not gauge invariant up to boundary terms under (\ref{eq_GT_A1A2A3A4+A1A2B3-A3-normal}).

Next, we assume that the gauge transformations are
\begin{equation}
\begin{alignedat}{1}A^{1}\rightarrow & A^{1}+d\chi^{1},\\
A^{2}\rightarrow & A^{2}+d\chi^{2},\\
A^{3}\rightarrow & A^{3}+d\chi^{3},\\
A^{4}\rightarrow & A^{4}+d\chi^{4}+X^{4},\\
B^{1}\rightarrow & B^{1}+dV^{1}+Y^{1},\\
B^{2}\rightarrow & B^{2}+dV^{2}+Y^{2},\\
B^{3}\rightarrow & B^{3}+dV^{3}+Y^{3},\\
B^{4}\rightarrow & B^{4}+dV^{4}.
\end{alignedat}
\label{eq_GT_A1A2A3A4+A1A2B3-B3-normal}
\end{equation}
Under (\ref{eq_GT_A1A2A3A4+A1A2B3-A3-normal}), the variation of
action is (boundary terms neglected)
\begin{equation}
\begin{alignedat}{1}\Delta S= & \int\sum_{i=1}^{3}\frac{N_{i}}{2\pi}Y^{i}dA^{i}+\frac{N_{4}}{2\pi}B^{4}dX^{4}\\
 & +\frac{q_{1234}}{\left(2\pi\right)^{3}}\left(d\chi^{1}A^{2}A^{3}A^{4}+A^{1}d\chi^{2}A^{3}A^{4}+A^{1}A^{2}d\chi^{3}A^{4}+A^{1}A^{2}A^{3}d\chi^{4}\right.\\
 & +d\chi^{1}d\chi^{2}A^{3}A^{4}+d\chi^{1}A^{2}d\chi^{3}A^{4}+A^{1}d\chi^{2}d\chi^{3}A^{4}+d\chi^{1}A^{2}A^{3}d\chi^{4}+A^{1}d\chi^{2}A^{3}d\chi^{4}+A^{1}A^{2}d\chi^{3}d\chi^{4}\\
 & +\left.A^{1}d\chi^{2}d\chi^{3}d\chi^{4}+d\chi^{1}A^{2}d\chi^{3}d\chi^{4}+d\chi^{1}d\chi^{2}A^{3}d\chi^{4}+d\chi^{1}d\chi^{2}d\chi^{3}A^{4}\right)\\
 & +\frac{q_{1234}}{\left(2\pi\right)^{3}}\left(A^{1}A^{2}A^{3}X^{4}+A^{1}d\chi^{2}A^{3}X^{4}+d\chi^{1}A^{2}A^{3}X^{4}+d\chi^{1}d\chi^{2}A^{3}X^{4}\right)\\
 & +\frac{q_{1234}}{\left(2\pi\right)^{3}}\left(A^{1}A^{2}d\chi^{3}X^{4}+A^{1}d\chi^{2}d\chi^{3}X^{4}+d\chi^{1}A^{2}d\chi^{3}X^{4}+d\chi^{1}d\chi^{2}d\chi^{3}X^{4}\right)\\
 & +\frac{p_{12,4}}{\left(2\pi\right)^{2}}\left(A^{1}d\chi^{2}B^{4}+d\chi^{1}A^{2}B^{4}+d\chi^{1}d\chi^{2}B^{4}\right)\\
 & +\frac{p_{12,4}}{\left(2\pi\right)^{2}}\left(A^{1}A^{2}dV^{3}+A^{1}d\chi^{2}dV^{3}+d\chi^{1}A^{2}dV^{3}\right)
\end{alignedat}
\end{equation}
We recognize that $d\chi^{1}d\chi^{2}d\chi^{3}A^{4}$ cannot be eliminated
by subtraction or be absorbed into a total derivative term. If we
want to eliminate $d\chi^{1}d\chi^{2}d\chi^{3}A^{4}$ by subtraction,
we could only expect that
\begin{equation}
d\chi^{1}d\chi^{2}d\chi^{3}A^{4}+d\chi^{1}d\chi^{2}d\chi^{3}X^{4}=d\chi^{1}d\chi^{2}d\chi^{3}\left(A^{4}-A^{4}+\cdots\right)
\end{equation}
which means that $X^{4}=-A^{4}+\cdots$, hence the gauge transformation
of $A^{4}$ is ill-defined:
\begin{equation}
A^{4}\rightarrow d\chi^{4}+\cdots.
\end{equation}
If we want to absorb $d\chi^{1}d\chi^{2}d\chi^{3}A^{4}$ into a total
derivative term, we need a $\chi^{1}d\chi^{2}d\chi^{3}dA^{4}$ term
in $\Delta S$ since
\begin{equation}
d\left(\chi^{1}d\chi^{2}d\chi^{3}A^{4}\right)=d\chi^{1}d\chi^{2}d\chi^{3}A^{4}+\chi^{1}d\chi^{2}d\chi^{3}dA^{4}.
\end{equation}
However, this $\chi^{1}d\chi^{2}d\chi^{3}dA^{4}$ term does not exist in
$\Delta S$. Therefore, due to the ``stubborn term'' $d\chi^{1}d\chi^{2}d\chi^{3}A^{4}$, the action~(\ref{eq_action_appendix_A1A2A3A4+A1A2B3}) is not
gauge invariant up to boundary terms under~(\ref{eq_GT_A1A2A3A4+A1A2B3-A3-normal}).

Finally, we can conclude that the action (\ref{eq_action_appendix_A1A2A3A4+A1A2B3})
could never be gauge invariant up to boundary terms under gauge transformations
which respect $\mathbb{Z}_{N_{i}}$ cyclic group structures, hence it is
not a legitimate TQFT action. Therefore $A^{1}A^{2}A^{3}A^{4}$ $\left(\mathsf{\Theta}_{1,2,3,4}^{\text{4L}}\right)$ is incompatible with
$A^{1}A^{2}B^{4}$ $\left(\mathsf{\Theta}_{1,2|4}^{\text{BR}}\right)$.

\twocolumngrid
%%\bibliography{3dbaa}
%%%%\bibliographystyle{abbrv}

\end{document}